\documentclass[traditabstract]{aa}

\usepackage{graphicx}
\pdfoutput=1
\usepackage{amsmath}
\usepackage{txfonts}
\usepackage{natbib}
\usepackage[usenames,dvipsnames]{color}
\usepackage{enumerate}
\usepackage{longtable, ltcaption}
\usepackage{lscape}
\definecolor{grey}{rgb}{0.4,0.4,0.4}
\definecolor{purple}{RGB}{139,0,139}
\definecolor{lightblue}{RGB}{0,0,255}

\newcommand{\Msun}{M_{\odot}}
\newcommand{\Rsun}{\mathrm{R}_{\odot}}

\newcommand{\Mprim}{M_{1,\mathrm{i}}}

\newcommand{\Msec}{M_{2,\mathrm{i}}}
\newcommand{\Mpmz}{M_{\mathrm{PMZ}}}

\newcommand{\Pin}{P_{\mathrm{i}}}
\newcommand{\Pf}{P_{\mathrm{f}}}
\newcommand{\Porb}{P_{\mathrm{orb}}}
\newcommand{\logg}{\log g}
\newcommand{\loggunits}{\log_{10}(g/\mathrm{cm}\,\mathrm{s}^{-2})}
\newcommand{\Teff}{T_{\mathrm{eff}}}

\newcommand{\Macc}{M_{\mathrm{acc}}}

\newcommand{\Hy}{\mathrm{H}}
\newcommand{\C}{\mathrm{C}}

\newcommand{\Cth}{^{13}\mathrm{C}}

\newcommand{\Nfo}{^{14}\mathrm{N}}
\newcommand{\Ox}{\mathrm{O}}

\newcommand{\Ne}{\mathrm{Ne}}

\newcommand{\Na}{\mathrm{Na}}

\newcommand{\Fe}{\mathrm{Fe}}

\newcommand{\Sr}{\mathrm{Sr}}
\newcommand{\Zr}{\mathrm{Zr}}
\newcommand{\Nb}{\mathrm{Nb}}

\newcommand{\Ba}{\mathrm{Ba}}
\newcommand{\La}{\mathrm{La}}
\newcommand{\Ce}{\mathrm{Ce}}
\newcommand{\Pb}{\mathrm{Pb}}
\newcommand{\Eu}{\mathrm{Eu}}

\newcommand{\Y}{\mathrm{Y}}
\newcommand{\ls}{\mathrm{ls}}
\newcommand{\hs}{\mathrm{hs}}

\newcommand{\errb}{\sigma_{i,\,\mathrm{obs}}}
\newcommand{\avres}{\overline{R}}
\newcommand{\averr}{\overline{\sigma}_{\mathrm{obs}}}
\newcommand{\chisq}{\chi^2}

\newcommand{\chimin}{\chi^2_{\mathrm{min}}}

\begin{document}
   \title{Carbon-enhanced metal-poor stars: a window\\
   on AGB nucleosynthesis and binary evolution.}
   \subtitle{II. Statistical analysis of a sample of 67 CEMP-$s$ stars.}
   \titlerunning{Carbon-enhanced metal-poor stars: a window on AGB nucleosynthesis and binary evolution. II.}

   \author{C. Abate
          \inst{1,2}
          \and
          O. R. Pols \inst{1} 
          \and
          R. G. Izzard \inst{2,3}
		  \and
		  A. I. Karakas \inst{4}
          }

   \institute{Department of Astrophysics/IMAPP, Radboud University Nijmegen, P.O. Box 9010, 6500 GL Nijmegen, The Netherlands
         \and
             Argelander Institut f\"ur Astronomie, Auf dem H\"ugel 71, D-53121 Bonn, Germany\\
              \email{cabate@uni-bonn.de}
         \and
        Institute of Astronomy, Madingley Road, Cambridge CB3 0HA, United Kingdom
         \and
         Research School of Astronomy \& Astrophysics, Mount Stromlo Observatory, Weston Creek ACT 2611, Australia
             }

   \date{Received ...; accepted ...}
 
  \abstract
   {
   {Many of the carbon-enhanced metal-poor (CEMP) stars that we observe in the Galactic halo are found in binary systems and show enhanced abundances of elements produced by the slow neutron-capture process ($s$-elements). The origin of the peculiar chemical abundances of these CEMP-$s$ stars is believed to be accretion in the past of enriched material from a primary star in the asymptotic giant branch (AGB) phase of its evolution.%
   }
   {We investigate the mechanism of mass transfer and the process of nucleosynthesis in low-metallicity AGB stars by modelling the binary systems in which the observed CEMP-$s$ stars were formed.%
   }
   {For this purpose we compare a sample of $67$ CEMP-$s$ stars with a grid of binary stars generated by our binary evolution and nucleosynthesis model. We classify our sample CEMP-$s$ stars in three groups based on the observed abundance of europium. %
   In CEMP$-s/r$ stars the europium-to-iron ratio is more than ten times higher than in the Sun, whereas it is lower than this threshold in CEMP$-s/nr$ stars. No measurement of europium is currently available for CEMP-$s/ur$ stars.%
   }
   {On average our models reproduce well the abundances observed in CEMP-$s/nr$ stars, whereas in CEMP-$s/r$ stars and CEMP-$s/ur$ stars the abundances of the light-$s$ elements (strontium, yttrium, zirconium) are systematically overpredicted by our models and in CEMP-$s/r$ stars the abundances of the heavy-$s$ elements (barium, lanthanum) are underestimated. In all stars our modelled abundances of sodium overestimate the observations. This discrepancy is reduced only in models that underestimate the abundances of most of the $s$-elements. Furthermore, the abundance of lead is underpredicted in most of our model stars, independent of the metallicity. These results point to the limitations of our AGB nucleosynthesis model, particularly in the predictions of the element-to-element ratios. In our models CEMP-$s$ stars are typically formed in wide systems with periods above $10,\!000$ days, while most of the observed CEMP-$s$ stars are found in relatively close orbits with periods below $5,\!000$ days. This evidence suggests that either the sample of CEMP-$s$ binary stars with known orbital parameters is biased towards short periods, or that our wind mass-transfer model requires more efficient accretion in close orbits.
   }
   }

\nopagebreak   
\maketitle

\section{Introduction}
\label{intro}
The very metal-poor stars observed in the Galactic halo are of low mass and exhibit abundances of iron of approximately $[\mathrm{Fe}/\mathrm{H}]\lesssim-2.0$. Very metal-poor stars carry the fingerprints of the early stages of evolution of the Milky Way and therefore have been extensively studied by different surveys in the past two decades, with particular focus on chemically peculiar stars \cite[e.g.:][]{BeersAJ1992, Christlieb2001, Frebel2006, Yanny2009}.
Among these, carbon-enhanced metal-poor (CEMP) stars are very metal-poor stars enriched in carbon. The observed fraction of CEMP stars in the halo increases with increasing galactic latitude and with decreasing metallicity, and varies between approximately $9\%$ at $[\Fe/\Hy]<-2$ and about $25\%$ at $[\Fe/\Hy]<-3$ \cite[][]{Cohen2005, Marsteller2005, Frebel2006, Lucatello2006, Lee2013, Yong2013II}.

In the literature CEMP stars are generally defined by the carbon excess $[\C/\Fe]>1.0$, and are classified in groups according to the observed abundances of barium and europium, two heavy elements produced by the slow ($s$-) and the rapid ($r$-) neutron-capture process, respectively. The exact definitions vary between different authors \cite[e.g.][]{BeersChristlieb2005, Jonsell2006, Aoki2007, Masseron2010}, and in this work we adopt the following classification scheme.
\begin{itemize}
\item [$i.$] CEMP-$s$ stars are CEMP stars that satisfy the criteria $[\Ba/\Fe] > 0.5$ and $[\Ba/\Eu] > 0$. 
\item [$ii.$] CEMP-$s/r$ stars are CEMP-$s$ stars enriched in europium, i.e. $[\Eu/\Fe] > 1$.
\item [$iii.$] CEMP-$r$ stars have $[\Eu/\Fe] > 1$ and $[\Ba/\Eu] < 0$.
\item [$iv.$] CEMP-no stars do not exhibit enhanced abundance of barium, i.e. $[\Ba/\Fe] < 0.5$.
\end{itemize}
It has been suggested that CEMP stars owe their abundances to mass transfer of carbon-rich material in the past from a thermally-pulsing asymptotic giant branch (TP-AGB) primary star that today is an unseen white dwarf \cite[e.g.][]{Wallerstein1998, Preston2001, BeersChristlieb2005, Ryan2005}. 

In this work we focus on CEMP-$s$ stars, for which there are stronger arguments in favour of the binary mass-transfer scenario. Qualitatively, in this scenario the primary star evolves to the AGB phase, produces carbon and heavy elements and subsequently transfers part of this material by wind mass transfer to the low-mass companion star, the star observed today. A quantitative model of this process depends on many aspects of stellar evolution, AGB nucleosynthesis and binary interaction that are not well understood. In AGB stars many uncertainties are related to the physics of mixing, which determines the stellar structure and chemical composition and which in turn influence the radius of the star, and hence its luminosity and mass loss rate \cite[][]{Herwig2005, Constantino2014, Fishlock2014}. The material that is expelled by the AGB star carries away angular momentum from the system and its fate depends on the mass transfer mechanism. Therefore the study of CEMP-$s$ stars provides us with a powerful tool to improve our understanding of AGB nucleosynthesis at low metallicity and of the mass-transfer process in binary systems.

In our previous paper \cite[][hereinafter Paper~I]{Abate2015-2} we analyse a sample of 15 CEMP binary stars with known orbital periods: through the comparison with the observed abundances while matching the measured periods we put new constraints on our models of binary stellar evolution and AGB nucleosynthesis. In most of the systems a combination of large mass accretion and efficient angular momentum loss is necessary to reproduce at the same time the observed chemical abundances and orbital periods. About half of the stars of the sample are not accurately reproduced by any of our models, regardless of the assumptions made about the dynamical evolution of the systems, and this points to the limitations in our AGB nucleosynthesis model, particularly on the maximum abundances of the heavy $s$-process elements and on the element-to-element ratios that are produced. 

In this work we extend the analysis of Paper I to a larger sample of $67$ CEMP-$s$ stars that includes systems without information about the orbital period. We model a grid of about $400,\!000$ binary stars with different masses and separations and we compare the modelled abundances with the observations. For each star in the sample we determine the model star that best matches the observed abundances, with the same procedure of $\chisq$ minimization followed in Paper I. In every star for each observed element we compute the residual as the difference between the observed and the modelled abundance and subsequently we analyse the distributions of the residuals for every element individually. The purpose of this analysis is to investigate if statistically our model reproduces the observations, even though discrepancies may occur for some elements in individual stars, and thus to test our models with a study complementary to that of Paper I.

The paper is organised as follows. In Sect. \ref{data} we describe our observational sample, and in Sect. \ref{model} we present our model of binary stellar evolution and nucleosynthesis.
In Sect. \ref{VMP} we discuss the set of initial abundances adopted in our model. In Sect. \ref{CEMPs} we present the results of the analysis of our sample CEMP-$s$ stars, we compare the modelled and observed abundances of every element individually and we discuss the distribution of the residuals. In Sect. \ref{init-param} we investigate the confidence limits on the initial parameters of our best fits and we study the distribution of these parameters. In Sect. \ref{discussion} we discuss our results while Sect. \ref{conclusions} concludes.

%
\begin{figure}[!t]
   \centering
   \includegraphics[width=0.488\textwidth]{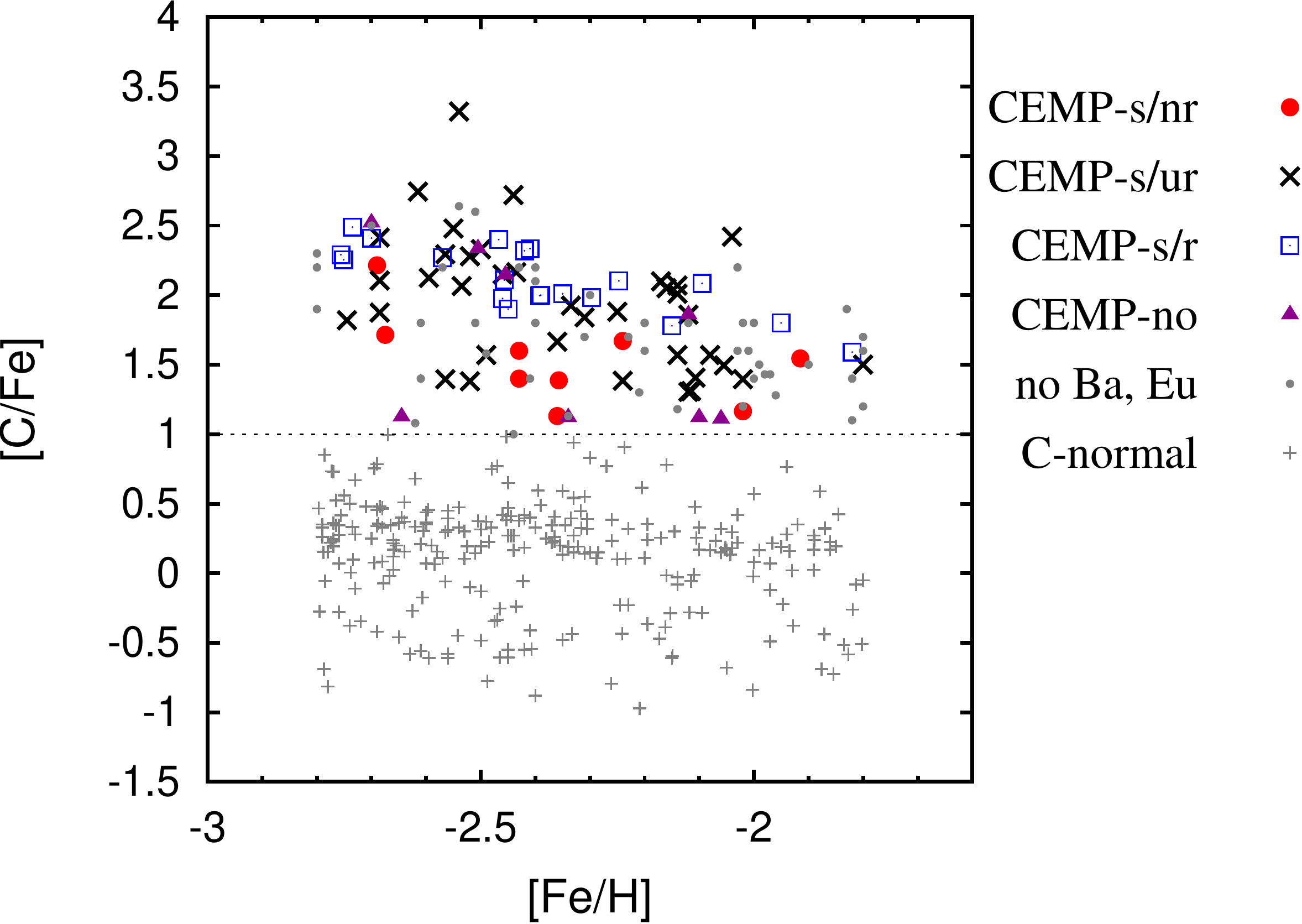}
      \caption{$[\C/\Fe]$ vs $[\Fe/\Hy]$ in the stars of our observed sample. Grey plus signs~(\textcolor{grey}{+}) are carbon-normal, very metal-poor stars; CEMP-$s$ stars are classified in three groups according to the abundance of europium, as described in the text: CEMP-$s/nr$ stars (\textcolor{Red}{$\bullet$}), CEMP-$s/ur$ stars ($\times$) and CEMP-$s/r$ stars (\textcolor{lightblue}{$\square$}). Purple triangles (\textcolor{purple}{$\blacktriangle$}) and grey dots (\textcolor{grey}{\scriptsize{$\bullet$}}) are CEMP-no stars and CEMP stars without measured abundances of heavy elements, respectively. The dotted line indicates $[\C/\Fe]=1$.}
         \label{fig:datasample}
\end{figure}

\section{Data sample}
\label{data}

Our database of observed very metal-poor stars is based on 580 stars catalogued in the SAGA observational database \citep[last updated in January 2015]{Suda2008, Suda2011} with iron abundance $-2.8\le[\Fe/\Hy]\le -1.8$. Among these objects we select the stars with observed abundances of carbon and barium and we ignore stars with only upper or lower limits. In some stars measurements of element abundances or stellar parameters are available from multiple sources. In most cases the measurements are consistent within the observational uncertainties and for the purpose of our study we use the arithmetic mean of the logarithm of the observed abundances and we adopt the largest observed error as the uncertainty on the measure. In case two measurements differ by more than the declared observational uncertainty and there is no obvious criterium to prefer one value, we compute the average and we adopt as the uncertainty half the difference between the two values. To this sample we add four CEMP stars studied by \cite{Masseron2010} with metallicity in the above range that were not present in the SAGA database.

This selection leaves us with a sample of $378$ very metal-poor stars, $67$ of which are classified as CEMP-$s$ stars, $8$ as CEMP-no stars, $46$ are CEMP stars with no information about the abundances of neutron-capture elements, and the remaining $257$ are carbon-normal very metal-poor stars. In Fig. \ref{fig:datasample} we show the carbon abundances of the stars in our observed sample as a function of $[\Fe/\Hy]$. CEMP-no stars are indicated as grey triangles. The dotted line represents the threshold carbon abundance $[\C/\Fe]=1$ above which the stars are defined CEMP stars.
We classify CEMP-$s$ stars in three groups based on the abundance of europium:
\begin{itemize}
\item[$\bullet$] In CEMP-$s/r$ stars the abundance of europium is enhanced, $[\Eu/\Fe]>1$ (open squares).
\item[$\bullet$] In CEMP-$s/nr$ stars the europium abundance is $[\Eu/\Fe]\le1$ (filled circles).
\item[$\bullet$] In CEMP-$s/ur$ stars the abundance of europium has not been determined (crosses), for example because the spectra have low signal-to-noise ratios, or the europium lines are blended.
\end{itemize}
Table \ref{tab:obs} summarises the surface gravities, temperatures, and abundances of iron, carbon, barium and europium (when available) of the $67$ CEMP-$s$ stars in our observed sample. For three stars in which barium is not observed and lanthanum is enhanced (HD$13826$, HD$198269$ and HD$201626$) the abundance of lanthanum is listed. The minimum uncertainty assumed in the chemical abundances and surface gravities is $0.1$ dex. We adopt an uncertainty of $100$ K in effective temperature unless differently stated.

\section{Models of binary evolution and nucleosynthesis}
\label{model}

As in Paper~I, in this study we use the code \texttt{binary\_c/nucsyn} that couples algorithms to compute the evolution of stars in binary systems with a model of stellar nucleosynthesis. The details of our code and the prescriptions used for the binary stellar evolution and nucleosynthesis are extensively discussed by \cite{Izzard2004, Izzard2006, Izzard2009}. Our default input physics is the same as in Paper~I, and we refer to this for a more complete description. In this section we summarise some important parameters adopted in our model, related to the wind mass-transfer process and the nucleosynthesis in the AGB phase (Sect. \ref{wind} and \ref{nucsyn}), and we describe the basic characteristics of our grid of model stars (Sect. \ref{grid}).

\subsection{Wind-accretion rate and angular momentum loss} %
\label{wind}
In this work we compare the results derived with three different model sets based on different assumptions about the wind-accretion rate and the mechanism of angular momentum loss. The model sets are listed in Table \ref{tab:models}. Model sets A and B are the same as in Paper~I. In our default model set A we calculate the wind-accretion rate according to the prescription for wind Roche-lobe overflow (WRLOF) with a dependence on the mass ratio, as presented by \citet[Eq. 9]{Abate2013}. We compute the angular momentum carried away by the wind material assuming a spherically symmetric wind \cite[Eq. 4 of][]{Abate2013}. %
The results of Paper~I show that a mechanism of wind accretion that is much more efficient at relatively short separations is necessary to reproduce the abundances observed in two thirds of the analysed CEMP-$s$ stars. To take this into account, in model set B we adopt the Bondi-Hoyle-Lyttleton (BHL) prescription as in Eq. (6) of \cite{BoffinJorissen1988} with $\alpha_{\mathrm{BHL}}=10$ (instead of $1\le\alpha_{\mathrm{BHL}}\le2$) to simulate a very efficient mass-transfer process. Model set B also adopts a prescription of efficient angular momentum loss, in which the material lost from the binary system carries away a multiple $\gamma = 2$ of the average specific orbital angular momentum \cite[as in Eq. 2 of Izzard et al., 2010, and Eq. 10 of][]{Abate2013}. %
Model set~C combines the WRLOF accretion rate as in model set A with the efficient angular momentum loss as in model set~B. In the analysis performed by \cite{Abate2013} this is the model set that predicts the largest fraction of CEMP stars.

\subsection{AGB nucleosynthesis}
\label{nucsyn}
We refer to the review papers by \cite{Busso1999} and \cite{Herwig2005} for a description of the nucleosynthesis process in the interior of AGB stars, the role of the $\Cth$-pocket at the top of the intershell region as a source of free neutrons for the production of slow neutron-capture elements ($s$-elements) in low-mass AGB stars, and the effect of the third dredge-up process (TDU hereinafter) that mixes the products of internal nucleosynthesis to the stellar surface.
In the models of \cite{Karakas2010} a $\Cth$-pocket is created by including a partial mixing zone (PMZ) at the deepest extent of each TDU, where protons are mixed in the intershell region and subsequently captured by the $^{12}\C$ nuclei to form a layer rich in $\Cth$. The mass of the PMZ is a free parameter in the models, and \cite{Lugaro2012} study the effects of different masses of the PMZ on the surface abundances in AGB stars with different initial masses.

We describe in Paper~I how in our models the amount of material dredged-up from the intershell region is determined in order to reproduce the evolution predicted in the detailed models of \cite{Karakas2010} and \cite{Lugaro2012}. The chemical composition of the intershell region is saved in a table as a function of three parameters: the mass of the star at the beginning of the TP-AGB phase, the thermal-pulse number, and the mass of the PMZ, $\Mpmz$. The surface abundances of an AGB star of mass $M_*$ evolve in time and are recalculated at every TDU, when material with the chemical composition of the intershell region taken from our table is mixed into the convective envelope.

\subsection{Grid of models}
\label{grid}
Our simulations are based on the same grid of models as in Paper~I, with $N$ binary-evolution models distributed in the $M_1 - M_2 - \log_{10}a - \Mpmz$ parameter space, where $M_{1,\,2}$ are the initial masses of the primary and secondary stars, respectively, $a$ is the initial separation of the system, and $\Mpmz$ is the mass of the partial mixing zone of any star of the binary system that becomes an AGB star. The grid resolution $N=N_{\mathrm{M1}}\times N_{\mathrm{M2}}\times N_a \times N_{\mathrm{PMZ}}$, 
where we choose $N_{\mathrm{M1}} = 34$, $N_{\mathrm{M2}} = 28$, $N_a = 30$, and $N_{\mathrm{PMZ}} = 10$. 
The initial parameters are chosen as follows:
\begin{itemize}
\item $M_1$ varies in the range [$0.9,6.0$] $\Msun$. The grid spacing is $\Delta \Mprim = 0.1\,\Msun$ up to $3\,\Msun$, and $\Delta \Mprim = 0.25\,\Msun$ otherwise.
\item $M_2$ is equally spaced in the range [$0.2,0.9$] $\Msun$, and by definition $\Msec \le \Mprim$.
\item $a$ varies between $10^2$ $\Rsun$ and $10^5$ $\Rsun$. The distribution of separations is flat in $\log_{10} a$. In the mass range considered here, at shorter separations the evolution of primary stars is interrupted before the AGB, because the systems undergo a common-envelope phase, whereas stars at wider separation do not interact in our models. The eccentricity is always zero.
\item When $\Mprim < 3\,\Msun$ we adopt $10$ different values for $\Mpmz$, namely $0$, $10^{-4}$, $2\times10^{-4}$, $5\times10^{-4}$, $6.66\times10^{-4}$, $10^{-3}$, $1.5\times10^{-3}$, $2\times10^{-3}$, $3\times10^{-3}$, and $4\times10^{-3}\,\Msun$. $\Mpmz$ is zero otherwise, in accordance with the detailed models of AGB nucleosynthesis of \cite{Karakas2010}.
\end{itemize}

\begin{table}
\caption{Models of the wind-accretion efficiency and angular momentum loss used in this study.}
\label{tab:models}
\centering
\begin{tabular}{ c c c }
\hline
\hline
Model set & wind accretion & angular momentum \\
		&	efficiency	&	loss\\
\hline
A	& WRLOF & spherically symmetric wind \\
B	& BHL, $\alpha_{\mathrm{BHL}}=10$	& $\Delta J/J = \gamma\,(\Delta M/M)$~,~~$\gamma=2$\\
C	& WRLOF & $\Delta J/J = \gamma\,(\Delta M/M)$~,~~$\gamma=2$\\
\hline
\end{tabular}
\tablefoot{
Our WRLOF model is calculated with Eq. (9) of \cite{Abate2013}. The BHL model is computed with Eq. (6) of \cite{BoffinJorissen1988}. Angular momentum loss due to wind ejection is computed alternatively with Eq. (4) of \cite{Abate2013} for a spherically symmetric wind or with $\gamma=2$ in Eq. (2) of \cite{Izzard2010} and Eq. (10) of \cite{Abate2013}.
}
\end{table}

%
\begin{figure*}[!t]
   \centering
   \includegraphics[width=0.99\textwidth]{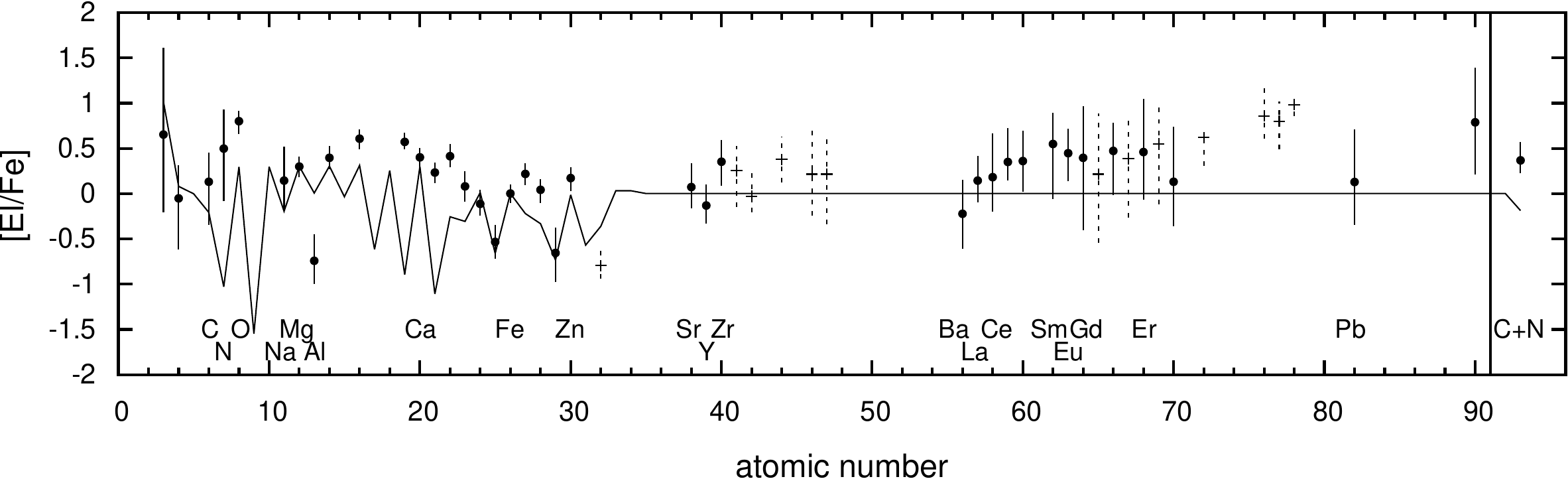}
   \caption{Average abundances observed in our sample of carbon-normal stars ($[\C/\Fe]<1$). Circles and plus signs represent elements observed in more, or fewer, than six stars, respectively. The solid and dashed bars enclose, respectively, $68.3\%$ and $100\%$ of the observations. The combined abundance of C+N is shown in the right panel. The solid line shows the initial-abundance mix adopted in our models.}
    \label{fig:vmp-abunds}
 \end{figure*}

All stars in our grid are formed in binary systems. In reality single stars or binaries in wider orbits may exist, and carbon-rich very metal-poor AGB stars have recently been observed \cite[][]{Boyer2015II}. To determine whether the abundances observed in the stars of our sample can be reproduced by a single-star model, we add to our grid $200$ synthetic single stars with masses uniformly spaced in the range $[0.4,\,1.0]\,\Msun$, and $\Mpmz$ chosen as above. 

We assume that the accreted material is instantaneously mixed throughout the entire secondary star, as in the default model of Paper~I. This approximation mimics the effect of efficient non-convective mixing processes, such as thermohaline mixing, and is reasonable in low-mass stars \cite[][]{Stancliffe2007}.

In our model we assume the same metallicity as in the detailed models of \cite{Lugaro2012}, $Z=10^{-4}$, that corresponds approximately to $[\Fe/\Hy]\approx-2.3$. As initial chemical composition of the stars in our grid we adopt the abundances predicted in the one-zone Galactic chemical-evolution model of \cite{Kobayashi2011} at $[\Fe/\Hy]=-2.3$ that includes the contributions of core-collapse and type Ia supernovae and AGB stars. Their results extend up to $^{76}$Ge and for heavier isotopes we adopt the solar abundance distribution derived by \cite{Asplund2009} scaled to metallicity $Z=10^{-4}$. Different assumptions on the initial composition of the stars negligibly affect AGB nucleosynthesis, as discussed by \cite{Lugaro2012}.
In the next section we discuss how representative our set of initial abundances is of the chemical composition of the very metal-poor stars in our sample.


\section{Comparison of the model initial abundances with carbon-normal metal-poor stars}
\label{VMP}
In this section we focus on the $257$ ``carbon-normal'' very metal-poor stars in our sample, i.e. with $[\C/\Fe]<1$, called C-normal stars hereinafter. %
C-normal stars do not exhibit evidence of duplicity and are not expected to have changed their initial surface composition. Consequently, their abundances are expected to follow the distribution predicted by Galactic chemical-evolution models at metallicity $Z\approx10^{-4}$. Some abundance variations may be introduced by internal mixing processes, such as rotational mixing, gravitational settling of heavy nuclei or thermohaline mixing. The only significant change in the abundances of stars with small surface gravity, $\loggunits\lesssim4$, is because of the first dredge-up that reduces the carbon-to-nitrogen ratio but leaves essentially unaltered the abundances of the other elements. Our purpose is to compare the set of initial abundances adopted in our models with the abundances observed in C-normal stars and find discrepancies that are relevant to our study of the chemical composition of CEMP-$s$ stars.

In Fig. \ref{fig:vmp-abunds} we show the initial abundances adopted in our models (solid line) and the abundances observed in the C-normal stars. For every observed element we compute the mean of the logarithm abundance relative to iron, [El/Fe] (filled circles and plus signs). Only elements observed in more than one star are shown. Elements observed in fewer than seven stars are plotted with dashed bars, which connect the minimum and the maximum observed value. For each element observed in at least seven stars we calculate the median of the values of [El/Fe] and we select the group of approximately $68\%$ stars, half of which have [El/Fe] larger than the median and the other half have smaller values. The solid bars in Fig. \ref{fig:vmp-abunds} indicate the interval of [El/Fe] that encloses the stars in this group. In the absence of significant nucleosythesis in C-normal stars, we expect that: (1) the mean and the median of the observed abundances coincide and are equal to the initial value predicted by the models of Galactic chemical evolution; (2) the deviations from the mean have a Gaussian distribution; (3) the solid bars are symmetrical with respect to the mean and correspond to the standard deviation in the observed abundances. However, Fig. \ref{fig:vmp-abunds} shows that these three statements are not always true, and several discrepancies occur between the predictions of the Galactic chemical evolution model and the observations. 

We briefly analyse the discrepancies that are most relevant for our purposes. The differences between the model abundances and the observations of carbon and $s$-elements (e.g. Sr, Y, Zr, Ba, La, Ce, Pb) are on average within a factor of two ($0.3$ dex), although in some cases the observations have a large spread, for example carbon and barium ($0.8$ dex), cerium ($0.9$ dex) and lead ($1.1$ dex). These uncertainties are expected to have a small effect on our study of CEMP-$s$ stars because typically in low-mass AGB stars ($M_*\le3\Msun$) the amount of carbon increases by more than two orders of magnitude and the abundances of $s$-elements by more than a factor of ten. The average abundance of nitrogen relative to iron is $1.5$ dex larger than the initial abundance assumed in our model, with a spread of $1$ dex. This discrepancy may alter the results of our best-fitting models. In stellar nucleosynthesis, nitrogen is made in the CN cycle when protons are mixed into regions of the star where carbon is abundant. For example, the mixing of protons can occur at the bottom of the convective envelope of red giants and AGB stars during first dredge-up and TDU, respectively. The increase in nitrogen abundance depends on the extent of the mixing, which is very uncertain \cite[e.g.:][]{Charbonnel1998, Boothroyd1999, Karakas2010-1}. However, the total amount of carbon plus nitrogen is conserved in the CN cycle and therefore when both the elements are observed it is convenient to compare the predictions of our models with the combined abundance C+N. In C-normal stars the average C+N (right panel of Fig. \ref{fig:vmp-abunds}) is underestimated by $0.5$ dex (approximately a factor of three). This offset should have a small effect on our fits of CEMP-$s$ stars, because the abundance of C+N increases by more than a factor of a hundred in AGB stars. 

Oxygen is underpredicted on average by a factor of three. The amount of oxygen increases by less than one order of magnitude during the AGB phase, hence an initial offset can substantially modify our results in the study for CEMP-$s$ stars. The observed abundances of sodium are underpredicted on average by $0.3$ dex and are spread over $0.7$ dex. A better model of the initial abundance of sodium is desirable because the amount of sodium produced in AGB nucleosynthesis depends strongly on the stellar mass and on the mass of the PMZ, and a reliable fit of its abundance can, in principle, provide constraints on these two parameters.

The elements between atomic numbers $13$ (aluminium) and $30$ (zinc) are not always consistent with our initial assumptions. These elements are not modified by AGB nucleosynthesis, with the exception of aluminium that is partly produced in massive AGB stars, hence the offset observed in C-normal stars is expected to be an issue also in CEMP-$s$ stars.

The elements with atomic numbers 62--71 and 75--80 are underestimated by $0.4$--$1$ dex, in many cases with large spreads in the observed abundances. In the solar system, more than $60\%$ of the total abundance of these elements is produced by the $r$-process, for example $69\%$ of samarium, $94\%$ of europium and $86\%$ of gadolinium \cite[][]{Arlandini1999, Bisterzo2011}. For brevity, these elements are hereinafter referred to as ``$r$-elements'', although some fraction of their total amount is produced by the $s$-process. However, during AGB nucleosynthesis their abundances generally increase by less than $1$--$1.5$ dex, and therefore the initial offset may affect our analysis of the CEMP-$s$ stars.

\section{Analysis of the abundances in carbon-enhanced metal-poor stars}
\label{CEMPs}

We compare our binary evolution and nucleosynthesis models with the abundances of the $67$ CEMP-$s$ stars in our observational sample. We also include $14$ of the $15$ binary stars with known orbital periods studied in Paper I (those with $-2.8\le[\Fe/\Hy]\le1.8$) but in the present work, for comparison, we focus only on the chemical abundances, and we ignore the constraints on the period of the systems. In general we find initial parameters of the best-fit model different from Paper~I. These differences give an estimate of the uncertainties in our present results caused by the lack of information on the orbital periods.

For each star in our sample we determine the model in our grid that best reproduces the observed chemical abundances with the same procedure as in Paper~I. Initially, to constrain the evolutionary stage, we select model stars that reproduce the measured surface gravity within the observational uncertainty, $\sigma_g$, at an age $10\le t\le13.7$ Gyr. Subsequently, for the stars that pass this selection, we compute the $\chisq$ of each model from
\begin{equation}
\chisq = \sum_i\frac{(A_{i,\mathrm{obs}} - A_{i,\mathrm{mod}})^2}{\errb^2}~~.\label{eq:chi}
\end{equation}
In Eq. (\ref{eq:chi}) every element $i$ has observed abundance $A_{i,\mathrm{obs}}=12+\log_{10} (N_{i,\mathrm{obs}}/N_{\Hy})$, where $N_{i,\mathrm{obs}}$ and $N_{\Hy}$ are the number densities of $i$ and hydrogen, respectively, $\errb$ is the observational error associated with $A_{i,\mathrm{obs}}$ and $A_{i,\mathrm{mod}}$ is the abundance predicted by the model. The minimum value of $\chisq$, $\chimin$,  determines the best model. 
In this procedure we do not include any constraint on the effective temperature, $\Teff$, because $\Teff$ depends strongly on the observed metallicity which varies by up to a factor of three in the stars of our sample compared to the value adopted in our model, $Z=10^{-4}\,$. All other parameters being equal, if the observed metallicity is lower (higher) than in our model we expect to find a model $\Teff$ lower (higher) than the observed.

As we noted in Paper I, because in our study we adopt a fixed metallicity for all our systems we ignore the scatter in observed [Fe/H] values. Qualitatively, a decrease in metallicity has two effects. First, the abundance relative to iron of the elements produced in AGB nucleosynthesis increases, because the initial amount of iron is lower, and iron is not produced in AGB stars. Second, because the production of neutrons is primary in AGB stars, the smaller the abundance of iron, the larger is the neutron-to-iron ratio, and consequently the abundances of the more neutron-rich elements (e.g. barium, lead) are enhanced. Opposite effects are caused by increasing metallicity. We restricted our sample to a $0.5$ dex range around $[\Fe/\Hy]=-2.3$ to have a sufficient number of stars for our analysis. It is important to consider that our approximation probably introduces bigger errors the larger the difference in [Fe/H] between observations and our model. An attempt to quantify this error is discussed in Sect. \ref{disc-1}.

We focus on the elements produced by nucleosynthesis in AGB stars. Therefore in Eq. (\ref{eq:chi}) we take into account C, N, O, F, Ne, Na, Mg, and all the heavy neutron-capture elements with atomic number in the range $[31,\, 82]$, including the light-$s$ elements (or ls, namely strontium, yttrium and zirconium), the heavy-s elements (or hs, namely barium, lanthanum and cerium) and lead. The abundance of nitrogen produced in AGB stars is uncertain. In AGB stars of mass above about $3\Msun$, carbon is efficiently converted to nitrogen at the base of the convective envelope \cite[hot bottom burning,][]{Lattanzio1991}. At lower mass, some form of deep mixing of the envelope material operates down to regions where hydrogen burning occurs, as discussed in Sect. \ref{VMP}, but the exact amount depends on the extent of mixing, which is very uncertain \cite[e.g.][]{Hollowell1988, Gallino1998, Goriely2000, Stancliffe2010, Lugaro2012}. However, the total amount of C+N is conserved, therefore when both elements are measured we consider their combined abundances. 

We exclude from Eq. (\ref{eq:chi}) the elements from aluminium to zinc because they are not involved in AGB nucleosynthesis (aluminium is produced only in massive AGB stars), and the differences between models and observations reflect a discrepancy with our set of initial abundances, as we noticed in the sample of C-normal stars. For comparison, in some CEMP-$s/r$ stars we take into account a smaller set of elements, i.e. only those that are mostly produced in AGB nucleosynthesis (C, Mg, Sr, Y, Zr, Ba, La, Ce, Pb), as will be explained in Sect. \ref{CEMP-rs}. Tables \ref{tab:best-WRLOFq-ssw},  \ref{tab:best-BoHo10-gamma2} and \ref{tab:best-WRLOFq-gamma2} in Appendix \ref{app:A} summarise our results (computed with model sets A, B and C, respectively) for the best-fitting models of 43 CEMP-$s$ stars with number of degrees of freedom $\nu\ge2$. The number of degrees of freedom is calculated as $\nu=N_{\mathrm{obs}}-4$, where $N_{\mathrm{obs}}$ is the number of observed elements used in Eq. (\ref{eq:chi}) and $4$ is the number of fitted parameters. Our results for the CEMP-$s$ stars with $\nu\le1$ are shown in Tables \ref{tab:few-WRLOFq-ssw}--\ref{tab:few-WRLOFq-gamma2}.

Some stars fail to match one or more elements within their observed uncertainty. To quantify the discrepancy between the observations and the model predictions, we calculate the residuals,
\begin{equation}
R_i = A_{i,\mathrm{obs}} - A_{i,\mathrm{mod}}~, \label{eq:residuals}
\end{equation}
where $A_{i,\mathrm{obs}}$ and $A_{i,\mathrm{mod}}$ are the observed and modelled number abundance of element $i$ of the best fitting model, as in Eq. (\ref{eq:chi}). 

For every element $i$, if our models reproduce correctly the nucleosynthesis and mass-transfer process, under the assumption that the measurements are only affected by Gaussian errors, the distribution of $R_i$ should resemble a Gaussian function centred on zero with a standard deviation approximately equal to the average observational error.
\begin{figure}[!t]
   \centering
   \includegraphics[width=0.488\textwidth]{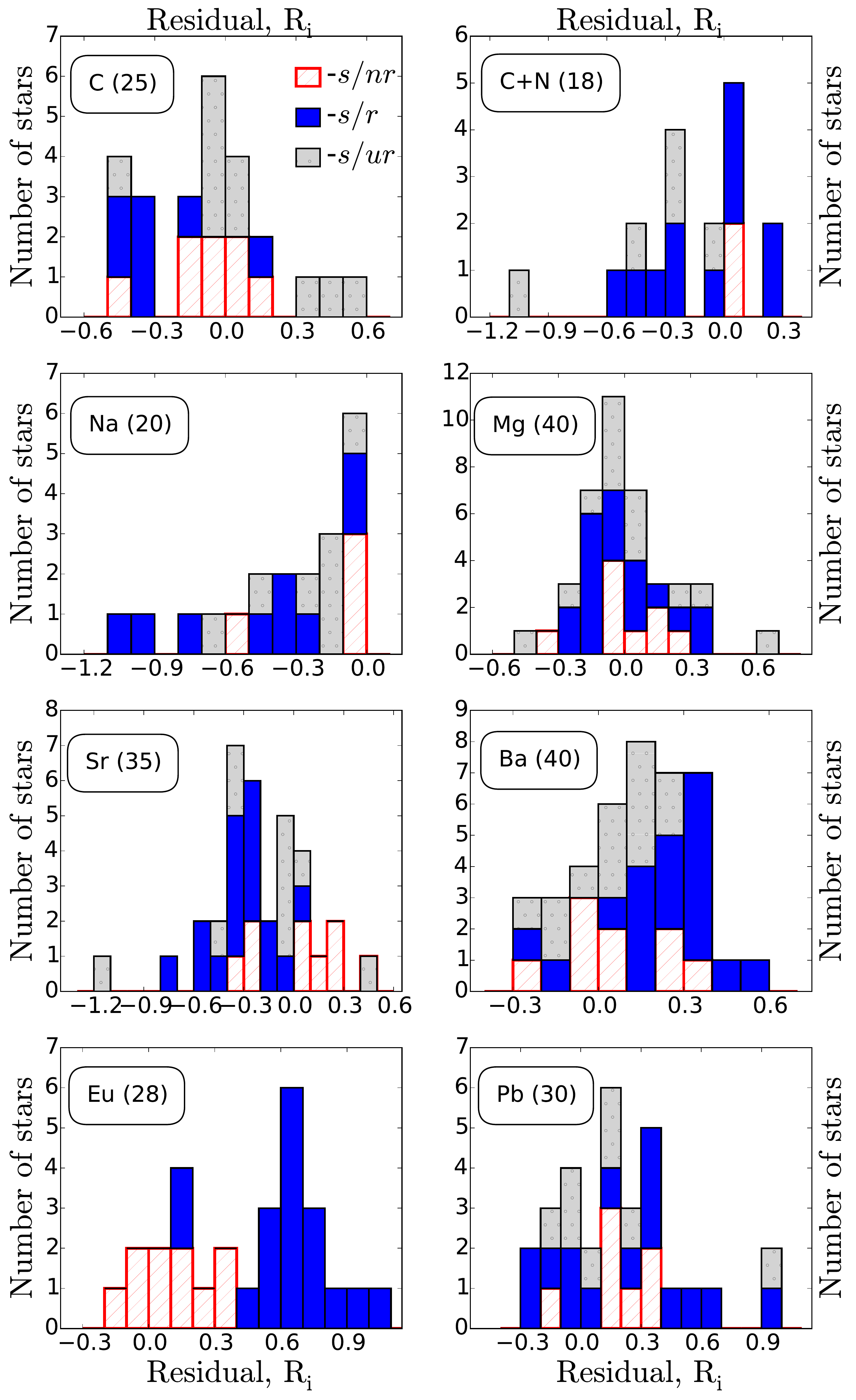}
   \caption{Residual distributions in the model-A CEMP-$s$ stars of our sample with $\nu\ge2$. The number of stars in which the element is measured is indicated in brackets. The red-hatched, blue-filled and grey-dotted histograms indicate the residual distributions for CEMP-$s/nr$, CEMP-$s/r$ and CEMP-$s/ur$ stars, respectively. In the top-left panel the residual distribution of carbon is shown for those stars without a nitrogen measurement.}
    \label{fig:histo_resid}
 \end{figure}
In Fig. \ref{fig:histo_resid} we plot the distributions of $R_i$ computed with the default model set A and binned in intervals of width $0.1$ for model stars with $\nu\ge2$. From the top left to the bottom right we show the distributions of the residuals computed for the abundances of carbon (for those stars without a nitrogen measurement), C+N, sodium, magnesium, strontium, barium, europium and lead. The subgroups of CEMP-$s/nr$, CEMP-$s/r$ and CEMP-$s/ur$ stars are indicated with hatched, filled and dotted histograms, respectively. In brackets we indicate the number of stars in which the element is measured.

\begin{figure*}[!t]
   \centering
   \includegraphics[width=0.99\textwidth]{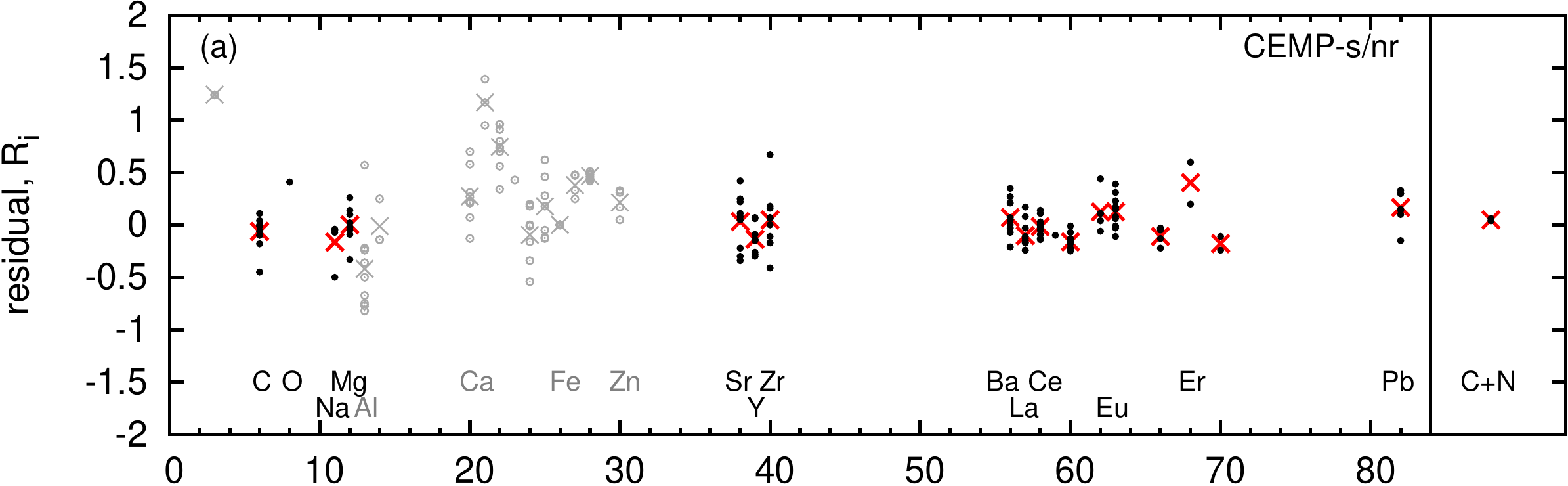}
   \includegraphics[width=0.99\textwidth]{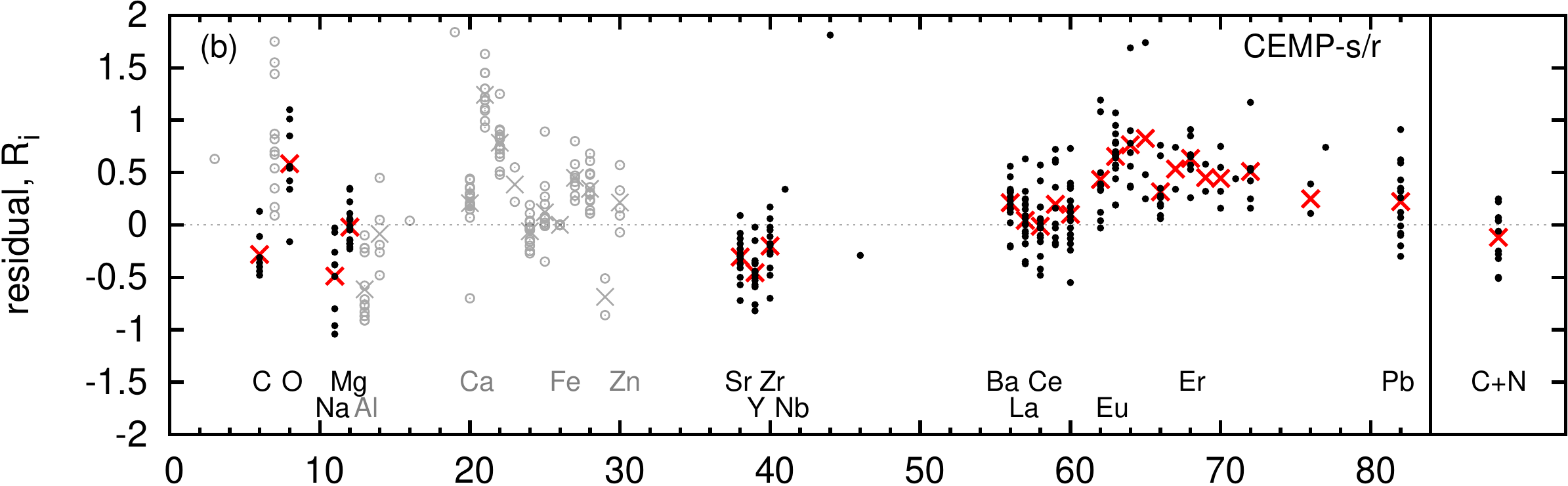}
   \includegraphics[width=0.99\textwidth]{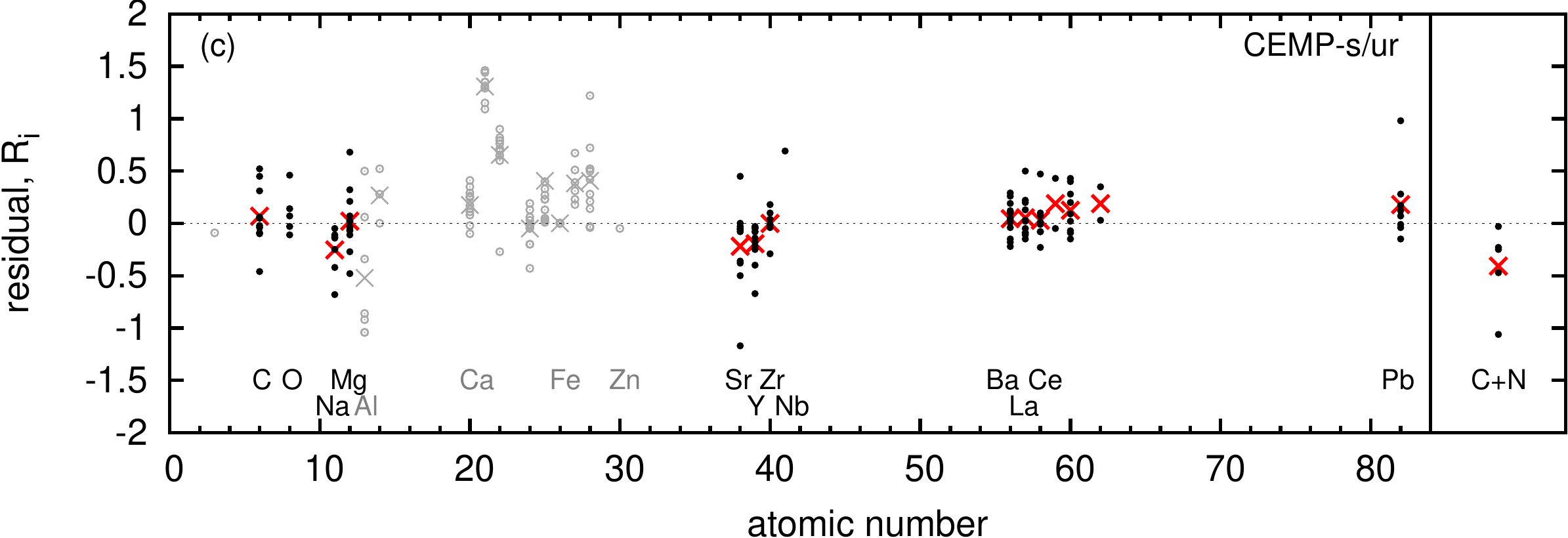}
   \caption{Residuals in our $43$ modelled CEMP-$s$ stars with a number of degrees of freedom $\nu\ge2$. Model set A is adopted. From top to bottom: CEMP-$s/nr$, CEMP-$s/r$ and CEMP-$s/ur$ stars. Black filled circles indicate the residuals of the elements included in Eq. (\ref{eq:chi}) to calculate $\chisq$, grey open circles indicate the other elements, crosses represent the average values of the residuals.}
    \label{fig:residCEMP-s}
 \end{figure*}

In Fig. \ref{fig:residCEMP-s} we plot, for each star with $\nu\ge2$, the residuals $R_i$ of every observed element. The elements considered in Eq. (\ref{eq:chi}) are shown as black filled circles, while the other elements are shown as grey open circles. The average values of the residuals of each element are shown as crosses. Panels (a)--(c) show the results obtained with our default model set A for CEMP-$s/nr$, CEMP-$s/r$ and CEMP-$s/ur$ stars, respectively.

%

\subsection{CEMP-s/nr stars}
\label{CEMP-s/nr}
Ten of the $67$ sampled CEMP-$s$ stars are classified as CEMP-$s/nr$, with $[\Ba/\Fe]>0.5$ and $[\Eu/\Fe]\le1$. For all the stars in this group we find good fits to the abundances.
Four stars of this group, CS22942--019, CS22964--161A,B and HD198269 belong to binary systems with known orbital periods. In Paper~I we find the models that best reproduce, at the same time, the observed abundances and the orbital periods of these stars. In the present work we ignore the constraint on the period in our models and we find essentially the same input parameters $\Mprim$, $\Msec$ and $\Mpmz$ as in the best models of Paper~I, whereas there are large differences in the initial and final periods.
These differences are due to the amount of mass $\Delta\Macc$ that the secondary star has to accrete to reproduce the observed $\logg$ and surface abundances. 
For example, the model star CS22942--019 accretes $\Delta\Macc\approx0.3\,\Msun$ with all the model sets and consequently the $\chimin$ of the fit is the same ($\chimin=21$). On the other hand, the initial periods vary by a factor of four ($\Pin=6.4\times10^4$ days with model sets A and C, $\Pin=1.6\times10^4$ with model set B), while the final periods are $\Pf=1.0\times10^5,\,3.5\times10^3,\,\Pf=1.5\times10^4$ days with model set A, B and C, respectively (observed $\Porb=2800$ days). Hence, only model set B reproduces the observed period within the uncertainty of our grid of models.

For binary star CS22964--161A,B, which probably formed in a triple system \cite[cf.][and Paper~I]{Thompson2008}, we find the same results as in Paper~I without significant differences between model sets A, B and C. A binary system of approximately $1.5\,\Msun$ accretes a small amount of material ($\Delta\Macc\approx0.06\,\Msun$) from an initially $1.6\,\Msun$ primary star in a very wide orbit ($\Pin=2.7\times10^5,\,1.3\times10^5,\,2.6\times10^6$ days according to model sets A, B and C, respectively). Note that the observed orbital period in Table \ref{tab:obs} does not correspond to the periods found in our models, which is the period of the unseen third star. The fit to the abundances observed in HD198269 improves compared to the result of Paper I if we ignore the orbital period ($\Porb=1295$ days). With essentially the same assumptions on the masses and longer orbital periods (see Tables \ref{tab:best-WRLOFq-ssw}--\ref{tab:best-WRLOFq-gamma2}) our model stars accrete $\Delta\Macc\approx0.09\,\Msun$ and we obtain $\chimin=9$, whereas in our best-fit model in Paper I (with model set B) we find $\Delta\Macc=0.13\,\Msun$ and $\chimin=18$.

Fig. \ref{fig:residCEMP-s}a shows that the abundances of almost all the elements involved in AGB nucleosynthesis are well reproduced and typically the mean of the residuals is close to zero. 
The abundance of sodium is well reproduced in three out of four CEMP-$s/nr$ stars, while it is overestimated by $0.5$ dex in CS22964--161B (in which $\sigma_{\Na}=0.3$ dex). The mean residual of yttrium (atomic number $39$) is approximately $-0.15$ dex, lower than for strontium and zirconium ($\avres\approx0$). The difference is smaller than the average error in the measurement of yttrium ($\averr=0.20$ dex) but indicates that the abundance of yttrium is systematically overestimated compared to strontium and zirconium. The abundances of erbium (atomic number $68$) are underestimated by $0.6$ and $0.2$ dex for stars CS22880--074 and HD196944, respectively, while the observational uncertainty is approximately $0.2$ dex in both stars. An offset of about $0.6$ dex is also found by \cite{Bisterzo2012} in their analysis of CS22880--074. In addition, for this star our best-fitting model and the models of \cite{Bisterzo2012} well reproduce other elements traditionally associated with the $r$-process, such as europium and dysprosium. This evidence possibly suggests that the abundance determination of erbium in CS22880--074 is affected by observational bias.

The abundances of lead are underestimated on average by $0.18$ dex, approximately the mean observational uncertainty of this element in CEMP-$s$ stars ($\averr=0.20$ dex). The distribution of the residuals in Fig. \ref{fig:histo_resid} (red-hatched histogram) shows that the residuals of lead are always within $2\averr$.
The distributions in Fig. \ref{fig:histo_resid} show that the abundances of the other elements are typically reproduced within $2\averr$. This indicates that on average our models of binary evolution and AGB nucleosynthesis reproduce reasonably well the abundances of CEMP-$s/nr$ stars, although the results should be interpreted with care because our sample is small and we do not constrain the orbital periods of the modelled systems.

\subsection{CEMP-s/r stars}
\label{CEMP-rs}

$20$ stars in our sample are classified as CEMP-$s/r$ stars. Two of these stars, BS$16080-175$ and BS$17436-058$, are not in Table \ref{tab:best-WRLOFq-ssw} because only $5$ elements are observed and therefore $\nu=1$. Four CEMP-$s/r$ stars (CS22948--027, CS29497--030, HD224959 and LP625--44) are analysed in Paper~I, and we find that to reproduce the large enhancements observed in most of the neutron-capture elements the secondary star has to accrete a large amount of mass, $\Delta\Macc>0.2\,\Msun$. At the observed orbital periods only model set B predicts efficient mass accretion, whereas without the constraint on the period also model sets A and C predict large mass accretion in much wider orbits, as shown in Tables \ref{tab:best-WRLOFq-ssw} and \ref{tab:best-WRLOFq-gamma2}.
Our best-fitting models have rather high $\chimin$ despite the large accretion efficiency, partly because we fail to reproduce the enhanced abundances of the $r$-elements, and this points to a limitation in our nucleosynthesis model, as noted also in Paper~I.

\begin{figure}[!t]
   	\includegraphics[width=0.488\textwidth]{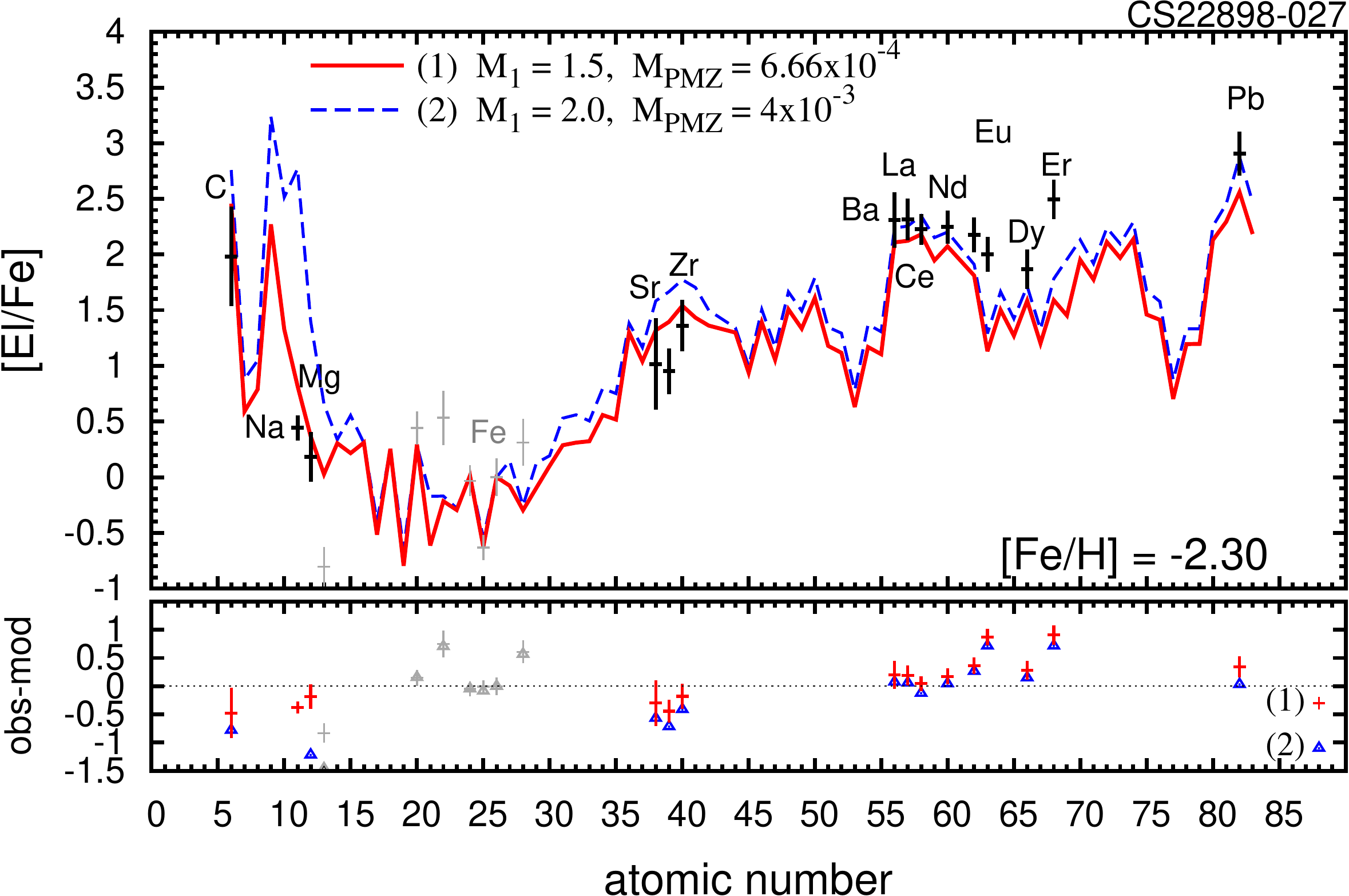}
   \caption{CEMP-$s/r$ star CS22898--027. {\it Points with error bars}: observed abundances, in black the elements adopted in Eq. (\ref{eq:chi}), in grey the other elements. {\it Red solid line}: the best-fitting model found with model set A, $\Mprim=1.5\,\Msun$ and $\Mpmz=6.66\times10^{-4}\,\Msun$. {\it Blue dashed line}: alternative model with $\Mprim=2\,\Msun$ and $\Mpmz=4\times10^{-3}\,\Msun$. {\it Lower panel}: the residuals of the two models, computed with Eq. (\ref{eq:residuals}), are shown as red plus signs with error bars and blue triangles, respectively.}
    \label{fig:CS22898-027}
 \end{figure}

For several CEMP-$s/r$ stars we find a particularly poor fit, with large reduced $\chisq$ ($\chimin/\nu\gtrsim5$, whereas at a visual inspection models appear to fit the observed abundances well if $\chimin/\nu\le3$). To verify if these results are caused only by the discrepancy in the $r$-elements we perform the $\chisq$ analysis taking into account only nine elements that are produced in large amounts by our AGB nucleosynthesis models, and that are frequently detected in our observed sample: C, Mg, Sr, Y, Zr, Ba, La, Ce, Pb.
This choice generally leads to a better fit of the abundances of these elements without significantly altering the results for the other elements. However, even with this choice, in most model stars the reduced $\chisq$ still exceeds $3$, with the exceptions of CS22948--027 and CS29497--030. The best-fitting models of these two stars, calculated taking into account only nine elements, have the same primary and PMZ masses as in the models determined including all the observed elements (for both stars, $\Mprim=1.5\,\Msun$ and $\Mpmz=2\times10^{-3}\,\Msun$, with model set~A). The secondary stars are slightly more massive (by approximately $0.05\,\Msun$ for both stars with all model sets).
In the other CEMP-$s/r$ stars large $\chimin$ are determined because in most of our models there is also an issue in reproducing the observed element-to-element ratios. In particular, it is difficult to reconcile the large enhancements of heavy-$s$ elements and lead with the relatively small abundances of lighter elements (carbon, sodium, magnesium, light-$s$ elements). %
As an example, we compare the modelled and observed abundances in star CS22898--027 in Fig. \ref{fig:CS22898-027}. %
A model of a $0.5\,\Msun$ secondary star that accretes $\Delta\Macc\approx0.3\,\Msun$ from a $2\,\Msun$ primary star with $\Mpmz=4\times10^{-3}\,\Msun$ (dotted line) reproduces the large enhancement of the heavy-$s$ elements and lead, but overestimates carbon by $0.8$ dex, sodium and magnesium by more than $1$ dex and the light-$s$ elements by at least $0.5$ dex (consequently, $\chimin=632$). The best compromise between the light and heavy elements is the model with $\Mprim = 1.5\,\Msun$ and $\Mpmz = 6.6\times10^{-4}\,\Msun$, shown as a solid line in Fig. \ref{fig:CS22898-027}, where all elements up to zirconium are overestimated, the neutron-rich elements between barium and lead are underestimated and $\chimin = 100\,(\nu=11)$. Similar results are found with all model sets.

\begin{figure}[!t]
   \centering
   	\includegraphics[width=0.488\textwidth]{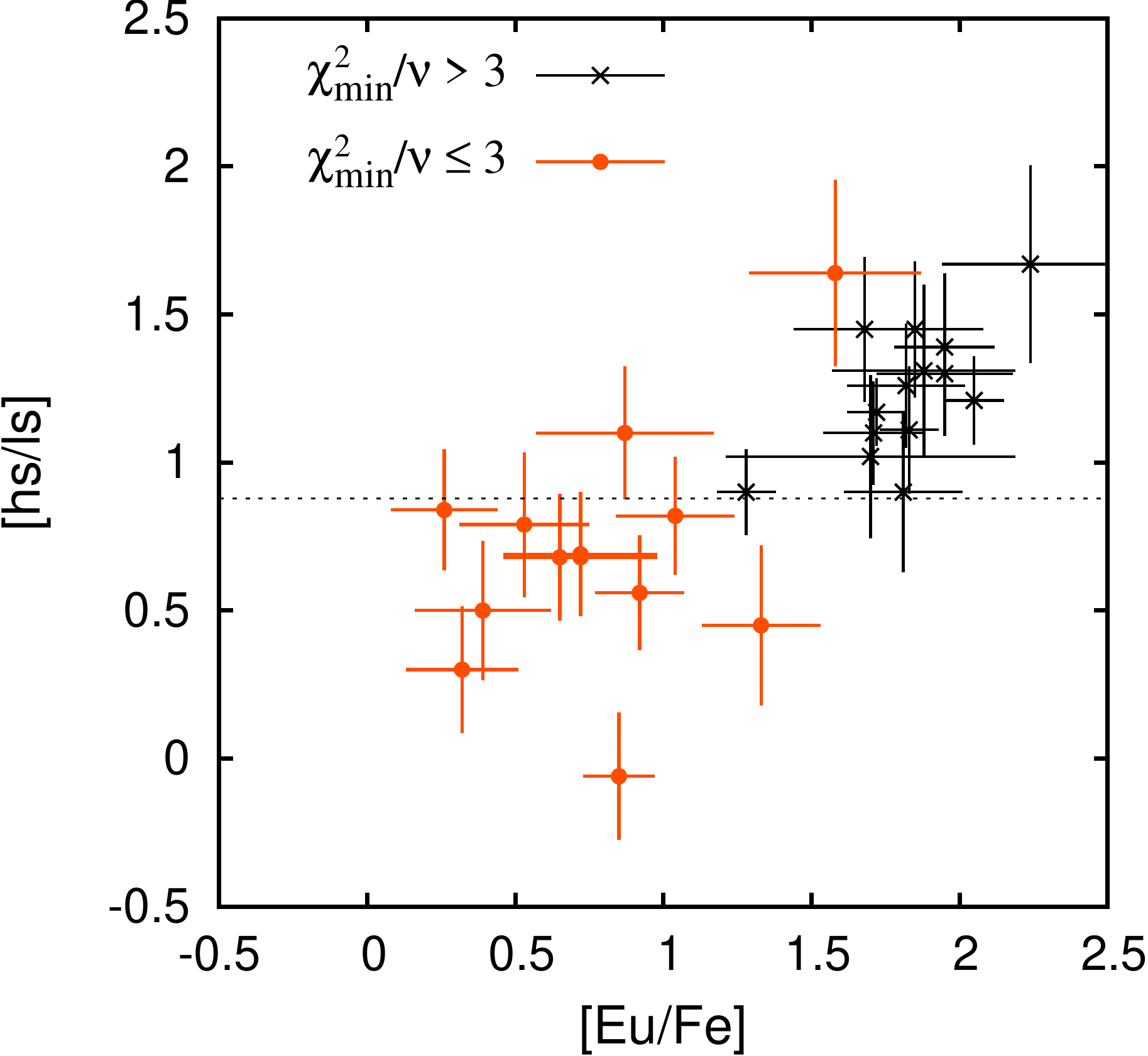}
   \caption{[hs/ls] vs [Eu/Fe] in our sample CEMP-$s/nr$ (those with $[\Eu/\Fe]<1$) and CEMP-$s/r$ stars. Orange filled circles indicate the stars for which we find a good fit ($\chimin/\nu\le3$). Black crosses indicate the stars poorly reproduced ($\chimin/\nu>3$). The dotted line indicates the maximum hs-to-ls ratio in our models, $[\hs/\ls]_{\mathrm{max}}=0.88$.}
    \label{fig:hsls}
 \end{figure}

In Fig. \ref{fig:hsls} we plot the hs-to-ls ratio versus the abundance of europium observed in our sample CEMP-$s/nr$ and CEMP-$s/r$ stars. The abundances of hs and ls are defined, respectively, as $[\hs/\Fe] = ([\Ba/\Fe]+[\La/\Fe]+[\Ce/\Fe])/3$ and $[\ls/\Fe] = ([\Sr/\Fe]+[\Y/\Fe]+[\Zr/\Fe])/3$. If one of these abundances is not available we average among the elements present in our database. Orange circles and black crosses indicate, respectively, the observed stars for which we find a fit with small reduced $\chisq$ ($\chimin/\nu\le3$) and those for which our best fit has $\chimin/\nu>3$. The dotted line indicates the maximum value of the hs-to-ls ratio predicted in any of our models, $[\hs/\ls]_{\mathrm{max}}=0.88$.
We notice that: 
\begin{itemize}
\item $[\hs/\ls]$ is generally larger for increasingly large [Eu/Fe], because the abundances of heavy-$s$ elements and europium correlate in metal-poor stars \cite[e.g.][]{Jonsell2006}, whereas the abundances of light-$s$ elements are essentially independent of europium.
\item $11$ stars have observed $[\hs/\ls]<[\hs/\ls]_{\mathrm{max}}$, and for all we find a model with $\chimin/\nu\le3$. 
Only $2$ out of $11$ stars are CEMP-$s/r$ stars (CS22881--036 and HD187861), and these have relatively low europium enhancements, $[\Eu/\Fe]<1.4$.
\item $17$ stars have observed $[\hs/\ls]>[\hs/\ls]_{\mathrm{max}}$; for only two of these stars we find a model with $\chimin/\nu\le3$ and one of these is a CEMP-$s/nr$ star (HD$198269$).
\item $16$ out of $18$ CEMP-$s/r$ stars have $[\hs/\ls]>[\hs/\ls]_{\mathrm{max}}$ (only the CEMP-$s/r$ stars with $\nu\ge2$ are shown in Fig. \ref{fig:hsls}). The value $[\hs/\ls]_{\mathrm{max}}$ is found in a model AGB star of mass $\Mprim\approx1.5\,\Msun$: consequently, an initial primary mass $\Mprim=1.5\,\Msun$ is selected in almost all our model CEMP-$s/r$ stars. However, only three model CEMP-$s/r$ stars reproduce the observed abundances well ($\chimin/\nu\le3$).
\end{itemize}

The distributions of the residuals of CEMP-$s/r$ stars in Fig.~\ref{fig:residCEMP-s}b reflect the difficulty in predicting the correct element-to-element ratios in most CEMP-$s/r$ stars. The abundances of carbon, sodium, strontium, yttrium, and zirconium are on average overestimated. The mean residual of carbon is $\avres=-0.3$ dex, whereas the mean residual of C+N is only $\avres=-0.1$ dex. This may be interpreted as the effect of extra mixing of protons in AGB stars converting carbon to nitrogen. However, the residuals of C+N have a large dispersion, therefore it is not possible to derive strong conclusions about the efficiency of the mixing process.
Oxygen is underestimated on average by $0.6$ dex, approximately the same as in C-normal metal-poor stars. Because the abundance of oxygen is not much affected by AGB nucleosynthesis, this result suggests that the discrepancy could be reduced by adopting a larger initial abundance of oxygen in our models. 

The abundance of sodium is always overestimated, on average by $0.5$ dex. In stars HE0338--3945 and HE1305+0007 the residuals of sodium are approximately $-1$ while the observational uncertainties are $0.1$ and $0.2$ dex, respectively. The abundance of sodium is increasingly larger the larger the mass of the PMZ. In our model CEMP-$s/r$ stars we mostly assume $\Mpmz\ge2\times10^{-3}\,\Msun$ to reproduce the observed abundances of heavy-$s$ elements and lead; consequently we overestimate sodium. An offset of approximately $0.8$ dex in the sodium abundance of HE1305+0007 is also found by \cite{Bisterzo2012} with models that reproduce the large enhancements of heavy-$s$ elements and lead. In contrast, the best fit of \cite{Bisterzo2011} matches the sodium abundance in HE0338--3945 within the observational uncertainty. This is probably because the abundances of all $r$-process isotopes are initially pre-enriched, $[r/\Fe]^{\mathrm{ini}}=2$ dex, and no dilution of the accreted material on the secondary star is considered. As a consequence, even a low-mass AGB model ($M^{\mathrm{AGB}}_{\mathrm{ini}}=1.3\,\Msun$) predicts large abundances of heavy elements, while it does not produce much sodium.

Magnesium is the only light element with mean residual approximately zero in our models. As shown in Fig. \ref{fig:histo_resid}, the maximum discrepancy between the observed abundances and our models is $0.4$ dex, within two times the average observational uncertainty ($\averr=0.2$ dex). 

The computed residuals of ls, hs and lead confirm the results obtained for the example star CS22898--027: the elements of the first $s$-peak are overestimated on average by $0.2-0.5$ dex, whereas lead and the elements of the second $s$-peak are underestimated by $0.1-0.2$ dex, with the exception of cerium and lanthanum ($\avres\approx0$). A systematic discrepancy, negative for strontium and positive for barium and lead, is present also in the distributions of the residuals in Fig. \ref{fig:histo_resid} (blue histograms). Because none of the model stars reproduces the large observed hs-to-ls ratio, to minimise $\chisq$ our model predicts larger abundances of strontium, and smaller abundances of barium and lead compared to the observations.

The yttrium residuals are typically lower than strontium and zirconium, analogously to the results obtained for the sample of CEMP-$s/nr$ stars. The abundance of niobium (atomic number $41$) is observed only in star CS29497--030 and is underestimated by $0.34$ dex (with observational uncertainty $0.2$ dex). This is peculiar because according to the models niobium is formed from the radioactive decay of $^{93}\Zr$, and the abundance of zirconium is well reproduced in this stars ($R\approx0$). This discrepancy is possibly related to the uncertain neutron-capture cross section of $^{93}\Zr$, as we discuss in Sect. \ref{disc-1}.

The $r$-elements are systematically underproduced, perhaps not surprisingly because in our nucleosynthesis model the $r$-process is not included. Our models typically produce $[\Ba/\Eu]$ close to 1, and the maximum europium enhancement is approximately $[\Eu/\Fe]=1.5$, whereas in most CEMP-$s/r$ stars the observed $[\Ba/\Eu]$ is below $0.6$ and $[\Eu/\Fe]>1.5$.

\subsection{CEMP-s/ur stars}
\label{CEMP-s/ur}

$37$ stars in our sample do not have an observed abundance of europium, and therefore are classified as CEMP-$s/ur$ stars, $15$ of which are listed in Tables \ref{tab:best-WRLOFq-ssw}--\ref{tab:best-WRLOFq-gamma2} because at least $6$ elements have been observed. Three of these $15$ stars have measured orbital periods and are discussed in Paper I.
One of them, BD$+04^{\circ}2466$, has a very wide orbit ($\Porb\approx4600$ days), and even if we do not consider the period constraint we find the same input parameters as in Paper~I. %
On the contrary, to reproduce the period of HD$201626$ ($407$ days), our model binary stars experience a common-envelope phase which shrinks the orbit. Without the period constraint the model progenitor system of HD$201626$ has a $1.4\,\Msun$ primary star that transfers $\Delta\Macc\approx0.18\,\Msun$ to its companion star in a wide orbit (cf. Tables \ref{tab:best-WRLOFq-ssw}--\ref{tab:best-WRLOFq-gamma2}). Consequently, the systems do not experience a common-envelope phase and therefore the final periods derived in the models are $20$ to $600$ times longer than that observed. Without the period constraint we find $\chisq\approx8$ ($\nu=5$), whereas in our best-fit model in Paper~I we find $\chisq\approx14$ ($\nu=6$). 
In Paper~I we find that the model for star HE0024--2523 needs to experience inefficient common-envelope ejection to reproduce the large enhancements of the heavy-$s$ elements. If we relax the period constraint the system does not enter a common-envelope phase and the final period of the system is $475$ days (with model set A), two orders of magnitude longer than the observed period of $3.14$ days. 

 \begin{figure*}[!t]
   \centering
   \includegraphics[height=0.36\textwidth]{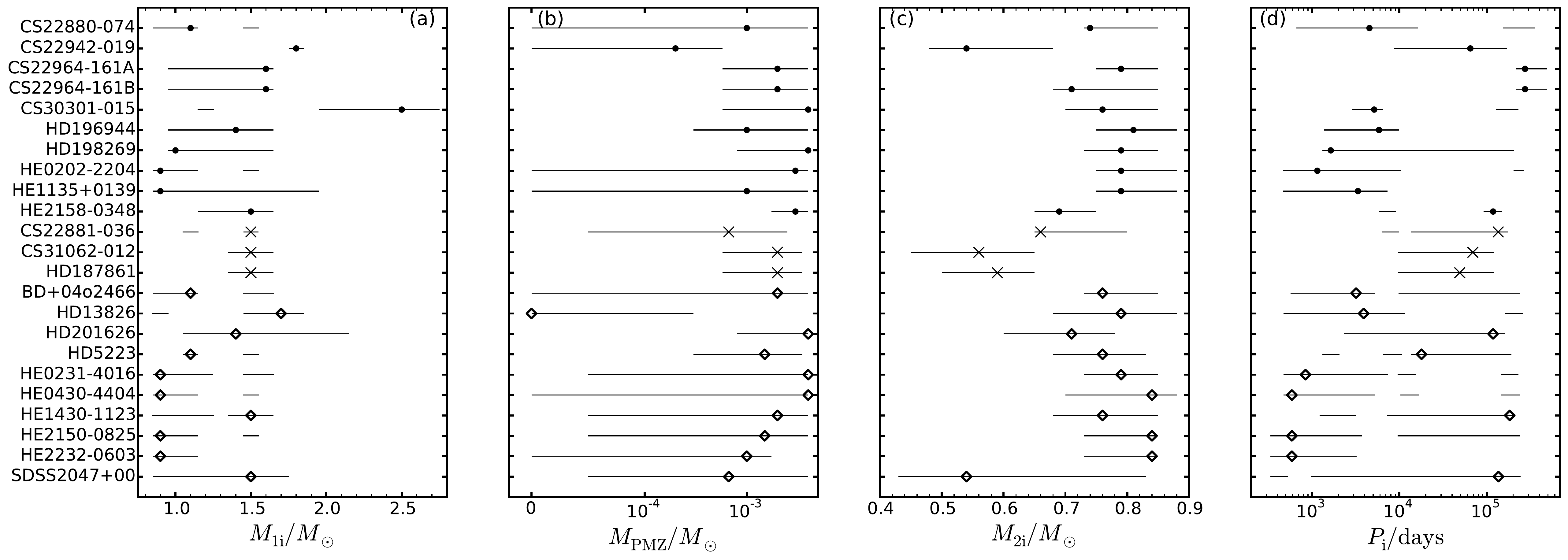}
   \caption{Initial parameters of our $23$ modelled stars with $\chimin/\nu\le3$ and $\nu\ge2$. CEMP-$s/nr$, CEMP-$s/r$ and CEMP-$s/ur$ stars are shown respectively with filled circles, crosses and open diamonds. The bars indicate the confidence intervals. From left to right: primary mass, $\Mprim$, mass of the PMZ, $\Mpmz$, secondary mass, $\Msec$, and orbital period, $\Pin$.}
    \label{fig:init_param}
 \end{figure*}

Fig. \ref{fig:residCEMP-s}c shows that carbon, magnesium, zirconium and the heavy-$s$ elements are reproduced within $0.1$ dex on average. The elements with the largest discrepancy with the observations are C+N ($\avres=-0.45$), sodium ($\avres=-0.25$), strontium and yttrium ($\avres=-0.2$) and lead ($\avres=0.25$). The abundance of nitrogen is observed only in five systems (BD$+04^{\circ}2466$, HD$13826$, HD$187216$, HD$201626$, HD$5223$) and the average results are greatly affected by the poor fit of star HD$187216$, which is possibly biased by large observational errors \cite[][]{Kipper1994}, and in which our model cannot reconcile the large enhancements of the $s$-elements with the solar ratios observed for nitrogen, sodium and magnesium. Our best fit reproduces the $s$-elements but overestimates C+N and sodium by $1.1$ and $0.7$ dex, respectively. The average residuals of C+N and sodium in CEMP-$s/ur$ stars decrease if we remove HD187216 from the sample ($\avres=-0.24$ and $0.18$, respectively). The distribution of strontium is greatly affected by the poor fit of star HE0212--0557, where the observed heavy-$s$ elements are enhanced by 2 dex whereas $[\Sr/\Fe]\approx0$, that is $1.2$ dex lower than predicted by our best model. Our model also overestimates the abundance of yttrium ($[\Y/\Fe]=0.7$) by $0.6$ dex, and the abundances of carbon and magnesium by $0.5$ dex. The models of \cite{Bisterzo2012} have similar discrepancies for these elements.
If we do not consider this star, the mean residuals of strontium and yttrium are $-0.1$ and $-0.15$, respectively. Although the differences are small, it may indicate the same systematic effect found in CEMP-$s/nr$ and CEMP-$s/r$ stars.
Fig. \ref{fig:histo_resid} shows that the abundances of barium and lead are reproduced within $0.3$ dex, that is $1.5$ times the average observational uncertainty ($\averr\approx0.2$), with the exception of the model for star HE0024--2523 that underestimates the observed $[\Pb/\Fe]=3.2$ by almost 1 dex.

\section{Confidence limits on the initials parameters}
\label{init-param}

The results presented in Sect. \ref{CEMPs} are based on the best-fitting models found from our $\chisq$-minimization procedure. The confidence intervals of the input parameters of our models (i.e. the initial primary and secondary masses, the initial period and the PMZ mass) are determined with the same procedure as described in Paper~I. In biref, on our grid of binary models and associated $\chisq$, we fix one of the input parameters $p$ and for each grid value of $p$ we determine the minimum $\chisq$ with respect to the other parameters, defining a function $\chisq(p)$. A confidence region is an interval of $p$ for which $\Delta\chisq=\chisq(p)-\chimin$ is below a certain threshold, where $\chimin$ is the $\chisq$ of the best fit. If the observational errors are Gaussian then the thresholds $\Delta\chisq=1$, $4$ and $9$ correspond to the confidence intervals with, respectively, $68.3\%$, $95.4\%$ and $99.7\%$ probability that the actual $p$ is in this interval. As we mention in Paper~I, the observational errors in our sample are not necessarily Gaussian, and they may in some cases be affected by systematic effects. Consequently, the thresholds $\Delta\chisq=1,\,4,\,9$ should not be used to calculate the theoretical Gaussian probabilities, and provide only an indication of the confidence levels. At a visual inspection models with $\chisq$ below the threshold $\Delta\chisq = 4$ predict very similar abundances to the best-fitting models. On the other hand, models with larger $\chisq$ are clearly distinct from the best model and the observations are not well reproduced.

Fig. \ref{fig:init_param} shows the input parameters of the $23$ model CEMP-$s$ stars with $\chimin/\nu<3$ and $\nu\ge2$ as determined with model set A of a spherically symmetric wind with WRLOF wind-accretion efficiency. CEMP-$s/nr$, CEMP-$s/r$ and CEMP-$s/ur$ stars are indicated as filled circles, crosses and open diamonds, respectively. The horizontal bars represent the confidence intervals determined with the threshold $\Delta\chisq<4$. Model stars with $\chimin/\nu>3$ are not included because they do not reproduce the observed abundances well enough, while stars with $\nu<2$ do not provide strong constraints on the models. The confidence intervals shown in Fig. \ref{fig:init_param} are summarised in Table \ref{tab:confi_WRLOFq-ssw}. In some stars we find multiple local minima in the $\chisq$ distribution. In these stars different combinations of initial parameters result in models with similar surface abundances and hence almost equal $\chisq$. As an example, for star CS22880--074 two local minima of $\chisq$ are found for models with primary mass between $0.9$ and $1.1\Msun$ and initial period between $600$ and $1.6\times10^4$ days, and another around $\Mprim=1.5\Msun$ and $\Pin=[1.6,\,3.5]\times10^5$ days. 

Most of the best-fitting models have primary mass below $2\,\Msun$. AGB models with masses in this range produce a significant amount of $s$-process elements regardless of the mass of the PMZ because of proton-ingestion events that occur in the first thermal pulses \cite[as described by][which they refer to as regime $4$ of neutron capture]{Lugaro2012}. As a consequence, the mass of the PMZ in our model stars is not well constrained, as indicated by the large confidence intervals in Fig. \ref{fig:init_param}b. Most model stars have $\Mpmz$ larger than $10^{-3}\,\Msun$ because in this range low-mass AGB models ($\Mprim<3\,\Msun$) produce positive [hs/ls] and [Pb/hs] as observed in the majority of our sample CEMP-$s$ stars.

 \begin{figure*}[!t]
   \centering
   \includegraphics[height=0.4\textwidth]{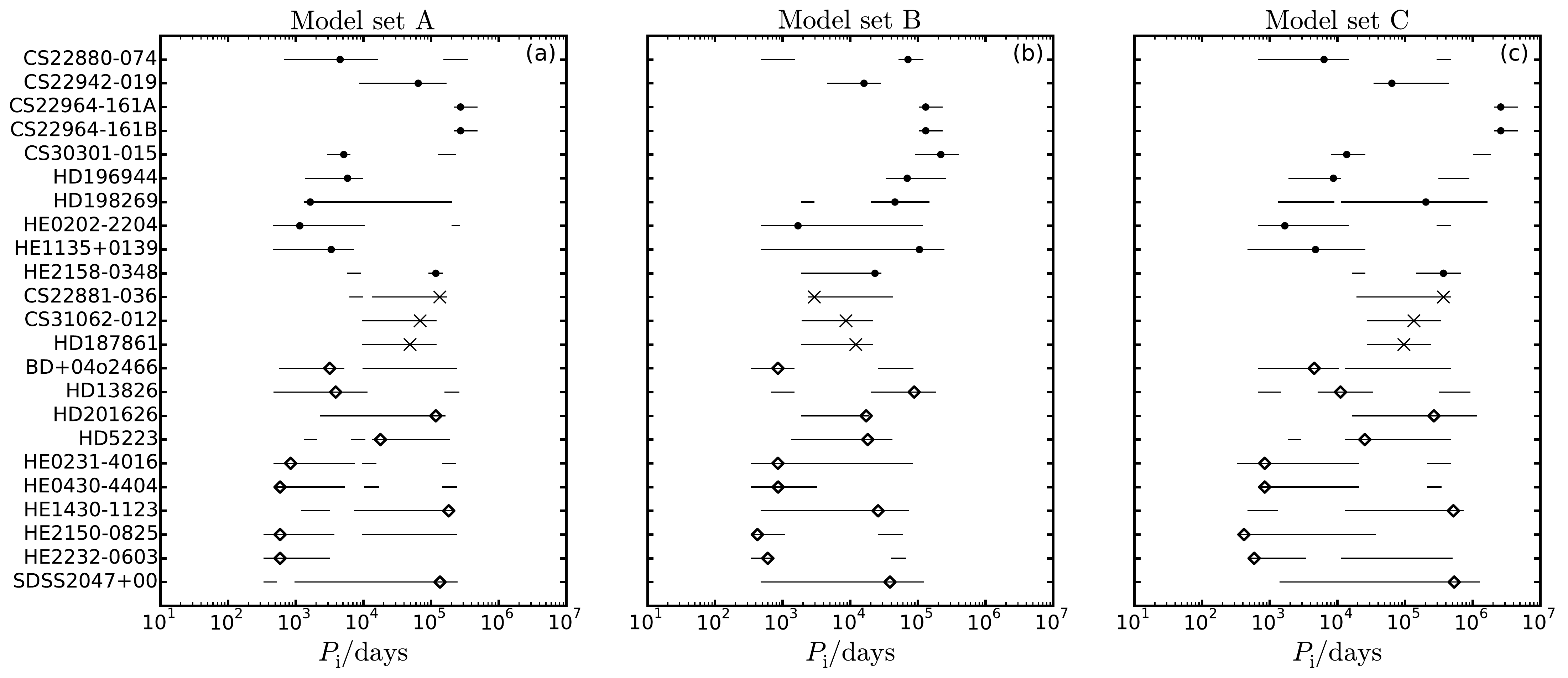}
   \includegraphics[height=0.4\textwidth]{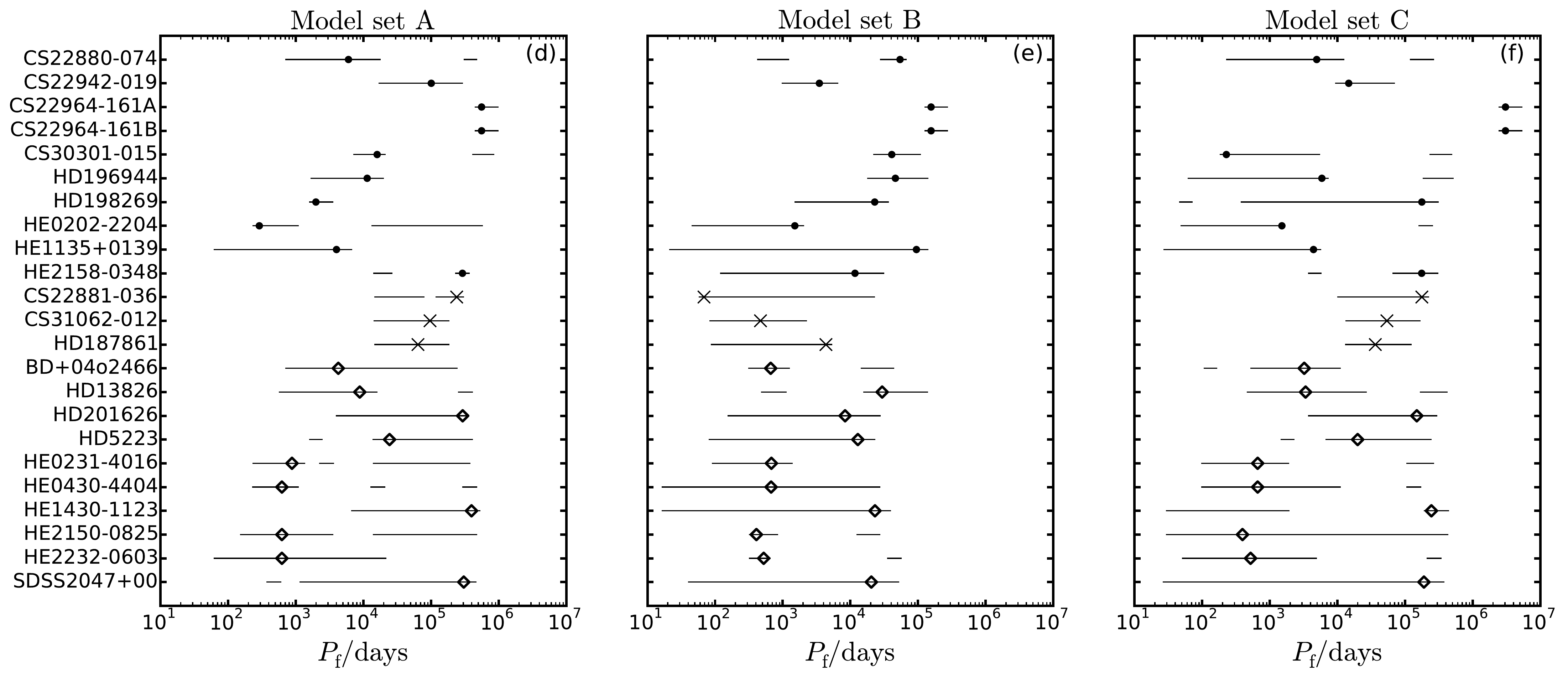}
   \caption{Initial and final periods, $\Pin$ and $\Pf$ (respectively, top and bottom panels), in our models with $\chimin/\nu\le3$ and $\nu\ge2$. Symbols are the same as Fig. \ref{fig:init_param}. From left to right model sets A, B and C are shown (panel a is the same as Fig. \ref{fig:init_param}d).}
    \label{fig:peri-Pf}
 \end{figure*}

Stars of mass around $1.5\,\Msun$ and with $\Mpmz\ge2\times 10^{-3}\,\Msun$ produce the largest $[\hs/\ls]$ value and $[\Pb/\hs]>0$. As a consequence, most observed stars with very enhanced heavy-$s$ elements and lead are best modelled with primary stars of masses between $\Mprim=1.4\,\Msun$ and $\Mprim=1.6\,\Msun$. CEMP-$s/r$ stars are mostly observed with $[\hs/\ls]$ close to one or higher, as discussed in Sect. \ref{CEMP-rs}, and therefore our best fits are found with $\Mprim=1.5\,\Msun$.
Two CEMP-$s/nr$ stars are modelled with $\Mprim$ in the mass range $[1.75,\,3]\,\Msun$, in which the $[\hs/\ls]$ ratio varies approximately between $-0.5$ and $0.5$ for $\Mpmz$ between $0$ and $2\times10^{-3}\,\Msun$ while [Pb/hs] is always positive \cite[regime 2 of][]{Lugaro2012}. 
The secondary mass (Fig. \ref{fig:init_param}c) in the models depends on the observed enhancements. 
The default assumption in our models, as in Paper~I, is that accreted material is efficiently mixed by thermohaline mixing. As a consequence, a large degree of accretion is required when the observed abundances are greatly enhanced. Hence, the secondary star has to be initially less massive to reach a mass of approximately $0.8-0.9\,\Msun$ after accretion, and still be visible at $t>10$ Gyr.

The input parameters $\Mprim$, $\Mpmz$ and $\Msec$ are similar in all model sets. On the contrary, there are several differences between the initial and final periods, as shown in the upper and lower panels of Fig. \ref{fig:peri-Pf}.
In model set C the binary systems typically start in wider orbits compared to the other model sets, because the dependence of the WRLOF process on the separation favours wider orbits compared to model set B and because of the more efficient angular momentum loss compared to model set A. The final periods computed with model set A and C are similar in many stars because the mass-transfer algorithm is the same in the two models, and therefore the binary stars need to have approximately the same separation during accretion to transfer the same amount of material. 
The majority of stars in model sets A and C have final periods longer than $4,\!600$ days, which is approximately the longest period measured in our sample. 
On the contrary, in model set B mass transfer is more efficient in close orbits, and consequently the majority of the systems are predicted at periods shorter than $10,\!000$ days. This is the only model set that predicts significant mass accretion in close orbits, as shown also in Paper I, and ten modelled stars have final periods less than $4,\!600$ days. The stars that need to accrete a small amount of mass to reproduce the observed abundances have longer periods because of the enhanced wind-accretion rate the and efficient angular momentum loss.

\section{Discussion}
\label{discussion}
\subsection{Comparison between observed and modelled abundances}
\label{disc-1}
Our best-fitting models are found with initial primary mass in the range between $0.9\Msun$ and $3\Msun$. In this mass range the neutron source is the $\Cth(\alpha,\, n)^{16}\Ox$ reaction that operates on $\Cth$ produced by the inclusion of a PMZ or by ingestion of protons in the stellar interior during the thermal pulses \cite[regimes $2-4$ in][]{Lugaro2012}. The abundances of $s$-elements predicted in the models depend on the mass of the AGB star and of its PMZ. We note that: ($i$) [Ba/Eu] is always close to 1 dex, ($ii$) the value of [hs/ls] varies from $[\hs/\ls]\approx-0.5$ for $\Mpmz=0\,\Msun$ up to $[\hs/\ls]\approx0.9$ for $\Mpmz\ge2\times10^{-3}\Msun$ and $\Mprim=1.5\Msun$, and ($iii$) [Pb/hs] is always positive, and in some cases greater than one, for $\Mprim$ above approximately $2\Msun$.

The success of our models in reproducing the observed abundances varies significantly for different classes of stars. CEMP-$s/nr$ stars typically exhibit $[\hs/\ls]\approx0.5$ dex and [Ba/Eu] between $0.5$ and $1$ dex, therefore our models generally well reproduce the abundances of the $s$-elements and also the $r$-elements.
On the other hand, most CEMP-$s/r$ stars have $[\hs/\ls]\gtrsim1$, and therefore our models are not able to reproduce the light-$s$ peak and the heavy-$s$ peak simultaneously. In most cases the model with the minimum $\chisq$ systematically overestimates the light-$s$ elements, on average by $0.3$ dex, and underestimates the heavy-$s$ elements, on average by $0.1$ dex. Furthermore, generally $[\Ba/\Eu]\lesssim1$ and the abundances of all $r$-elements are underestimated by a factor of two to up to one hundred. In our CEMP-$s/ur$ stars the abundances of the heavy-$s$ elements are typically reproduced within the observational uncertainty, as in CEMP-$s/nr$ stars, whereas the light-$s$ elements are systematically overestimated on average by $0.15$ dex, similar to CEMP-$s/r$ stars. These results are probably related to the reason why the abundance of europium is indeterminate. This group is likely to contain a mixture of CEMP-$s/nr$ stars and CEMP-$s/r$ stars. Hence, in some stars the abundance of europium is actually low, whereas in other stars europium is enhanced but it is not detected, for example because the europium lines are blended.

The abundance ratio between carbon and heavy-$s$ elements is generally positive in our AGB-nucleosynthesis model, $[\C/\hs]\gtrsim0$. Consequently, in our synthetic stars high abundances of heavy-$s$ elements are associated with large enhancements of carbon. A similar correlation is also found in our observed sample: compared to CEMP-$s/nr$ stars, CEMP-$s/r$ stars typically exhibit higher abundances of carbon (Fig.~\ref{fig:datasample}) and heavy-$s$ elements (as indicated by the increase of [hs/ls] with [Eu/Fe] in Fig.~\ref{fig:hsls}). However, our models overestimate the carbon-to-hs ratio in CEMP-$s/r$ stars. As a consequence, in the majority of these stars the carbon abundance is overestimated by our models with high [hs/Fe]. A similar problem is also encountered by \citet[][e.g. Fig. 13b]{Izzard2009} and in some models of \cite{Bisterzo2011, Bisterzo2012}. This probably indicates that the AGB-nucleosynthesis models should be able to produce higher hs abundances for a given amount of carbon, as we also discuss in a forthcoming paper \cite[][\emph{in press}]{Abate2015-3}.

Some discrepancies between models and observations are found in all three classes of CEMP-$s$ stars. The abundance of nitrogen is underestimated in $15$ out of $16$ stars by $0.1$ to $1.8$ dex. The difficulty in reproducing the observed abundance of nitrogen is a well known issue of AGB nucleosynthesis models. In AGB stars of mass below approximately $3\Msun$ some nitrogen is produced from carbon by a deep mixing process that operates at the bottom of the convective envelope, but the exact amount is very uncertain \cite[][]{Milam2009, Stancliffe2010, Karakas2010-1}. If we take into account the combined abundance of carbon and nitrogen our results improve significantly, and the observations are always reproduced within at most twice the observational uncertainty.

Oxygen is underestimated in almost all stars of our sample, on average by $0.4$ dex, approximately the same amount as in C-normal stars (cf. Sect. \ref{VMP}). The abundance of oxygen does not change much during AGB nucleosynthesis, and most likely the discrepancy could be solved by adopting a larger initial abundance without significantly affecting the abundances of the other elements.

The abundance of sodium is overpredicted in all our model stars, and this points to a general issue in our nucleosynthesis model. Sodium is produced in the intershell region of the AGB star due to proton and neutron capture on the abundant $^{22}\Ne$. The abundance of sodium increases rapidly with increasing PMZ mass and therefore models with a smaller PMZ predict lower abundances of sodium at each stellar mass. However, also the abundance of magnesium is sensitive to the mass of the PMZ and a large PMZ is required in most stars to reproduce its abundance, as well as the abundance of heavy-$s$ elements. \cite{Lugaro2012} suggest three effects that may help lower the predicted sodium abundance: ($i$) in the detailed models of AGB nucleosynthesis the density profile of protons introduced to make the $\Cth$-pocket could decrease more rapidly with depth \cite[e.g.,][]{Goriely2000}, ($ii$) a smaller $^{22}\Ne(p,\,\gamma)^{23}\Na$ reaction rate or ($iii$) a larger $^{23}\Ne(p,\,\alpha)^{20}\Ne$ reaction rate, both of which are very uncertain \cite[][]{Iliadis2010}.

In many CEMP-$s$ stars the abundance of yttrium is systematically lower than strontium and zirconium by $0.1-0.2$ dex, whereas in our models these elements are generally produced in approximately the same amount for every stellar mass and PMZ. This suggest that less yttrium should be produced in the models. Yttrium is produced by the reaction $^{88}\Sr(n,\,\gamma)^{89}\Sr$ and the subsequent decay of $^{89}\Sr$ to $^{89}\Y$ with a half life of $51$ days. The neutron-capture cross section of $^{89}\Sr$ is uncertain by a factor of two \cite[][]{Bao2000}. Qualitatively, a larger cross section implies that before the decay of $^{89}\Sr$ the reaction $^{89}\Sr(n,\,\gamma)^{90}\Sr$ may occur and after two consecutive $\beta$-decays $^{90}\Zr$ is produced while $^{89}\Y$ is skipped and this effect may help to reduce the discrepancy.

Niobium is observed in the stars CS29497--030 and HD$187216$ and it is underestimated by $0.35$ dex and $0.69$ dex, respectively \cite[although the abundances of HD$187216$ are very uncertain,][]{Kipper1994}. However, $^{93}\Nb$ is formed by $\beta$-decay of $^{93}\Zr$ and in the two stars zirconium is well reproduced. A similar problem is found by \cite{Kashiv2010} in the context of presolar SiC grains, in which the Nb/Zr ratio is systematically higher than in CI chondrites. These authors suggest that a cross section of $^{93}\Zr$ reduced by a factor of two removes the discrepancy and we note that this solution would help to reduce the difference observed in the two CEMP-$s$ stars. New measurements of the $^{93}\Zr$ cross section are currently being analysed and will provide a more accurate determination of the reaction rate \cite[][]{Lugaro2014}.

The abundances of barium are underestimated by our models in all but two CEMP-$s/r$ stars and in the majority of the CEMP-$s/ur$ stars. However, we note that the observed barium abundances might be affected by large spectroscopic uncertainties because most of the observed barium lines are strong resonance lines, often saturated and sensitive to non-LTE effects \cite[][]{Busso1995, Andrievsky2009, Masseron2010}.

The observed abundance of lead in most CEMP-$s$ stars is larger than predicted by the models. The lead abundance increases with increasing stellar mass ([Pb/Fe] is maximum for $\Mprim$ between $2\Msun$ and $3\Msun$) and PMZ mass. In most of the stars in our sample the choice of a more massive primary star implies that the abundances of sodium, magnesium, and the light-$s$ elements are overestimated. This suggest that nucleosynthesis models of low-mass stars ($\Mprim<2\Msun$) should produce higher [Pb/Na], [Pb/Mg] and [Pb/ls]. On the other hand, the determination of lead abundances generally rely on one single line blended by CH-lines, therefore the observed lead abundances may be affected by systematic errors \cite[][]{Masseron2010}.

Lead is one of the most neutron-rich elements produced by AGB stars, and therefore its abundance is strongly metallicity-dependent \cite[][]{Travaglio2001}.
To assess the effect of metallicity on our results, we use the theoretical models presented in Fig. 1 of \cite{Travaglio2001} to construct a table of correction factors which account for the differences in lead abundance predicted by their models in the range $-2.8\le[\Fe/\Hy]\le-1.8$ and our model at $[\Fe/\Hy]=-2.3$. We correct the lead abundances predicted in our models by interpolating in this table the iron abundances observed in CEMP-$s$ stars. However, the distribution of the residuals of [Pb/Fe] does not improve with this analysis, even if we vary the correction factors based on different models of \cite{Travaglio2001}. The residuals of $[\Pb/\Fe]$ computed with our models do not exhibit a dependence on $[\Fe/\Hy]$. Consequently, even correcting for different metallicities, the discrepancies between our model predictions and the observations are not reduced. Hence, as yet it is not possible to quantify the errors introduced in our analysis by the assumption of models with fixed metallicity. A larger sample of reliable abundances of lead (and other $s$-elements) in CEMP-$s$ stars will help to address this issue.

\subsection{Comparison with previous results}

We compare our results with the analysis performed by \cite{Bisterzo2012} on $94$ CEMP-$s$ stars in the range of metallicity $-3.5 \le [\Fe/\Hy] \le -1.0\,$, $60$ of which have $-2.8 \le [\Fe/\Hy] \le -1.8\,$ and are in common with our sample. One CEMP-$s/r$ star (HE2258--6358, analysed by \citealp{Placco2013}) and six CEMP-$s/ur$ stars are only included in our sample. In five of these stars the abundances of only a few elements have been determined (hence $\nu\le1$), whereas HD187216 was not included in the study of \cite{Bisterzo2012} because of the possibly large observational biases.

\cite{Bisterzo2012} compare the observations with the results of AGB nucleosynthesis models of single stars with masses between $1.2\,\Msun$ and $2\,\Msun$ in a range of metallicities and $\Cth$-pockets.
A quantitative comparison with their results is not straighforward because of the intrinsic differences in the methods used in the two studies. For example, the treatment of the $\Cth$-pocket/PMZ is not the same in the AGB models by \cite{Bisterzo2012} and by \citet[on which our model is based]{Karakas2010}. In the former, the parameter that determines the efficiency of the $s$-process is the abundance of $\Cth$ in the pocket, which is added at the top of the intershell region at each third dredge-up.
In the latter, the size of the PMZ determines the mass of a layer of free protons that are mixed with the intershell. The protons react with the elements in the intershell and the mass fractions of $\Cth$, $\Nfo$ and other isotopes are calculated consistently. The standard mass of the $\Cth$-pocket in the models by \cite{Bisterzo2012} corresponds roughly to $\Mpmz=2\times10^{-3}\Msun$ in our models.
Another important difference of our study is that we explicitly compute the binary evolution, wind mass-transfer process and the mixing effects while in the paper by \cite{Bisterzo2012} these aspects are mimicked by including a dilution factor, i.e. the ratio between the amount of accreted material and the envelope mass of the accreting star. 

Despite these differences, in $46$ of the $60$ CEMP-$s$ stars common to both studies we find primary masses that are consistent with the results of \cite{Bisterzo2012} within the confidence intervals of the models. Among these stars are $21$ of the $23$ stars for which we find a good fit (with $\chimin/\nu\le3$ and $\nu\ge2$). The two exceptions are HD5223 and HE2232--0603. Compared to the results of \cite{Bisterzo2012}, in these CEMP-$s/ur$ stars our models with low primary mass ($0.9-1.1\,\Msun$) better match the observed carbon and lead (in both stars), fluorine, sodium and magnesium (in HD$5223$), whereas larger $\Mprim$ overestimate the observed abundances.

Of the stars for which there is no agreement with the masses predicted by \cite{Bisterzo2012}, ten are CEMP-$s/r$ stars, while the other four are CEMP-$s/ur$ stars. One possible explanation of these discrepancies is in the way \cite{Bisterzo2012} account for the contribution of the $r$-process in these stars. The authors assume that CEMP-$s/r$ stars were formed from gas polluted by the explosion of a Type~II supernova, and therefore adopt in their models an initial enrichment of the $r$-process component of the elements heavier than iron, between $[r/\Fe]^{\mathrm{ini}}=0.5$ and $[r/\Fe]^{\mathrm{ini}}=2$. In CEMP-$s/r$ stars this initial enhancement is chosen to reproduce the observed $[\Eu/\Fe]$.
The enhanced abundance of the $r$-process component does not alter significantly the abundance of the $s$-process elements, except in low-mass stars ($\Mprim\lesssim1.7\Msun$) with $[r/\Fe]^{\mathrm{ini}}=2$, in which case the heavy-$s$ elements are enriched by an extra factor of a few tenths of a dex \cite[e.g., Fig. 6 of][]{Lugaro2012}. As a consequence, the $r$-enriched models of \cite{Bisterzo2012} with low initial masses ($M=1.2-1.4\Msun$) and relatively small $\Cth$-pockets, corresponding roughly to $\Mpmz\le10^{-3}\Msun$, reproduce the abundance of the heavy-$s$ elements, $r$-elements and lead without greatly overestimating the abundances of the light-$s$ elements. In contrast, in our models neutron-capture elements are entirely produced by the $s$-process during the AGB phase. Consequently, our models of CEMP-$s/r$ stars are constrained towards relatively massive primary stars, $1.5\Msun\le\Mprim\le2.9\,\Msun$, with $\Mpmz\ge2\times10^{-3}\,\Msun$, which predict large abundances of elements heavier than barium but typically overestimate the observed abundances of lighter elements.

The origin of the $r$-elements in CEMP-$s/r$ stars is unclear. Many hypotheses have been suggested that can be classified in three main classes. In the first class of scenarios the binary systems were formed in a molecular cloud already enriched in $r$-elements by the nearby explosion of one or more Type II supernovae \cite[][]{Cohen2003, Jonsell2006, Sneden2008, Bisterzo2011, Bisterzo2012}. In the second class of scenarios, the primary star of the binary system is relatively massive and undergoes the AGB phase, producing the $s$-elements, and then explodes as an electron-capture or a Type 1.5 supernova, providing the $r$-elements \cite[][]{Zijlstra2004, Wanajo2006, Jonsell2006}. If the $s$-process and the $r$-process enrichments are independent, as suggested in the first class of scenarios, it is difficult to explain the correlation between the abundances of barium and europium that is observed in CEMP-$s/nr$ stars and in CEMP-$s/r$ stars. Moreover, the massive AGB stars invoked in the second class of scenarios are rare in a solar-neighbourhood initial mass function \cite[][]{Pols2012}. Also, massive AGB stars produce more nitrogen than carbon \cite[e.g.][]{Karakas2007} and would therefore produce nitrogen-enhanced metal-poor stars rather than CEMP stars. For a thorough discussion of these scenarios we refer to \cite{Jonsell2006} and \cite{Lugaro2009}. 
\cite{Lugaro2009} suggest a third speculative scenario in which $r$- and $s$-elements are both formed in low-mass extremely metal-poor AGB stars ($[\Fe/\Hy]<-3$), that may be able to produce very large densities of free neutrons when protons are ingested in the region of the He-flash \cite[][]{Campbell2007, Campbell2008, Herwig2011, Reifarth2014}.

\subsection{Constraints on binary mass transfer}

We considered several model sets with different combinations of assumptions for the models of wind mass-accretion rate and angular momentum loss. 
Because in this study we do not consider constraints on the orbital period there are no significant differences between the results obtained with the three model sets, with the exception of the distributions of initial and final orbital periods of the binary systems.
Without the constraint on the orbital period our fitting procedure mostly selects model systems with periods longer than $5,\!000$ days, which is more than the longest period currently observed.
In our default model set A relatively long periods are necessary to accrete large amounts of mass. Therefore, when large mass accretion is required to reproduce the observed abundances we find modelled CEMP-$s$ stars with periods approximately $20-500$ times longer than predicted in Paper~I, in which the observed orbital periods are used to constrain our models (e.g. the stars CS$22942-019$ and CS$29497-030$). Shorter periods are found in a few cases if the observed abundances are reproduced by a model with a low-mass primary star ($\Mprim\le1.1\Msun$) and small accreted mass. For these stars we compute orbital periods of a few thousand days, consistent with the results of Paper I (e.g. the stars BD$+04^{\circ}2466$ and HD$198269$).

With our model set A we find that our best-fitting model systems have orbital periods between $600$ days and $400,\!000$ days with an average period of $90,\!000$ days. With model sets B and C we obtain on average $\Pf=17,\!000$ and $\Pf=60,\!000$ days, respectively. 
On the other hand, the average orbital period of the systems analysed in Paper I is approximately $1,\!500$ days. Furthermore, a recent statistical study of currently available radial velocity variations in CEMP-$s$ stars indicates that these stars are expected to have a maximum period of about $10,\!000$ days and an average period of $400$ days \cite[][]{Starkenburg2014}.
This suggests that a large proportion of the CEMP-$s$ stars in our sample have periods of less than a few thousand days, i.e. in closer orbits than we find in our best fits. If this hypothesis is correct, then AGB stars in binary systems have to transfer wind material with great efficiency at relatively short separations and this process is currently possible in our models only with some ad hoc assumptions, namely highly efficient mass-transfer and efficient angular momentum loss. This hypothesis is supported by the period distribution derived in barium stars, the equivalent of CEMP-$s$ stars at solar metallicity \cite[e.g.][]{Jorissen1998}, and by observations of ongoing mass transfer in post-AGB binaries \cite[e.g.][]{Gorlova2012}.

\section{Summary and conclusions}
\label{conclusions}
In this work we have used our model of binary evolution and nucleosynthesis to calculate the best-fitting models to the surface abundances observed in a sample of $67$ CEMP-$s$ stars. We distinguish three classes of CEMP-$s$ stars based on the europium abundance. In CEMP-$s/nr$ stars the abundance of europium relative to iron is up to ten times as high as in the Sun. In CEMP-$s/r$ stars the europium-to-iron ratio is higher than ten times the solar ratio. In CEMP-$s/ur$ stars the abundance of europium is indeterminate.
From the comparison between the observed and modelled abundances of these stars we draw the following conclusions.
\begin{itemize}
\item The model stars that provide the best fit to the abundances observed in CEMP-$s$ stars have low initial mass (up to $2.5\Msun$). In this range, the main neutron source for the $s$-process is the $\Cth(\alpha,\,n)^{16}\Ox$ reaction.
\item The chemical properties observed in CEMP-$s/nr$ and CEMP-$s/r$ stars are fundamentally different. 
The results of our models are consistent with the abundances observed in CEMP-$s/nr$ stars.
On the contrary, most of the model CEMP-$s/r$ stars overestimate the observed abundances of light-$s$ elements and underestimate the abundances of heavy-$s$ elements, $r$-elements and lead. In CEMP-$s/r$ stars the ratio of heavy-$s$ elements to light-$s$ elements is too high to be reproduced in our models. This result points to a different nucleosynthesis process at the origin of the abundances in CEMP-$s/r$ stars, that enhances the heaviest elements without affecting the abundances of elements such as sodium, magnesium and light-$s$ elements.
\item The abundance of sodium is always overpredicted in our models. This discrepancy likely points to a problem in the numerical treatment of the partial mixing zone. A proton profile in the partial mixing zone weighted towards a low proton abundance may contribute to reducing this discrepancy.
\item The orbital periods predicted in this work are significantly longer than the results of Paper I, in which our models are constrained to reproduce the observed orbital periods of the systems. This indicates that either the currently known periods of CEMP-$s$ stars are biased towards close orbits and are not representative of the whole population, or our model of wind accretion in binary stars needs to efficiently transfer mass and lose angular momentum in close orbits.
\end{itemize}

\begin{acknowledgements}
We are grateful to Dr. T. Suda for his kind support and help to access the SAGA database. We thank Dr. M. Lugaro, Dr. R.J. Stancliffe, Prof. R. Gallino and Dr. T. Masseron for many useful comments and constructive criticism of our work. CA is grateful for financial support from the Netherlands Organisation for Scientific Research (NWO) under grant 614.000.901. RGI thanks the Humboldt Foundation and the Science and Technology Facilities Council (STFC) for funding support. We thank the referee for her/his helpful comments that improved the clarity of this paper.
\end{acknowledgements}


\Online
\begin{appendix}
\section{Tables}
\label{app:A}
In this section we summarise the observed and modelled properties of the $67$ CEMP-$s$ stars in our sample.
Table \ref{tab:obs} shows the observed temperatures, surface gravities and abundances of selected elements.
The parameters of the best-fitting models to the CEMP-$s$ stars with at least of two degrees of freedom ($\nu\ge2$), as computed with model sets A, B and C, are shown in Tables \ref{tab:best-WRLOFq-ssw}--\ref{tab:best-WRLOFq-gamma2}, as follows. 
\begin{itemize}
\item[] Columns 2--5: the fitted parameters $\Mprim$, $\Mpmz$, $\Msec$ and $\Pin$.
\item[] Columns 6--8: the mass accreted by the secondary star, $\Delta\Macc$, the orbital period of the binary when the secondary star best reproduces the observed $\logg$ and surface abundances, $\Pf$, and for stars with observed orbital periods, the final period determined in Paper~I.
\item[] Columns 9--11: $\chimin$, the number of degrees of freedom of the fit, $\nu$, and the reduced $\chisq$ of the best fit, i.e. $\chimin/\nu=\chisq/\nu$. 
\end{itemize}
In Tables \ref{tab:few-WRLOFq-ssw}--\ref{tab:few-WRLOFq-gamma2} the physical parameters of the best-fitting models of stars with $\nu\le1$ are shown.The confidence intervals of the input parameters of our models with $\chimin/\nu\le3$ and $\nu\ge2$ computed with model set A are shown in Table \ref{tab:confi_WRLOFq-ssw}.

\begin{table*}[!ht]
\caption{Surface gravities, temperatures and chemical properties observed in the $67$ CEMP-$s$ stars of our sample.}
\label{tab:obs}
\centering
\tiny
\begin{tabular}{ l  c c c  c  c c c c }
\hline
\hline
\hspace{0.7cm}ID & $\loggunits$ & $\Teff/\mathrm{K}$ & $\Porb/\mathrm{days}$ & number of & $[\Fe/\Hy]$ & $[\C/\Fe]$ & $[s/\Fe]$ & $[\Eu/\Fe]$ \\ & & & & observed elements & & & & \\
\hline
 & & & & & & & & \\ 
CEMP-$s/nr$ stars & & & & & & & & \\ 
\hline
CS22880$-$074 & $3.9\pm0.13$ & $5917$ & $ $ & $21$ & $-1.9$ & $1.5$ & $1.3$ & $0.6$ \\ 
CS22942$-$019 & $2.2\pm0.4$ & $4967$ & $2800$ & $18$ & $-2.7$ & $2.2$ & $1.8$ & $0.8$ \\ 
CS22964$-$161A & $3.7\pm0.2$ & $6050$ & $252.481$ & $21$ & $-2.4$ & $1.6$ & $1.4$ & $0.7$ \\ 
CS22964$-$161B & $4.1\pm0.4$ & $5850$ & $252.481$ & $22$ & $-2.4$ & $1.4$ & $1.3$ & $0.7$ \\ 
CS30301$-$015 & $0.8\pm0.1$ & $4750$ & $ $ & $17$ & $-2.7$ & $1.7$ & $1.5$ & $0.3$ \\ 
HD196944 & $1.8\pm0.1$ & $5234$ & $ $ & $27$ & $-2.4$ & $1.4$ & $1.2$ & $0.3$ \\ 
HD198269 & $1.3\pm0.25$ & $4800$ & $1295$ & $11$ & $-2.2$ & $1.7$ & $1.7$ & $0.9$ \\ 
HE0202$-$2204 & $1.6\pm0.1$ & $5280$ & $ $ & $21$ & $-2.0$ & $1.2$ & $1.4$ & $0.5$ \\ 
HE1135$+$0139 & $1.8\pm0.1$ & $5487$ & $ $ & $20$ & $-2.4$ & $1.1$ & $1.1$ & $0.4$ \\ 
HE2158$-$0348 & $2.5\pm0.1$ & $5215$ & $ $ & $18$ & $-2.8$ & $2.1$ & $1.6$ & $0.9$ \\ 
\hline
 & & & & & & & & \\ 
CEMP-$s/r$ stars & & & & & & & & \\ 
\hline
BS16080$-$175 & $3.7\pm0.2$ & $6240$ & $ $ & $6$ & $-1.9$ & $1.8$ & $1.6$ & $1.1$ \\ 
BS17436$-$058 & $2.7\pm0.2$ & $5690$ & $ $ & $7$ & $-1.8$ & $1.6$ & $1.7$ & $1.2$ \\ 
CS22881$-$036 & $4.0\pm0.1$ & $6200$ & $ $ & $14$ & $-2.1$ & $2.1$ & $1.9$ & $1.0$ \\ 
CS22898$-$027 & $3.7\pm0.28$ & $6110$ & $ $ & $22$ & $-2.3$ & $2.0$ & $2.3$ & $2.0$ \\ 
CS22948$-$027 & $1.8\pm0.4$ & $4800$ & $426.5$ & $21$ & $-2.5$ & $2.4$ & $2.4$ & $1.9$ \\ 
CS29497$-$030 & $4.0\pm0.5$ & $6966$ & $344$ & $33$ & $-2.5$ & $2.4$ & $2.3$ & $1.7$ \\ 
CS29526$-$110 & $3.2\pm0.1$ & $6500$ & $ $ & $18$ & $-2.4$ & $2.3$ & $2.1$ & $1.8$ \\ 
CS31062$-$012 & $4.2\pm0.38$ & $6099$ & $ $ & $24$ & $-2.8$ & $2.3$ & $2.1$ & $1.6$ \\ 
CS31062$-$050 & $2.9\pm0.24$ & $5489$ & $ $ & $37$ & $-2.5$ & $1.9$ & $2.4$ & $2.0$ \\ 
HD187861 & $2.0\pm0.35$ & $4960$ & $ $ & $14$ & $-2.4$ & $2.0$ & $1.9$ & $1.3$ \\ 
HD224959 & $1.9\pm0.25$ & $5050$ & $1273$ & $14$ & $-2.1$ & $1.8$ & $2.2$ & $1.7$ \\ 
HE0131$-$3953 & $3.8\pm0.1$ & $5928$ & $ $ & $16$ & $-2.7$ & $2.5$ & $2.2$ & $1.7$ \\ 
HE0143$-$0441 & $4.0\pm0.35$ & $6305$ & $ $ & $22$ & $-2.4$ & $2.0$ & $2.4$ & $1.7$ \\ 
HE0338$-$3945 & $4.1\pm0.33$ & $6161$ & $ $ & $32$ & $-2.5$ & $2.1$ & $2.4$ & $2.0$ \\ 
HE1105$+$0027 & $3.5\pm0.1$ & $6132$ & $ $ & $16$ & $-2.5$ & $2.0$ & $2.4$ & $1.9$ \\ 
HE1305$+$0007 & $1.5\pm0.5$ & $4655$ & $ $ & $21$ & $-2.2$ & $2.1$ & $2.6$ & $2.2$ \\ 
HE2148$-$1247 & $3.9\pm0.1$ & $6380$ & $ $ & $25$ & $-2.4$ & $2.0$ & $2.4$ & $2.0$ \\ 
HE2258$-$6358 & $1.6\pm0.1$ & $4900$ & $ $ & $31$ & $-2.7$ & $2.4$ & $2.3$ & $1.7$ \\ 
LP625$-$44 & $2.6\pm0.3$ & $5500$ & $>4383$ & $31$ & $-2.8$ & $2.3$ & $2.8$ & $1.9$ \\ 
SDSSJ0912$+$0216 & $4.5\pm0.1$ & $6500$ & $ $ & $28$ & $-2.6$ & $2.3$ & $1.6$ & $1.3$ \\ 
\hline
 & & & & & & & & \\ 
CEMP-$s/ur$ stars & & & & & & & & \\ 
\hline
BD$+$04$^{\circ}$2466 & $1.8\pm0.2$ & $5032$ & $4592.7$ & $20$ & $-2.1$ & $1.3$ & $1.6$ & $ $ \\ 
CS22945$-$017 & $3.8\pm0.2$ & $6400$ & $ $ & $6$ & $-2.5$ & $2.3$ & $0.6$ & $ $ \\ 
CS22956$-$028 & $3.9\pm0.1$ & $6900$ & $1290$ & $11$ & $-2.1$ & $1.9$ & $0.4$ & $ $ \\ 
CS29509$-$027 & $4.2\pm0.1$ & $7050$ & $196$ & $5$ & $-2.1$ & $1.5$ & $1.3$ & $ $ \\ 
CS30338$-$089 & $2.1\pm0.1$ & $5000$ & $ $ & $10$ & $-2.5$ & $2.1$ & $2.3$ & $ $ \\ 
HD13826 & $0.1\pm0.35$ & $3540$ & $ $ & $9$ & $-2.5$ & $1.6$ & $1.4$ & $ $ \\ 
HD187216 & $0.4\pm0.4$ & $3500$ & $ $ & $14$ & $-2.5$ & $1.4$ & $2.3$ & $ $ \\ 
HD201626 & $2.2\pm0.25$ & $5190$ & $407$ & $11$ & $-2.1$ & $2.1$ & $2.0$ & $ $ \\ 
HD5223 & $1.0\pm0.25$ & $4500$ & $ $ & $18$ & $-2.1$ & $1.6$ & $1.9$ & $ $ \\ 
HE0012$-$1441 & $3.5\pm0.1$ & $5730$ & $ $ & $7$ & $-2.7$ & $1.9$ & $1.3$ & $ $ \\ 
HE0024$-$2523 & $4.3\pm0.1$ & $6625$ & $3.14$ & $17$ & $-2.7$ & $2.1$ & $1.6$ & $ $ \\ 
HE0206$-$1916 & $2.7\pm0.1$ & $5200$ & $ $ & $8$ & $-2.2$ & $2.1$ & $2.0$ & $ $ \\ 
HE0212$-$0557 & $2.1\pm0.4$ & $5075$ & $ $ & $15$ & $-2.3$ & $1.9$ & $2.2$ & $ $ \\ 
HE0231$-$4016 & $3.6\pm0.1$ & $5972$ & $ $ & $17$ & $-2.1$ & $1.3$ & $1.5$ & $ $ \\ 
HE0322$-$1504 & $0.8\pm0.1$ & $4460$ & $ $ & $5$ & $-2.0$ & $2.4$ & $2.8$ & $ $ \\ 
HE0430$-$4404 & $4.3\pm0.1$ & $6214$ & $ $ & $15$ & $-2.1$ & $1.3$ & $1.6$ & $ $ \\ 
HE0441$-$0652 & $1.4\pm0.1$ & $4900$ & $ $ & $11$ & $-2.6$ & $1.4$ & $1.2$ & $ $ \\ 
HE0507$-$1430 & $0.8\pm0.1$ & $4560$ & $446$ & $5$ & $-2.4$ & $2.7$ & $1.3$ & $ $ \\ 
HE0534$-$4548 & $1.5\pm0.3$ & $4250$ & $ $ & $4$ & $-1.8$ & $1.5$ & $0.6$ & $ $ \\ 
HE1045$-$1434 & $1.8\pm0.1$ & $4950$ & $ $ & $5$ & $-2.5$ & $3.3$ & $3.0$ & $ $ \\ 
HE1157$-$0518 & $2.0\pm0.1$ & $4900$ & $ $ & $8$ & $-2.4$ & $2.2$ & $2.2$ & $ $ \\ 
HE1429$-$0551 & $1.5\pm0.1$ & $4700$ & $ $ & $7$ & $-2.6$ & $2.3$ & $1.7$ & $ $ \\ 
HE1430$-$1123 & $3.8\pm0.1$ & $5915$ & $ $ & $15$ & $-2.7$ & $1.8$ & $1.8$ & $ $ \\ 
HE1434$-$1442 & $3.1\pm0.4$ & $5420$ & $ $ & $15$ & $-2.5$ & $2.1$ & $1.3$ & $ $ \\ 
HE1443$+$0113 & $1.9\pm0.1$ & $4945$ & $ $ & $5$ & $-2.2$ & $2.0$ & $1.4$ & $ $ \\ 
HE1447$+$0102 & $1.7\pm0.1$ & $5100$ & $ $ & $7$ & $-2.5$ & $2.5$ & $2.8$ & $ $ \\ 
HE1523$-$1155 & $1.6\pm0.1$ & $4800$ & $ $ & $6$ & $-2.2$ & $1.9$ & $1.8$ & $ $ \\ 
HE1528$-$0409 & $1.8\pm0.1$ & $5000$ & $ $ & $6$ & $-2.7$ & $2.4$ & $2.4$ & $ $ \\ 
HE2150$-$0825 & $3.7\pm0.1$ & $5960$ & $ $ & $18$ & $-2.0$ & $1.4$ & $1.7$ & $ $ \\ 
HE2221$-$0453 & $0.4\pm0.1$ & $4400$ & $ $ & $7$ & $-2.3$ & $1.8$ & $1.8$ & $ $ \\ 
HE2227$-$4044 & $3.9\pm0.1$ & $5811$ & $ $ & $14$ & $-2.4$ & $1.7$ & $1.4$ & $ $ \\ 
HE2228$-$0706 & $2.6\pm0.1$ & $5100$ & $ $ & $8$ & $-2.5$ & $2.3$ & $2.6$ & $ $ \\ 
HE2232$-$0603 & $3.5\pm0.1$ & $5750$ & $ $ & $17$ & $-2.1$ & $1.6$ & $1.6$ & $ $ \\ 
HE2240$-$0412 & $4.3\pm0.1$ & $5852$ & $ $ & $13$ & $-2.2$ & $1.4$ & $1.4$ & $ $ \\ 
SDSS0924$+$40 & $4.0\pm0.3$ & $6200$ & $ $ & $12$ & $-2.6$ & $2.7$ & $1.9$ & $ $ \\ 
SDSS1707$+$58 & $4.2\pm0.3$ & $6700$ & $ $ & $8$ & $-2.6$ & $2.1$ & $3.5$ & $ $ \\ 
SDSS2047$+$00 & $4.5\pm0.3$ & $6600$ & $ $ & $12$ & $-2.1$ & $2.0$ & $1.6$ & $ $ \\ 
\hline
\end{tabular}
\tablefoot{Barium abundance is used as indicator of $s$-elements. In HD198269, HD13826 and HD201626 lanthanum is used because barium is not observed.\\
}
\end{table*}%


\begin{table*}[!ht]
\caption{Physical parameters of the model CEMP-$s$ stars with number of degrees of freedom $\nu\ge2$ computed with model set A  with WRLOF wind-accretion efficiency and spherically symmetric wind.}
\label{tab:best-WRLOFq-ssw}
\centering
\begin{tabular}{ l  c c c c  c c c  c c c }
\hline
\hline
\hspace{1cm}ID & $\Mprim$ & $\Mpmz$ & $\Msec$ & $\Pin$ & $\Delta\Macc$ & $\Pf$ & $\Pf-$Paper~I & $\chimin$ & $\nu$ & $\chimin/\nu$ \\
\hline
 & & & & & & & & & & \\ 
CEMP-$s/nr$ stars & & & & & & & & & & \\ 
\hline
CS22880$-$074 & $1.1$ & $1.00 \times 10^{-3}$ & $0.74$ & $4.54 \times 10^{3}$ & $0.09$ & $6.03 \times 10^{3}$ &   & $16.8$ & $10$ & 1.7 \\ 
CS22942$-$019 & $1.8$ & $2.00 \times 10^{-4}$ & $0.54$ & $6.38 \times 10^{4}$ & $0.30$ & $1.03 \times 10^{5}$ & $ 2.87\times 10^3 $ & $21.1$ & $9$ & 2.3 \\ 
CS22964$-$161A & $1.6$ & $2.00 \times 10^{-3}$ & $0.79$ & $2.74 \times 10^{5}$ & $0.04$ & $5.60 \times 10^{5}$ & $ 5.60\times 10^5 $ & $19.5$ & $8$ & 2.5 \\ 
CS22964$-$161B & $1.6$ & $2.00 \times 10^{-3}$ & $0.71$ & $2.74 \times 10^{5}$ & $0.02$ & $5.60 \times 10^{5}$ & $ 5.60\times 10^5 $ & $11.8$ & $9$ & 1.3 \\ 
CS30301$-$015 & $2.5$ & $4.00 \times 10^{-3}$ & $0.76$ & $4.81 \times 10^{3}$ & $0.06$ & $1.68 \times 10^{4}$ &   & $11.9$ & $8$ & 1.5 \\ 
HD196944 & $1.4$ & $1.00 \times 10^{-3}$ & $0.81$ & $5.84 \times 10^{3}$ & $0.06$ & $1.14 \times 10^{4}$ &   & $13.8$ & $12$ & 1.2 \\ 
HD198269 & $1.0$ & $4.00 \times 10^{-3}$ & $0.79$ & $1.63 \times 10^{3}$ & $0.09$ & $1.97 \times 10^{3}$ & $ 1.43\times 10^3 $ & $9.2$ & $5$ & 1.8 \\ 
HE0202$-$2204 & $0.9$ & $3.00 \times 10^{-3}$ & $0.79$ & $1.19 \times 10^{3}$ & $0.06$ & $1.43 \times 10^{3}$ &   & $5.2$ & $7$ & 0.7 \\ 
HE1135$+$0139 & $0.9$ & $1.00 \times 10^{-3}$ & $0.79$ & $3.35 \times 10^{3}$ & $0.08$ & $4.01 \times 10^{3}$ &   & $2.8$ & $6$ & 0.5 \\ 
HE2158$-$0348 & $1.5$ & $3.00 \times 10^{-3}$ & $0.69$ & $1.32 \times 10^{5}$ & $0.17$ & $2.52 \times 10^{5}$ &   & $16.4$ & $7$ & 2.3 \\ 
\hline
 & & & & & & & & & & \\ 
CEMP-$s/r$ stars & & & & & & & & & & \\ 
\hline
CS22881$-$036 & $1.5$ & $6.66 \times 10^{-4}$ & $0.66$ & $1.32 \times 10^{5}$ & $0.15$ & $2.57 \times 10^{5}$ &   & $10.3$ & $6$ & 1.7 \\ 
CS22898$-$027 & $1.5$ & $6.66 \times 10^{-4}$ & $0.51$ & $2.44 \times 10^{4}$ & $0.31$ & $2.89 \times 10^{4}$ &   & $99.8$ & $11$ & 9.1 \\ 
CS22948$-$027 & $1.5$ & $2.00 \times 10^{-3}$ & $0.64$ & $9.42 \times 10^{4}$ & $0.24$ & $1.58 \times 10^{5}$ & $ 4.76\times 10^2 $ & $29.5$ & $9$ & 3.3 \\ 
CS29497$-$030 & $1.5$ & $2.00 \times 10^{-3}$ & $0.46$ & $2.47 \times 10^{4}$ & $0.28$ & $2.87 \times 10^{4}$ & $ 4.04\times 10^2 $ & $126.0$ & $16$ & 7.9 \\ 
CS29526$-$110 & $1.5$ & $2.00 \times 10^{-3}$ & $0.54$ & $2.42 \times 10^{4}$ & $0.32$ & $2.92 \times 10^{4}$ &   & $22.4$ & $6$ & 3.7 \\ 
CS31062$-$012 & $1.5$ & $2.00 \times 10^{-3}$ & $0.56$ & $6.79 \times 10^{4}$ & $0.25$ & $1.00 \times 10^{5}$ &   & $17.5$ & $8$ & 2.2 \\ 
CS31062$-$050 & $1.5$ & $2.00 \times 10^{-3}$ & $0.54$ & $2.42 \times 10^{4}$ & $0.32$ & $2.92 \times 10^{4}$ &   & $96.6$ & $20$ & 4.8 \\ 
HD187861 & $1.5$ & $2.00 \times 10^{-3}$ & $0.59$ & $4.78 \times 10^{4}$ & $0.31$ & $6.49 \times 10^{4}$ &   & $16.6$ & $8$ & 2.1 \\ 
HD224959 & $1.5$ & $2.00 \times 10^{-3}$ & $0.54$ & $2.42 \times 10^{4}$ & $0.32$ & $2.96 \times 10^{4}$ & $ 1.42\times 10^3 $ & $64.6$ & $8$ & 8.1 \\ 
HE0131$-$3953 & $1.5$ & $2.00 \times 10^{-3}$ & $0.61$ & $6.71 \times 10^{4}$ & $0.28$ & $9.88 \times 10^{4}$ &   & $13.2$ & $4$ & 3.3 \\ 
HE0143$-$0441 & $1.5$ & $2.00 \times 10^{-3}$ & $0.59$ & $4.78 \times 10^{4}$ & $0.31$ & $6.43 \times 10^{4}$ &   & $37.2$ & $7$ & 5.3 \\ 
HE0338$-$3945 & $1.5$ & $6.66 \times 10^{-4}$ & $0.49$ & $2.45 \times 10^{4}$ & $0.30$ & $2.88 \times 10^{4}$ &   & $183.7$ & $17$ & 10.8 \\ 
HE1105$+$0027 & $1.5$ & $2.00 \times 10^{-3}$ & $0.56$ & $3.40 \times 10^{4}$ & $0.32$ & $4.28 \times 10^{4}$ &   & $23.0$ & $4$ & 5.8 \\ 
HE1305$+$0007 & $1.5$ & $6.66 \times 10^{-4}$ & $0.54$ & $2.42 \times 10^{4}$ & $0.32$ & $2.97 \times 10^{4}$ &   & $56.0$ & $10$ & 5.6 \\ 
HE2148$-$1247 & $1.5$ & $2.00 \times 10^{-3}$ & $0.51$ & $2.44 \times 10^{4}$ & $0.31$ & $2.89 \times 10^{4}$ &   & $204.7$ & $11$ & 18.6 \\ 
HE2258$-$6358 & $1.5$ & $6.66 \times 10^{-4}$ & $0.59$ & $4.78 \times 10^{4}$ & $0.31$ & $6.62 \times 10^{4}$ &   & $293.5$ & $18$ & 16.3 \\ 
LP625$-$44 & $1.7$ & $4.00 \times 10^{-3}$ & $0.49$ & $2.34 \times 10^{4}$ & $0.34$ & $2.86 \times 10^{4}$ & $ 4.03\times 10^3 $ & $173.2$ & $17$ & 10.2 \\ 
SDSSJ0912$+$0216 & $1.5$ & $2.00 \times 10^{-3}$ & $0.46$ & $2.47 \times 10^{4}$ & $0.28$ & $2.87 \times 10^{4}$ &   & $425.0$ & $14$ & 30.4 \\ 
\hline
 & & & & & & & & & & \\ 
CEMP-$s/ur$ stars & & & & & & & & & & \\ 
\hline
BD$+$04$^{\circ}$2466 & $1.1$ & $2.00 \times 10^{-3}$ & $0.76$ & $3.19 \times 10^{3}$ & $0.09$ & $4.27 \times 10^{3}$ & $ 4.29\times 10^3 $ & $13.4$ & $7$ & 1.9 \\ 
HD13826 & $1.7$ & $0$ & $0.79$ & $3.90 \times 10^{3}$ & $0.07$ & $8.85 \times 10^{3}$ &   & $3.7$ & $3$ & 1.2 \\ 
HD187216 & $1.5$ & $5.00 \times 10^{-4}$ & $0.56$ & $1.71 \times 10^{4}$ & $0.32$ & $2.33 \times 10^{4}$ &   & $28.7$ & $4$ & 7.2 \\ 
HD201626 & $1.4$ & $4.00 \times 10^{-3}$ & $0.71$ & $1.34 \times 10^{5}$ & $0.18$ & $2.33 \times 10^{5}$ & $ 4.75\times 10^2 $ & $7.5$ & $5$ & 1.5 \\ 
HD5223 & $1.1$ & $1.50 \times 10^{-3}$ & $0.76$ & $1.79 \times 10^{4}$ & $0.17$ & $2.44 \times 10^{4}$ &   & $15.0$ & $9$ & 1.7 \\ 
HE0024$-$2523 & $0.9$ & $2.00 \times 10^{-3}$ & $0.81$ & $4.19 \times 10^{2}$ & $0.12$ & $4.75 \times 10^{2}$ & $ 3.33 $ & $35.4$ & $4$ & 8.8 \\ 
HE0212$-$0557 & $1.5$ & $2.00 \times 10^{-3}$ & $0.59$ & $4.78 \times 10^{4}$ & $0.31$ & $6.51 \times 10^{4}$ &   & $67.9$ & $5$ & 13.6 \\ 
HE0231$-$4016 & $0.9$ & $4.00 \times 10^{-3}$ & $0.79$ & $8.42 \times 10^{2}$ & $0.07$ & $8.81 \times 10^{2}$ &   & $1.4$ & $4$ & 0.4 \\ 
HE0430$-$4404 & $0.9$ & $4.00 \times 10^{-3}$ & $0.84$ & $5.88 \times 10^{2}$ & $0.09$ & $6.25 \times 10^{2}$ &   & $1.0$ & $2$ & 0.5 \\ 
HE1430$-$1123 & $1.5$ & $2.00 \times 10^{-3}$ & $0.76$ & $1.83 \times 10^{5}$ & $0.08$ & $3.97 \times 10^{5}$ &   & $4.6$ & $2$ & 2.3 \\ 
HE1434$-$1442 & $1.0$ & $3.00 \times 10^{-3}$ & $0.79$ & $2.06 \times 10^{5}$ & $0.05$ & $2.73 \times 10^{5}$ &   & $17.7$ & $3$ & 5.9 \\ 
HE2150$-$0825 & $0.9$ & $1.50 \times 10^{-3}$ & $0.84$ & $5.88 \times 10^{2}$ & $0.09$ & $6.25 \times 10^{2}$ &   & $1.1$ & $5$ & 0.2 \\ 
HE2232$-$0603 & $0.9$ & $1.00 \times 10^{-3}$ & $0.84$ & $5.88 \times 10^{2}$ & $0.09$ & $6.25 \times 10^{2}$ &   & $9.7$ & $4$ & 2.4 \\ 
SDSS0924$+$40 & $1.5$ & $2.00 \times 10^{-3}$ & $0.51$ & $2.44 \times 10^{4}$ & $0.31$ & $2.89 \times 10^{4}$ &   & $14.0$ & $2$ & 7.0 \\ 
SDSS2047$+$00 & $1.5$ & $6.66 \times 10^{-4}$ & $0.54$ & $1.36 \times 10^{5}$ & $0.07$ & $3.05 \times 10^{5}$ &   & $0.7$ & $3$ & 0.2 \\ 
\hline
\end{tabular}
\tablefoot{Masses are expressed in units of $\Msun$, periods in days.\\
}
\end{table*}

\begin{table*}[!ht]
\caption{Physical parameters of the model CEMP-$s$ stars with number of degrees of freedom $\nu\ge2$ computed with model set B  in which we adopt an efficient BHL model of wind mass transfer and efficient angular momentum loss.}
\label{tab:best-BoHo10-gamma2}
\centering
\begin{tabular}{ l  c c c c  c c c  c c c }
\hline
\hline
\hspace{1cm}ID & $\Mprim$ & $\Mpmz$ & $\Msec$ & $\Pin$ & $\Delta\Macc$ & $\Pf$ & $\Pf-$Paper~I & $\chimin$ & $\nu$ & $\chimin/\nu$ \\
\hline
 & & & & & & & & & & \\ 
CEMP-$s/nr$ stars & & & & & & & & & & \\ 
\hline
CS22880$-$074 & $1.1$ & $1.00 \times 10^{-3}$ & $0.79$ & $7.10 \times 10^{4}$ & $0.05$ & $5.42 \times 10^{4}$ &   & $16.6$ & $10$ & 1.7 \\ 
CS22942$-$019 & $1.8$ & $2.00 \times 10^{-4}$ & $0.56$ & $1.59 \times 10^{4}$ & $0.32$ & $3.48 \times 10^{3}$ & $ 3.23\times 10^3 $ & $21.0$ & $9$ & 2.3 \\ 
CS22964$-$161A & $1.6$ & $2.00 \times 10^{-3}$ & $0.79$ & $1.30 \times 10^{5}$ & $0.04$ & $1.56 \times 10^{5}$ & $ 1.30\times 10^5 $ & $19.6$ & $8$ & 2.5 \\ 
CS22964$-$161B & $1.6$ & $2.00 \times 10^{-3}$ & $0.71$ & $1.30 \times 10^{5}$ & $0.02$ & $1.56 \times 10^{5}$ & $ 1.30\times 10^5 $ & $11.8$ & $9$ & 1.3 \\ 
CS30301$-$015 & $2.4$ & $4.00 \times 10^{-3}$ & $0.79$ & $2.17 \times 10^{5}$ & $0.04$ & $4.09 \times 10^{4}$ &   & $11.0$ & $8$ & 1.4 \\ 
HD196944 & $1.2$ & $1.00 \times 10^{-3}$ & $0.79$ & $6.92 \times 10^{4}$ & $0.04$ & $4.63 \times 10^{4}$ &   & $13.6$ & $12$ & 1.1 \\ 
HD198269 & $1.5$ & $3.00 \times 10^{-3}$ & $0.79$ & $4.56 \times 10^{4}$ & $0.08$ & $2.29 \times 10^{4}$ & $ 1.17\times 10^3 $ & $9.4$ & $5$ & 1.9 \\ 
HE0202$-$2204 & $0.9$ & $4.00 \times 10^{-3}$ & $0.79$ & $1.68 \times 10^{3}$ & $0.07$ & $1.51 \times 10^{3}$ &   & $5.1$ & $7$ & 0.7 \\ 
HE1135$+$0139 & $0.9$ & $6.66 \times 10^{-4}$ & $0.81$ & $1.05 \times 10^{5}$ & $0.03$ & $9.47 \times 10^{4}$ &   & $3.2$ & $6$ & 0.5 \\ 
HE2158$-$0348 & $1.5$ & $3.00 \times 10^{-3}$ & $0.74$ & $2.31 \times 10^{4}$ & $0.15$ & $1.17 \times 10^{4}$ &   & $16.5$ & $7$ & 2.4 \\ 
\hline
 & & & & & & & & & & \\ 
CEMP-$s/r$ stars & & & & & & & & & & \\ 
\hline
CS22881$-$036 & $1.5$ & $6.66 \times 10^{-4}$ & $0.71$ & $2.93 \times 10^{3}$ & $0.14$ & $6.86 \times 10^{1}$ &   & $9.4$ & $6$ & 1.6 \\ 
CS22898$-$027 & $1.1$ & $1.00 \times 10^{-3}$ & $0.64$ & $3.30 \times 10^{3}$ & $0.22$ & $2.24 \times 10^{3}$ &   & $114.0$ & $11$ & 10.4 \\ 
CS22948$-$027 & $1.5$ & $2.00 \times 10^{-3}$ & $0.61$ & $5.98 \times 10^{3}$ & $0.25$ & $2.56 \times 10^{2}$ & $ 3.69\times 10^2 $ & $29.5$ & $9$ & 3.3 \\ 
CS29497$-$030 & $1.6$ & $2.00 \times 10^{-3}$ & $0.51$ & $1.19 \times 10^{4}$ & $0.30$ & $2.31 \times 10^{3}$ & $ 3.00\times 10^2 $ & $145.8$ & $16$ & 9.1 \\ 
CS29526$-$110 & $1.6$ & $2.00 \times 10^{-3}$ & $0.54$ & $1.19 \times 10^{4}$ & $0.30$ & $3.07 \times 10^{3}$ &   & $25.2$ & $6$ & 4.2 \\ 
CS31062$-$012 & $1.5$ & $2.00 \times 10^{-3}$ & $0.54$ & $8.60 \times 10^{3}$ & $0.23$ & $4.69 \times 10^{2}$ &   & $17.6$ & $8$ & 2.2 \\ 
CS31062$-$050 & $1.6$ & $2.00 \times 10^{-3}$ & $0.54$ & $1.19 \times 10^{4}$ & $0.30$ & $3.07 \times 10^{3}$ &   & $115.9$ & $20$ & 5.8 \\ 
HD187861 & $1.5$ & $2.00 \times 10^{-3}$ & $0.59$ & $1.20 \times 10^{4}$ & $0.26$ & $4.35 \times 10^{3}$ &   & $17.3$ & $8$ & 2.2 \\ 
HD224959 & $1.8$ & $4.00 \times 10^{-3}$ & $0.54$ & $1.60 \times 10^{4}$ & $0.31$ & $2.59 \times 10^{3}$ & $ 1.41\times 10^3 $ & $70.5$ & $8$ & 8.8 \\ 
HE0131$-$3953 & $1.5$ & $2.00 \times 10^{-3}$ & $0.61$ & $5.98 \times 10^{3}$ & $0.25$ & $2.59 \times 10^{2}$ &   & $13.6$ & $4$ & 3.4 \\ 
HE0143$-$0441 & $1.5$ & $2.00 \times 10^{-3}$ & $0.54$ & $1.21 \times 10^{4}$ & $0.24$ & $3.65 \times 10^{3}$ &   & $39.5$ & $7$ & 5.6 \\ 
HE0338$-$3945 & $1.1$ & $1.50 \times 10^{-3}$ & $0.59$ & $3.35 \times 10^{3}$ & $0.22$ & $2.05 \times 10^{3}$ &   & $211.8$ & $17$ & 12.5 \\ 
HE1105$+$0027 & $1.5$ & $2.00 \times 10^{-3}$ & $0.59$ & $1.20 \times 10^{4}$ & $0.26$ & $4.38 \times 10^{3}$ &   & $24.2$ & $4$ & 6.0 \\ 
HE1305$+$0007 & $1.1$ & $1.00 \times 10^{-3}$ & $0.66$ & $3.28 \times 10^{3}$ & $0.22$ & $2.30 \times 10^{3}$ &   & $64.2$ & $10$ & 6.4 \\ 
HE2148$-$1247 & $1.6$ & $2.00 \times 10^{-3}$ & $0.54$ & $1.19 \times 10^{4}$ & $0.30$ & $3.07 \times 10^{3}$ &   & $243.8$ & $11$ & 22.2 \\ 
HE2258$-$6358 & $1.5$ & $1.00 \times 10^{-3}$ & $0.59$ & $1.20 \times 10^{4}$ & $0.26$ & $4.28 \times 10^{3}$ &   & $301.9$ & $18$ & 16.8 \\ 
LP625$-$44 & $1.8$ & $4.00 \times 10^{-3}$ & $0.54$ & $1.60 \times 10^{4}$ & $0.31$ & $2.61 \times 10^{3}$ & $ 3.65\times 10^3 $ & $188.9$ & $17$ & 11.1 \\ 
SDSSJ0912$+$0216 & $1.5$ & $2.00 \times 10^{-3}$ & $0.54$ & $8.60 \times 10^{3}$ & $0.23$ & $4.69 \times 10^{2}$ &   & $454.2$ & $14$ & 32.4 \\ 
\hline
 & & & & & & & & & & \\ 
CEMP-$s/ur$ stars & & & & & & & & & & \\ 
\hline
BD$+$04$^{\circ}$2466 & $0.9$ & $2.00 \times 10^{-3}$ & $0.76$ & $8.49 \times 10^{2}$ & $0.12$ & $6.64 \times 10^{2}$ & $ 4.25\times 10^3 $ & $13.5$ & $7$ & 1.9 \\ 
HD13826 & $1.7$ & $0$ & $0.76$ & $8.78 \times 10^{4}$ & $0.05$ & $2.95 \times 10^{4}$ &   & $4.0$ & $3$ & 1.3 \\ 
HD187216 & $1.1$ & $2.00 \times 10^{-4}$ & $0.66$ & $3.28 \times 10^{3}$ & $0.22$ & $2.07 \times 10^{3}$ &   & $23.8$ & $4$ & 5.9 \\ 
HD201626 & $1.4$ & $4.00 \times 10^{-3}$ & $0.66$ & $1.71 \times 10^{4}$ & $0.16$ & $8.35 \times 10^{3}$ & $ 3.55\times 10^2 $ & $7.8$ & $5$ & 1.6 \\ 
HD5223 & $1.1$ & $1.50 \times 10^{-3}$ & $0.74$ & $1.81 \times 10^{4}$ & $0.12$ & $1.29 \times 10^{4}$ &   & $14.6$ & $9$ & 1.6 \\ 
HE0024$-$2523 & $0.9$ & $4.00 \times 10^{-3}$ & $0.76$ & $6.01 \times 10^{2}$ & $0.16$ & $5.23 \times 10^{2}$ & $ 3.48 $ & $37.0$ & $4$ & 9.3 \\ 
HE0212$-$0557 & $1.5$ & $2.00 \times 10^{-3}$ & $0.59$ & $1.20 \times 10^{4}$ & $0.26$ & $4.34 \times 10^{3}$ &   & $70.1$ & $5$ & 14.0 \\ 
HE0231$-$4016 & $0.9$ & $1.50 \times 10^{-3}$ & $0.76$ & $8.49 \times 10^{2}$ & $0.12$ & $6.80 \times 10^{2}$ &   & $1.1$ & $4$ & 0.3 \\ 
HE0430$-$4404 & $0.9$ & $3.00 \times 10^{-3}$ & $0.74$ & $8.55 \times 10^{2}$ & $0.11$ & $6.73 \times 10^{2}$ &   & $0.7$ & $2$ & 0.4 \\ 
HE1430$-$1123 & $1.0$ & $4.00 \times 10^{-3}$ & $0.81$ & $2.57 \times 10^{4}$ & $0.09$ & $2.31 \times 10^{4}$ &   & $4.5$ & $2$ & 2.2 \\ 
HE1434$-$1442 & $1.0$ & $3.00 \times 10^{-3}$ & $0.84$ & $7.19 \times 10^{4}$ & $0.05$ & $6.34 \times 10^{4}$ &   & $18.3$ & $3$ & 6.1 \\ 
HE2150$-$0825 & $0.9$ & $4.00 \times 10^{-3}$ & $0.79$ & $4.22 \times 10^{2}$ & $0.20$ & $4.08 \times 10^{2}$ &   & $0.8$ & $5$ & 0.2 \\ 
HE2232$-$0603 & $0.9$ & $6.66 \times 10^{-4}$ & $0.76$ & $6.01 \times 10^{2}$ & $0.16$ & $5.23 \times 10^{2}$ &   & $8.8$ & $4$ & 2.2 \\ 
SDSS0924$+$40 & $1.6$ & $2.00 \times 10^{-3}$ & $0.51$ & $1.19 \times 10^{4}$ & $0.30$ & $2.31 \times 10^{3}$ &   & $16.0$ & $2$ & 8.0 \\ 
SDSS2047$+$00 & $1.1$ & $1.50 \times 10^{-3}$ & $0.51$ & $3.85 \times 10^{4}$ & $0.05$ & $2.04 \times 10^{4}$ &   & $0.8$ & $3$ & 0.3 \\ 
\hline
\end{tabular}
\tablefoot{Masses are expressed in units of $\Msun$, periods in days.\\
}
\end{table*}

\begin{table*}[!ht]
\caption{Physical parameters of the model CEMP-$s$ stars with number of degrees of freedom $\nu\ge2$ computed with model set C in which we
adopt the WRLOF wind-accretion efficiency and efficient angular momentum loss.}
\label{tab:best-WRLOFq-gamma2}
\centering
\begin{tabular}{ l  c c c c  c c  c c c }
\hline
\hline
\hspace{1cm}ID & $\Mprim$ & $\Mpmz$ & $\Msec$ & $\Pin$ & $\Delta\Macc$ & $\Pf$ & $\chimin$ & $\nu$ & $\chimin/\nu$ \\
\hline
 & & & & & & & & & \\ 
CEMP-$s/nr$ stars & & & & & & & & & \\ 
\hline
CS22880$-$074 & $1.1$ & $1.00 \times 10^{-3}$ & $0.79$ & $6.33 \times 10^{3}$ & $0.09$ & $4.94 \times 10^{3}$ & $16.8$ & $10$ & 1.7 \\ 
CS22942$-$019 & $1.8$ & $2.00 \times 10^{-4}$ & $0.54$ & $6.38 \times 10^{4}$ & $0.29$ & $1.47 \times 10^{4}$ & $20.9$ & $9$ & 2.3 \\ 
CS22964$-$161A & $1.6$ & $2.00 \times 10^{-3}$ & $0.79$ & $2.60 \times 10^{6}$ & $0.04$ & $3.05 \times 10^{6}$ & $19.7$ & $8$ & 2.5 \\ 
CS22964$-$161B & $1.6$ & $2.00 \times 10^{-3}$ & $0.71$ & $2.60 \times 10^{6}$ & $0.03$ & $3.05 \times 10^{6}$ & $11.5$ & $9$ & 1.3 \\ 
CS30301$-$015 & $2.4$ & $2.00 \times 10^{-3}$ & $0.81$ & $1.37 \times 10^{4}$ & $0.05$ & $2.28 \times 10^{2}$ & $11.0$ & $8$ & 1.4 \\ 
HD196944 & $1.2$ & $1.00 \times 10^{-3}$ & $0.79$ & $8.71 \times 10^{3}$ & $0.07$ & $5.89 \times 10^{3}$ & $13.9$ & $12$ & 1.2 \\ 
HD198269 & $1.0$ & $2.00 \times 10^{-3}$ & $0.84$ & $2.03 \times 10^{5}$ & $0.07$ & $1.77 \times 10^{5}$ & $9.2$ & $5$ & 1.8 \\ 
HE0202$-$2204 & $0.9$ & $2.00 \times 10^{-3}$ & $0.81$ & $1.67 \times 10^{3}$ & $0.07$ & $1.51 \times 10^{3}$ & $5.1$ & $7$ & 0.7 \\ 
HE1135$+$0139 & $0.9$ & $1.00 \times 10^{-3}$ & $0.79$ & $4.74 \times 10^{3}$ & $0.09$ & $4.43 \times 10^{3}$ & $2.9$ & $6$ & 0.5 \\ 
HE2158$-$0348 & $1.5$ & $3.00 \times 10^{-3}$ & $0.71$ & $3.69 \times 10^{5}$ & $0.15$ & $1.76 \times 10^{5}$ & $15.4$ & $7$ & 2.2 \\ 
\hline
 & & & & & & & & & \\ 
CEMP-$s/r$ stars & & & & & & & & & \\ 
\hline
CS22881$-$036 & $1.5$ & $6.66 \times 10^{-4}$ & $0.71$ & $3.69 \times 10^{5}$ & $0.15$ & $1.77 \times 10^{5}$ & $10.1$ & $6$ & 1.7 \\ 
CS22898$-$027 & $1.5$ & $6.66 \times 10^{-4}$ & $0.54$ & $4.84 \times 10^{4}$ & $0.31$ & $1.75 \times 10^{4}$ & $101.6$ & $11$ & 9.2 \\ 
CS22948$-$027 & $1.5$ & $2.00 \times 10^{-3}$ & $0.64$ & $1.88 \times 10^{5}$ & $0.24$ & $8.29 \times 10^{4}$ & $29.6$ & $9$ & 3.3 \\ 
CS29497$-$030 & $1.5$ & $2.00 \times 10^{-3}$ & $0.54$ & $4.84 \times 10^{4}$ & $0.31$ & $1.75 \times 10^{4}$ & $129.6$ & $16$ & 8.1 \\ 
CS29526$-$110 & $1.5$ & $2.00 \times 10^{-3}$ & $0.54$ & $4.84 \times 10^{4}$ & $0.31$ & $1.75 \times 10^{4}$ & $22.5$ & $6$ & 3.8 \\ 
CS31062$-$012 & $1.5$ & $2.00 \times 10^{-3}$ & $0.59$ & $1.35 \times 10^{5}$ & $0.26$ & $5.39 \times 10^{4}$ & $17.5$ & $8$ & 2.2 \\ 
CS31062$-$050 & $1.5$ & $2.00 \times 10^{-3}$ & $0.54$ & $4.84 \times 10^{4}$ & $0.31$ & $1.75 \times 10^{4}$ & $98.3$ & $20$ & 4.9 \\ 
HD187861 & $1.5$ & $2.00 \times 10^{-3}$ & $0.56$ & $9.59 \times 10^{4}$ & $0.28$ & $3.63 \times 10^{4}$ & $16.6$ & $8$ & 2.1 \\ 
HD224959 & $1.5$ & $2.00 \times 10^{-3}$ & $0.54$ & $4.84 \times 10^{4}$ & $0.31$ & $1.74 \times 10^{4}$ & $65.6$ & $8$ & 8.2 \\ 
HE0131$-$3953 & $1.5$ & $2.00 \times 10^{-3}$ & $0.61$ & $1.34 \times 10^{5}$ & $0.27$ & $5.79 \times 10^{4}$ & $13.3$ & $4$ & 3.3 \\ 
HE0143$-$0441 & $1.5$ & $2.00 \times 10^{-3}$ & $0.56$ & $9.59 \times 10^{4}$ & $0.28$ & $3.65 \times 10^{4}$ & $37.2$ & $7$ & 5.3 \\ 
HE0338$-$3945 & $1.5$ & $6.66 \times 10^{-4}$ & $0.54$ & $4.84 \times 10^{4}$ & $0.31$ & $1.75 \times 10^{4}$ & $188.6$ & $17$ & 11.1 \\ 
HE1105$+$0027 & $1.5$ & $2.00 \times 10^{-3}$ & $0.56$ & $6.79 \times 10^{4}$ & $0.31$ & $2.66 \times 10^{4}$ & $23.0$ & $4$ & 5.8 \\ 
HE1305$+$0007 & $1.5$ & $6.66 \times 10^{-4}$ & $0.54$ & $4.84 \times 10^{4}$ & $0.31$ & $1.73 \times 10^{4}$ & $56.7$ & $10$ & 5.7 \\ 
HE2148$-$1247 & $1.5$ & $2.00 \times 10^{-3}$ & $0.56$ & $4.81 \times 10^{4}$ & $0.32$ & $1.92 \times 10^{4}$ & $208.9$ & $11$ & 19.0 \\ 
HE2258$-$6358 & $1.5$ & $6.66 \times 10^{-4}$ & $0.59$ & $6.75 \times 10^{4}$ & $0.32$ & $2.81 \times 10^{4}$ & $293.4$ & $18$ & 16.3 \\ 
LP625$-$44 & $1.7$ & $4.00 \times 10^{-3}$ & $0.54$ & $6.52 \times 10^{4}$ & $0.35$ & $1.86 \times 10^{4}$ & $177.6$ & $17$ & 10.4 \\ 
SDSSJ0912$+$0216 & $1.5$ & $2.00 \times 10^{-3}$ & $0.49$ & $6.92 \times 10^{4}$ & $0.27$ & $2.03 \times 10^{4}$ & $429.1$ & $14$ & 30.7 \\ 
\hline
 & & & & & & & & & \\ 
CEMP-$s/ur$ stars & & & & & & & & & \\ 
\hline
BD$+$04$^{\circ}$2466 & $1.1$ & $1.50 \times 10^{-3}$ & $0.74$ & $4.54 \times 10^{3}$ & $0.10$ & $3.23 \times 10^{3}$ & $13.2$ & $7$ & 1.9 \\ 
HD13826 & $1.7$ & $0$ & $0.76$ & $1.11 \times 10^{4}$ & $0.06$ & $3.38 \times 10^{3}$ & $3.8$ & $3$ & 1.3 \\ 
HD187216 & $1.5$ & $5.00 \times 10^{-4}$ & $0.56$ & $4.81 \times 10^{4}$ & $0.32$ & $1.70 \times 10^{4}$ & $28.7$ & $4$ & 7.2 \\ 
HD201626 & $1.4$ & $4.00 \times 10^{-3}$ & $0.71$ & $2.67 \times 10^{5}$ & $0.18$ & $1.49 \times 10^{5}$ & $7.5$ & $5$ & 1.5 \\ 
HD5223 & $1.1$ & $1.50 \times 10^{-3}$ & $0.76$ & $2.54 \times 10^{4}$ & $0.16$ & $1.99 \times 10^{4}$ & $15.1$ & $9$ & 1.7 \\ 
HE0024$-$2523 & $1.1$ & $0$ & $0.69$ & $7.29 \times 10^{4}$ & $0.14$ & $5.39 \times 10^{4}$ & $47.7$ & $4$ & 11.9 \\ 
HE0212$-$0557 & $1.5$ & $2.00 \times 10^{-3}$ & $0.59$ & $9.53 \times 10^{4}$ & $0.29$ & $3.91 \times 10^{4}$ & $67.9$ & $5$ & 13.6 \\ 
HE0231$-$4016 & $0.9$ & $1.50 \times 10^{-3}$ & $0.79$ & $8.42 \times 10^{2}$ & $0.08$ & $6.62 \times 10^{2}$ & $1.1$ & $4$ & 0.3 \\ 
HE0430$-$4404 & $0.9$ & $4.00 \times 10^{-3}$ & $0.79$ & $8.42 \times 10^{2}$ & $0.08$ & $6.62 \times 10^{2}$ & $0.7$ & $2$ & 0.3 \\ 
HE1430$-$1123 & $1.5$ & $2.00 \times 10^{-3}$ & $0.74$ & $5.18 \times 10^{5}$ & $0.09$ & $2.45 \times 10^{5}$ & $4.5$ & $2$ & 2.2 \\ 
HE1434$-$1442 & $1.0$ & $4.00 \times 10^{-3}$ & $0.79$ & $2.90 \times 10^{5}$ & $0.05$ & $2.46 \times 10^{5}$ & $17.8$ & $3$ & 5.9 \\ 
HE2150$-$0825 & $0.9$ & $3.00 \times 10^{-3}$ & $0.84$ & $4.16 \times 10^{2}$ & $0.14$ & $3.95 \times 10^{2}$ & $2.2$ & $5$ & 0.4 \\ 
HE2232$-$0603 & $0.9$ & $6.66 \times 10^{-4}$ & $0.84$ & $5.88 \times 10^{2}$ & $0.11$ & $5.20 \times 10^{2}$ & $10.1$ & $4$ & 2.5 \\ 
SDSS0924$+$40 & $1.5$ & $2.00 \times 10^{-3}$ & $0.54$ & $4.84 \times 10^{4}$ & $0.31$ & $1.75 \times 10^{4}$ & $14.4$ & $2$ & 7.2 \\ 
SDSS2047$+$00 & $1.5$ & $6.66 \times 10^{-4}$ & $0.61$ & $5.33 \times 10^{5}$ & $0.08$ & $1.90 \times 10^{5}$ & $0.7$ & $3$ & 0.2 \\ 
\hline
\end{tabular}
\tablefoot{Masses are expressed in units of $\Msun$, periods in days.\\
}
\end{table*}

%
\begin{table*}[!ht]
\caption{Physical parameters of the model CEMP-$s$ stars with number of degrees of freedom $\nu\le1$ computed with model set A.}
\label{tab:few-WRLOFq-ssw}
\centering
\begin{tabular}{ l  c c c c  c c c  c c c }
\hline
\hline
\hspace{1cm}ID & $\Mprim$ & $\Mpmz$ & $\Msec$ & $\Pin$ & $\Delta\Macc$ & $\Pf$ & $\Pf-$Paper~I & $\chimin$ & $\nu$ & $\chimin/\nu$ \\
\hline
 & & & & & & & & & & \\ 
CEMP-$s/r$ stars & & & & & & & & & & \\ 
\hline
BS16080$-$175 & $2.9$ & $4.00 \times 10^{-3}$ & $0.59$ & $9.29 \times 10^{3}$ & $0.26$ & $2.55 \times 10^{4}$ &   & $20.7$ & $1$ & 20.7 \\ 
BS17436$-$058 & $1.1$ & $4.00 \times 10^{-3}$ & $0.74$ & $3.61 \times 10^{4}$ & $0.15$ & $4.73 \times 10^{4}$ &   & $18.4$ & $1$ & 18.4 \\ 
\hline
 & & & & & & & & & & \\ 
CEMP-$s/ur$ stars & & & & & & & & & & \\ 
\hline
CS22945$-$017 & $2.0$ & $2.00 \times 10^{-4}$ & $0.74$ & $1.66 \times 10^{5}$ & $0.13$ & $4.50 \times 10^{5}$ &   & $3.7$ & $-1$ &    \\ 
CS22956$-$028 & $1.3$ & $0$ & $0.64$ & $1.25 \times 10^{4}$ & $0.20$ & $1.77 \times 10^{4}$ & $ 1.37\times 10^3 $ & $14.8$ & $1$ & 14.8 \\ 
CS29509$-$027 & $2.9$ & $1.50 \times 10^{-3}$ & $0.69$ & $3.25 \times 10^{3}$ & $0.11$ & $5.28 \times 10^{3}$ & $ 1.65\times 10^2 $ & $0.8$ & $-1$ &    \\ 
CS30338$-$089 & $1.5$ & $6.66 \times 10^{-4}$ & $0.59$ & $4.78 \times 10^{4}$ & $0.31$ & $6.53 \times 10^{4}$ &   & $6.1$ & $-1$ &    \\ 
HE0012$-$1441 & $2.6$ & $4.00 \times 10^{-3}$ & $0.76$ & $4.74 \times 10^{3}$ & $0.06$ & $1.58 \times 10^{4}$ &   & $0.2$ & $-1$ &    \\ 
HE0206$-$1916 & $1.5$ & $6.66 \times 10^{-4}$ & $0.64$ & $9.42 \times 10^{4}$ & $0.24$ & $1.57 \times 10^{5}$ &   & $2.5$ & $-1$ &    \\ 
HE0322$-$1504 & $1.8$ & $4.00 \times 10^{-3}$ & $0.49$ & $2.29 \times 10^{4}$ & $0.36$ & $3.13 \times 10^{4}$ &   & $1.7$ & $-1$ &    \\ 
HE0441$-$0652 & $1.2$ & $3.00 \times 10^{-3}$ & $0.79$ & $1.55 \times 10^{3}$ & $0.05$ & $1.90 \times 10^{2}$ &   & $0.2$ & $-1$ &    \\ 
HE0507$-$1430 & $6.0$ & $0$ & $0.56$ & $6.77 \times 10^{3}$ & $0.26$ & $3.83 \times 10^{4}$ & $ 5.16\times 10^2 $ & $0.9$ & $-1$ &    \\ 
HE0534$-$4548 & $2.3$ & $1.00 \times 10^{-3}$ & $0.81$ & $2.20 \times 10^{5}$ & $0.02$ & $9.10 \times 10^{5}$ &   & $0.0$ & $-2$ &    \\ 
HE1045$-$1434 & $6.0$ & $0$ & $0.49$ & $1.36 \times 10^{4}$ & $0.37$ & $5.56 \times 10^{4}$ &   & $17.1$ & $-1$ &    \\ 
HE1157$-$0518 & $1.5$ & $6.66 \times 10^{-4}$ & $0.59$ & $4.78 \times 10^{4}$ & $0.31$ & $6.55 \times 10^{4}$ &   & $1.7$ & $-1$ &    \\ 
HE1429$-$0551 & $1.3$ & $1.00 \times 10^{-3}$ & $0.71$ & $1.37 \times 10^{5}$ & $0.17$ & $2.29 \times 10^{5}$ &   & $0.5$ & $-1$ &    \\ 
HE1443$+$0113 & $1.7$ & $0$ & $0.74$ & $2.79 \times 10^{3}$ & $0.09$ & $4.67 \times 10^{3}$ &   & $0.1$ & $-1$ &    \\ 
HE1447$+$0102 & $1.7$ & $6.66 \times 10^{-4}$ & $0.49$ & $2.34 \times 10^{4}$ & $0.34$ & $2.94 \times 10^{4}$ &   & $73.7$ & $-1$ &    \\ 
HE1523$-$1155 & $1.7$ & $4.00 \times 10^{-3}$ & $0.76$ & $1.75 \times 10^{5}$ & $0.11$ & $4.35 \times 10^{5}$ &   & $1.2$ & $-1$ &    \\ 
HE1528$-$0409 & $1.5$ & $1.50 \times 10^{-3}$ & $0.54$ & $2.42 \times 10^{4}$ & $0.32$ & $2.99 \times 10^{4}$ &   & $12.2$ & $-1$ &    \\ 
HE2221$-$0453 & $1.8$ & $4.00 \times 10^{-3}$ & $0.74$ & $1.73 \times 10^{5}$ & $0.11$ & $5.33 \times 10^{5}$ &   & $1.3$ & $-1$ &    \\ 
HE2227$-$4044 & $1.0$ & $3.00 \times 10^{-3}$ & $0.81$ & $4.57 \times 10^{3}$ & $0.10$ & $5.55 \times 10^{3}$ &   & $0.3$ & $1$ & 0.3 \\ 
HE2228$-$0706 & $1.7$ & $2.00 \times 10^{-3}$ & $0.51$ & $2.33 \times 10^{4}$ & $0.36$ & $2.87 \times 10^{4}$ &   & $7.6$ & $-1$ &    \\ 
HE2240$-$0412 & $1.0$ & $3.00 \times 10^{-3}$ & $0.79$ & $6.50 \times 10^{3}$ & $0.09$ & $7.96 \times 10^{3}$ &   & $1.2$ & $-1$ &    \\ 
SDSS1707$+$58 & $1.8$ & $4.00 \times 10^{-3}$ & $0.49$ & $2.29 \times 10^{4}$ & $0.36$ & $2.87 \times 10^{4}$ &   & $22.3$ & $1$ & 22.3 \\ 
\hline
\end{tabular}
\tablefoot{Masses are expressed in units of $\Msun$, periods in days.
}
\end{table*}

\begin{table*}[!ht]
\caption{Physical parameters of the model CEMP-$s$ stars with number of degrees of freedom $\nu\le1$ computed with model set B.}
\label{tab:few-BoHo10-gamma2}
\centering
\begin{tabular}{ l  c c c c  c c c  c c c }
\hline
\hline
\hspace{1cm}ID & $\Mprim$ & $\Mpmz$ & $\Msec$ & $\Pin$ & $\Delta\Macc$ & $\Pf$ & $\Pf-$Paper~I & $\chimin$ & $\nu$ & $\chimin/\nu$ \\
\hline
 & & & & & & & & & & \\ 
CEMP-$s/r$ stars & & & & & & & & & & \\ 
\hline
BS16080$-$175 & $3.0$ & $0$ & $0.59$ & $3.64 \times 10^{4}$ & $0.23$ & $2.86 \times 10^{2}$ &   & $20.8$ & $1$ & 20.8 \\ 
BS17436$-$058 & $1.5$ & $1.50 \times 10^{-3}$ & $0.74$ & $2.91 \times 10^{3}$ & $0.15$ & $7.30 \times 10^{1}$ &   & $18.4$ & $1$ & 18.4 \\ 
\hline
 & & & & & & & & & & \\ 
CEMP-$s/ur$ stars & & & & & & & & & & \\ 
\hline
CS22945$-$017 & $2.0$ & $2.00 \times 10^{-4}$ & $0.71$ & $3.74 \times 10^{3}$ & $0.16$ & $3.65 \times 10^{1}$ &   & $3.4$ & $-1$ &    \\ 
CS22956$-$028 & $1.3$ & $0$ & $0.69$ & $1.23 \times 10^{4}$ & $0.19$ & $7.36 \times 10^{3}$ & $ 1.05\times 10^3 $ & $14.3$ & $1$ & 14.3 \\ 
CS29509$-$027 & $3.0$ & $0$ & $0.66$ & $6.41 \times 10^{3}$ & $0.15$ & $2.08 \times 10^{1}$ & $ 1.67\times 10^2 $ & $0.4$ & $-1$ &    \\ 
CS30338$-$089 & $1.1$ & $1.50 \times 10^{-3}$ & $0.66$ & $4.63 \times 10^{3}$ & $0.22$ & $3.35 \times 10^{3}$ &   & $5.8$ & $-1$ &    \\ 
HE0012$-$1441 & $2.6$ & $4.00 \times 10^{-3}$ & $0.76$ & $2.12 \times 10^{5}$ & $0.05$ & $3.38 \times 10^{4}$ &   & $0.3$ & $-1$ &    \\ 
HE0206$-$1916 & $1.1$ & $1.50 \times 10^{-3}$ & $0.74$ & $6.41 \times 10^{3}$ & $0.21$ & $5.29 \times 10^{3}$ &   & $1.9$ & $-1$ &    \\ 
HE0322$-$1504 & $1.8$ & $4.00 \times 10^{-3}$ & $0.54$ & $1.60 \times 10^{4}$ & $0.31$ & $2.40 \times 10^{3}$ &   & $2.3$ & $-1$ &    \\ 
HE0441$-$0652 & $1.5$ & $3.00 \times 10^{-3}$ & $0.79$ & $1.29 \times 10^{5}$ & $0.03$ & $6.07 \times 10^{4}$ &   & $0.1$ & $-1$ &    \\ 
HE0507$-$1430 & $6.0$ & $0$ & $0.59$ & $2.14 \times 10^{5}$ & $0.27$ & $2.51 \times 10^{2}$ & $ 3.58\times 10^2 $ & $1.0$ & $-1$ &    \\ 
HE0534$-$4548 & $2.8$ & $6.66 \times 10^{-4}$ & $0.81$ & $5.76 \times 10^{5}$ & $0.02$ & $8.47 \times 10^{4}$ &   & $0.0$ & $-2$ &    \\ 
HE1045$-$1434 & $6.0$ & $0$ & $0.56$ & $3.02 \times 10^{5}$ & $0.29$ & $2.61 \times 10^{3}$ &   & $20.9$ & $-1$ &    \\ 
HE1157$-$0518 & $1.1$ & $1.50 \times 10^{-3}$ & $0.66$ & $3.28 \times 10^{3}$ & $0.22$ & $2.30 \times 10^{3}$ &   & $1.7$ & $-1$ &    \\ 
HE1429$-$0551 & $1.3$ & $1.00 \times 10^{-3}$ & $0.66$ & $1.24 \times 10^{4}$ & $0.18$ & $6.78 \times 10^{3}$ &   & $0.5$ & $-1$ &    \\ 
HE1443$+$0113 & $1.3$ & $6.66 \times 10^{-4}$ & $0.74$ & $2.42 \times 10^{4}$ & $0.11$ & $1.45 \times 10^{4}$ &   & $0.3$ & $-1$ &    \\ 
HE1447$+$0102 & $1.2$ & $2.00 \times 10^{-3}$ & $0.61$ & $3.24 \times 10^{3}$ & $0.35$ & $1.39 \times 10^{3}$ &   & $79.1$ & $-1$ &    \\ 
HE1523$-$1155 & $1.7$ & $4.00 \times 10^{-3}$ & $0.76$ & $2.78 \times 10^{3}$ & $0.12$ & $3.76 \times 10^{1}$ &   & $1.2$ & $-1$ &    \\ 
HE1528$-$0409 & $1.6$ & $1.00 \times 10^{-3}$ & $0.54$ & $1.19 \times 10^{4}$ & $0.30$ & $3.01 \times 10^{3}$ &   & $13.1$ & $-1$ &    \\ 
HE2221$-$0453 & $1.8$ & $4.00 \times 10^{-3}$ & $0.74$ & $4.33 \times 10^{4}$ & $0.12$ & $1.34 \times 10^{4}$ &   & $1.3$ & $-1$ &    \\ 
HE2227$-$4044 & $1.0$ & $1.50 \times 10^{-3}$ & $0.81$ & $7.24 \times 10^{4}$ & $0.05$ & $6.26 \times 10^{4}$ &   & $0.2$ & $1$ & 0.2 \\ 
HE2228$-$0706 & $1.7$ & $3.00 \times 10^{-3}$ & $0.59$ & $1.62 \times 10^{4}$ & $0.26$ & $4.49 \times 10^{3}$ &   & $9.7$ & $-1$ &    \\ 
HE2240$-$0412 & $1.5$ & $2.00 \times 10^{-3}$ & $0.74$ & $2.06 \times 10^{3}$ & $0.04$ & $3.67 \times 10^{1}$ &   & $0.8$ & $-1$ &    \\ 
SDSS1707$+$58 & $1.8$ & $4.00 \times 10^{-3}$ & $0.54$ & $1.60 \times 10^{4}$ & $0.31$ & $2.61 \times 10^{3}$ &   & $24.9$ & $1$ & 24.9 \\ 
\hline
\end{tabular}
\tablefoot{Masses are expressed in units of $\Msun$, periods in days.
}
\end{table*}

\begin{table*}[!ht]
\caption{Physical parameters of model CEMP-$s$ with number of degrees of freedom $\nu\le1$ computed with model set C}
\label{tab:few-WRLOFq-gamma2}
\centering
\begin{tabular}{ l  c c c c  c c  c c c }
\hline
\hline
\hspace{1cm}ID & $\Mprim$ & $\Mpmz$ & $\Msec$ & $\Pin$ & $\Delta\Macc$ & $\Pf$ & $\chimin$ & $\nu$ & $\chimin/\nu$ \\
\hline
 & & & & & & & & & \\ 
CEMP-$s/r$ stars & & & & & & & & & \\ 
\hline
BS16080$-$175 & $2.9$ & $1.00 \times 10^{-3}$ & $0.56$ & $2.09 \times 10^{5}$ & $0.26$ & $1.57 \times 10^{4}$ & $20.7$ & $1$ & 20.7 \\ 
BS17436$-$058 & $1.1$ & $4.00 \times 10^{-3}$ & $0.71$ & $2.57 \times 10^{4}$ & $0.16$ & $2.00 \times 10^{4}$ & $18.4$ & $1$ & 18.4 \\ 
\hline
 & & & & & & & & & \\ 
CEMP-$s/ur$ stars & & & & & & & & & \\ 
\hline
CS22945$-$017 & $2.0$ & $2.00 \times 10^{-4}$ & $0.69$ & $2.98 \times 10^{4}$ & $0.15$ & $7.49 \times 10^{3}$ & $4.0$ & $-1$ &    \\ 
CS22956$-$028 & $1.3$ & $0$ & $0.64$ & $2.49 \times 10^{4}$ & $0.19$ & $1.37 \times 10^{4}$ & $14.8$ & $1$ & 14.8 \\ 
CS29509$-$027 & $3.0$ & $0$ & $0.69$ & $3.60 \times 10^{4}$ & $0.11$ & $5.09 \times 10^{2}$ & $0.5$ & $-1$ &    \\ 
CS30338$-$089 & $1.5$ & $6.66 \times 10^{-4}$ & $0.59$ & $9.53 \times 10^{4}$ & $0.29$ & $3.90 \times 10^{4}$ & $6.1$ & $-1$ &    \\ 
HE0012$-$1441 & $2.6$ & $4.00 \times 10^{-3}$ & $0.74$ & $1.89 \times 10^{4}$ & $0.07$ & $2.55 \times 10^{2}$ & $0.3$ & $-1$ &    \\ 
HE0206$-$1916 & $1.5$ & $1.00 \times 10^{-3}$ & $0.66$ & $2.35 \times 10^{4}$ & $0.18$ & $1.04 \times 10^{4}$ & $2.4$ & $-1$ &    \\ 
HE0322$-$1504 & $1.8$ & $4.00 \times 10^{-3}$ & $0.54$ & $9.01 \times 10^{4}$ & $0.36$ & $2.09 \times 10^{4}$ & $1.9$ & $-1$ &    \\ 
HE0441$-$0652 & $1.5$ & $3.00 \times 10^{-3}$ & $0.81$ & $2.86 \times 10^{3}$ & $0.04$ & $7.12 \times 10^{1}$ & $0.1$ & $-1$ &    \\ 
HE0507$-$1430 & $4.0$ & $0$ & $0.54$ & $1.02 \times 10^{6}$ & $0.32$ & $2.76 \times 10^{4}$ & $1.1$ & $-1$ &    \\ 
HE0534$-$4548 & $1.9$ & $6.66 \times 10^{-4}$ & $0.84$ & $1.32 \times 10^{6}$ & $0.02$ & $4.51 \times 10^{5}$ & $0.0$ & $-2$ &    \\ 
HE1045$-$1434 & $2.0$ & $4.00 \times 10^{-3}$ & $0.51$ & $1.23 \times 10^{5}$ & $0.35$ & $2.10 \times 10^{4}$ & $24.6$ & $-1$ &    \\ 
HE1157$-$0518 & $1.5$ & $6.66 \times 10^{-4}$ & $0.59$ & $6.75 \times 10^{4}$ & $0.32$ & $2.83 \times 10^{4}$ & $1.7$ & $-1$ &    \\ 
HE1429$-$0551 & $1.3$ & $1.00 \times 10^{-3}$ & $0.69$ & $2.75 \times 10^{5}$ & $0.15$ & $1.55 \times 10^{5}$ & $0.7$ & $-1$ &    \\ 
HE1443$+$0113 & $1.7$ & $0$ & $0.74$ & $7.01 \times 10^{5}$ & $0.09$ & $2.60 \times 10^{5}$ & $0.1$ & $-1$ &    \\ 
HE1447$+$0102 & $1.7$ & $6.66 \times 10^{-4}$ & $0.54$ & $6.52 \times 10^{4}$ & $0.35$ & $1.82 \times 10^{4}$ & $74.5$ & $-1$ &    \\ 
HE1523$-$1155 & $1.5$ & $4.00 \times 10^{-3}$ & $0.74$ & $3.67 \times 10^{5}$ & $0.15$ & $1.81 \times 10^{5}$ & $1.3$ & $-1$ &    \\ 
HE1528$-$0409 & $1.5$ & $1.50 \times 10^{-3}$ & $0.54$ & $4.84 \times 10^{4}$ & $0.31$ & $1.72 \times 10^{4}$ & $12.3$ & $-1$ &    \\ 
HE2221$-$0453 & $1.8$ & $4.00 \times 10^{-3}$ & $0.76$ & $6.84 \times 10^{5}$ & $0.12$ & $2.22 \times 10^{5}$ & $1.3$ & $-1$ &    \\ 
HE2227$-$4044 & $1.5$ & $1.00 \times 10^{-3}$ & $0.76$ & $5.78 \times 10^{3}$ & $0.06$ & $3.89 \times 10^{2}$ & $0.2$ & $1$ & 0.2 \\ 
HE2228$-$0706 & $1.7$ & $2.00 \times 10^{-3}$ & $0.54$ & $6.52 \times 10^{4}$ & $0.35$ & $1.85 \times 10^{4}$ & $7.8$ & $-1$ &    \\ 
HE2240$-$0412 & $1.5$ & $2.00 \times 10^{-3}$ & $0.76$ & $4.09 \times 10^{3}$ & $0.05$ & $1.51 \times 10^{2}$ & $1.0$ & $-1$ &    \\ 
SDSS1707$+$58 & $1.8$ & $4.00 \times 10^{-3}$ & $0.51$ & $9.06 \times 10^{4}$ & $0.34$ & $2.04 \times 10^{4}$ & $23.2$ & $1$ & 23.2 \\ 
\hline
\end{tabular}
\tablefoot{Masses are expressed in units of $\Msun$, periods in days.
}
\end{table*}
%
\begin{table*}[!ht]
\caption{Confidence ranges of the input parameters of the modelled CEMP-$s$ stars with $\chisq/\nu\le 3$ and $\nu\ge2$ computed with model set A  with WRLOF wind-accretion efficiency and spherically symmetric wind.}
\label{tab:confi_WRLOFq-ssw}
\centering
\begin{tabular}{  l | c c c | c c c | c c c | c c c  }
\hline
\hline
\hspace{0.8cm} ID & \multicolumn{3}{c|}{$\Mprim$} & \multicolumn{3}{c|}{$\Mpmz/10^{-3}$} &  \multicolumn{3}{c|}{$\Msec$} & \multicolumn{3}{c}{$\Pin$} \\ 
 & best & min & max & best & min & max & best & min & max & best & min & max \\
\hline
 & & & & & & & & & & & & \\ 
CEMP-$s/nr$ & & & & & & & & & & & & \\ 
\hline
CS22880$-$074 & $1.10$ & $0.85$ & $1.15$ & $1.00$ & $0.00$ & $4.00$ & $0.74$ & $0.73$ & $0.85$  & $4.54 \times 10^{3}$  & $6.64 \times 10^{2}$  & $1.64 \times 10^{4}$ \\
 &  & $1.45$  & $1.55$  &  &  &  &  &  &  &  & $1.56 \times 10^{5}$  & $3.47 \times 10^{5}$ \\ 
CS22942$-$019 & $1.80$ & $1.75$ & $1.85$ & $0.20$ & $0.00$ & $0.58$ & $0.54$ & $0.48$ & $0.68$  & $6.48 \times 10^{4}$  & $8.73 \times 10^{3}$  & $1.70 \times 10^{5}$ \\
CS22964$-$161A & $1.60$ & $0.95$ & $1.65$ & $2.00$ & $0.58$ & $4.00$ & $0.79$ & $0.75$ & $0.85$  & $2.74 \times 10^{5}$  & $2.18 \times 10^{5}$  & $4.87 \times 10^{5}$ \\
CS22964$-$161B & $1.60$ & $0.95$ & $1.65$ & $2.00$ & $0.58$ & $4.00$ & $0.71$ & $0.68$ & $0.85$  & $2.74 \times 10^{5}$  & $2.18 \times 10^{5}$  & $4.87 \times 10^{5}$ \\
CS30301$-$015 & $2.50$ & $1.95$ & $2.75$ & $4.00$ & $0.58$ & $4.00$ & $0.76$ & $0.70$ & $0.85$  & $5.14 \times 10^{3}$  & $2.90 \times 10^{3}$  & $6.47 \times 10^{3}$ \\
 &  & $1.15$  & $1.25$  &  &  &  &  &  &  &  & $1.30 \times 10^{5}$  & $2.27 \times 10^{5}$ \\ 
HD196944 & $1.40$ & $0.95$ & $1.65$ & $1.00$ & $0.30$ & $4.00$ & $0.81$ & $0.75$ & $0.88$  & $5.84 \times 10^{3}$  & $1.38 \times 10^{3}$  & $9.92 \times 10^{3}$ \\
HD198269 & $1.00$ & $0.95$ & $1.65$ & $4.00$ & $0.80$ & $4.00$ & $0.79$ & $0.73$ & $0.85$  & $1.64 \times 10^{3}$  & $1.31 \times 10^{3}$  & $2.05 \times 10^{5}$ \\
 &  & $1.95$  & $2.15$  &  &  &  &  &  &  &  &  & \\ 
HE0202$-$2204 & $0.90$ & $0.85$ & $1.15$ & $3.00$ & $0.00$ & $4.00$ & $0.79$ & $0.75$ & $0.88$  & $1.15 \times 10^{3}$  & $4.67 \times 10^{2}$  & $1.05 \times 10^{4}$ \\
 &  & $1.45$  & $1.55$  &  &  &  &  &  &  &  & $2.06 \times 10^{5}$  & $2.60 \times 10^{5}$ \\ 
HE1135$+$0139 & $0.90$ & $0.85$ & $1.95$ & $1.00$ & $0.00$ & $4.00$ & $0.79$ & $0.75$ & $0.88$  & $3.35 \times 10^{3}$  & $4.67 \times 10^{2}$  & $7.29 \times 10^{3}$ \\
HE2158$-$0348 & $1.50$ & $1.15$ & $1.65$ & $3.00$ & $1.75$ & $4.00$ & $0.69$ & $0.65$ & $0.75$  & $1.18 \times 10^{5}$  & $5.79 \times 10^{3}$  & $9.17 \times 10^{3}$ \\
 &  &  &  &  &  &  &  &  &  &  & $9.34 \times 10^{4}$  & $1.48 \times 10^{5}$ \\
\hline
 & & & & & & & & & & & & \\ 
CEMP-$s/r$ & & & & & & & & & & & & \\ 
\hline
CS22881$-$036 & $1.50$ & $1.45$ & $1.55$ & $0.67$ & $0.05$ & $2.50$ & $0.66$ & $0.65$ & $0.80$  & $1.35 \times 10^{5}$  & $6.25 \times 10^{3}$  & $9.90 \times 10^{3}$ \\
 &  & $1.05$  & $1.15$  &  &  &  &  &  &  &  & $1.38 \times 10^{4}$  & $1.71 \times 10^{5}$ \\ 
CS31062$-$012 & $1.50$ & $1.35$ & $1.65$ & $2.00$ & $0.58$ & $3.50$ & $0.56$ & $0.45$ & $0.65$  & $6.92 \times 10^{4}$  & $9.65 \times 10^{3}$  & $1.21 \times 10^{5}$ \\
HD187861 & $1.50$ & $1.35$ & $1.65$ & $2.00$ & $0.58$ & $3.50$ & $0.59$ & $0.50$ & $0.65$  & $4.90 \times 10^{4}$  & $9.59 \times 10^{3}$  & $1.21 \times 10^{5}$ \\
\hline
 & & & & & & & & & & & & \\ 
CEMP-$s/ur$ & & & & & & & & & & & & \\ 
\hline
BD$+$04$^{\circ}$2466 & $1.10$ & $0.85$ & $1.15$ & $2.00$ & $0.00$ & $4.00$ & $0.76$ & $0.73$ & $0.85$  & $3.19 \times 10^{3}$  & $5.70 \times 10^{2}$  & $5.27 \times 10^{3}$ \\
 &  & $1.45$  & $1.65$  &  &  &  &  &  &  &  & $9.90 \times 10^{3}$  & $2.37 \times 10^{5}$ \\ 
HD13826 & $1.70$ & $1.45$ & $1.85$ & $0.00$ & $0.00$ & $0.30$ & $0.79$ & $0.68$ & $0.88$  & $3.90 \times 10^{3}$  & $4.70 \times 10^{2}$  & $1.16 \times 10^{4}$ \\
 &  & $0.85$  & $0.95$  &  &  &  &  &  &  &  & $1.62 \times 10^{5}$  & $2.57 \times 10^{5}$ \\ 
HD201626 & $1.40$ & $1.05$ & $2.15$ & $4.00$ & $0.80$ & $4.00$ & $0.71$ & $0.60$ & $0.78$  & $1.18 \times 10^{5}$  & $2.31 \times 10^{3}$  & $1.63 \times 10^{5}$ \\
HD5223 & $1.10$ & $1.05$ & $1.15$ & $1.50$ & $0.30$ & $3.50$ & $0.76$ & $0.68$ & $0.83$  & $1.79 \times 10^{4}$  & $1.31 \times 10^{3}$  & $2.07 \times 10^{3}$ \\
 &  & $1.45$  & $1.55$  &  &  &  &  &  &  &  & $6.63 \times 10^{3}$  & $1.05 \times 10^{4}$ \\ 
 &  &  &  &  &  &  &  &  &  &  & $1.38 \times 10^{4}$  & $1.88 \times 10^{5}$ \\ 
HE0231$-$4016 & $0.90$ & $0.85$ & $1.25$ & $4.00$ & $0.05$ & $4.00$ & $0.79$ & $0.73$ & $0.85$  & $8.42 \times 10^{2}$  & $4.70 \times 10^{2}$  & $7.44 \times 10^{3}$ \\
 &  & $1.45$  & $1.65$  &  &  &  &  &  &  &  & $9.65 \times 10^{3}$  & $1.53 \times 10^{4}$ \\ 
 &  &  &  &  &  &  &  &  &  &  & $1.49 \times 10^{5}$  & $2.27 \times 10^{5}$ \\ 
HE0430$-$4404 & $0.90$ & $0.85$ & $1.15$ & $4.00$ & $0.00$ & $4.00$ & $0.84$ & $0.70$ & $0.88$  & $5.88 \times 10^{2}$  & $4.70 \times 10^{2}$  & $5.30 \times 10^{3}$ \\
 &  & $1.45$  & $1.55$  &  &  &  &  &  &  &  & $1.04 \times 10^{4}$  & $1.66 \times 10^{4}$ \\ 
 &  &  &  &  &  &  &  &  &  &  & $1.49 \times 10^{5}$  & $2.37 \times 10^{5}$ \\ 
HE1430$-$1123 & $1.50$ & $1.35$ & $1.65$ & $2.00$ & $0.05$ & $4.00$ & $0.76$ & $0.68$ & $0.85$  & $1.83 \times 10^{5}$  & $7.26 \times 10^{3}$  & $2.05 \times 10^{5}$ \\
 &  & $0.85$  & $1.25$  &  &  &  &  &  &  &  & $1.24 \times 10^{3}$  & $3.17 \times 10^{3}$ \\ 
HE2150$-$0825 & $0.90$ & $0.85$ & $1.15$ & $1.50$ & $0.05$ & $4.00$ & $0.84$ & $0.73$ & $0.85$  & $5.88 \times 10^{2}$  & $3.33 \times 10^{2}$  & $3.75 \times 10^{3}$ \\
 &  & $1.45$  & $1.55$  &  &  &  &  &  &  &  & $9.65 \times 10^{3}$  & $2.37 \times 10^{5}$ \\ 
HE2232$-$0603 & $0.90$ & $0.85$ & $1.15$ & $1.00$ & $0.00$ & $1.75$ & $0.84$ & $0.73$ & $0.85$  & $5.88 \times 10^{2}$  & $3.33 \times 10^{2}$  & $3.24 \times 10^{3}$ \\
SDSS2047$+$00 & $1.50$ & $0.85$ & $1.75$ & $0.67$ & $0.05$ & $4.00$ & $0.54$ & $0.43$ & $0.83$  & $1.36 \times 10^{5}$  & $3.33 \times 10^{2}$  & $5.28 \times 10^{2}$ \\
 &  &  &  &  &  &  &  &  &  &  & $9.82 \times 10^{2}$  & $2.41 \times 10^{5}$ \\ 
\hline
\end{tabular}
\tablefoot{Masses are expressed in units of $\Msun$, periods in days.\\
}
\end{table*}

\end{appendix}


\begin{thebibliography}{72}
\expandafter\ifx\csname natexlab\endcsname\relax\def\natexlab#1{#1}\fi

\bibitem[{{Abate} {et~al.}(2013){Abate}, {Pols}, {Izzard}, {Mohamed}, \& {de
  Mink}}]{Abate2013}
{Abate}, C., {Pols}, O.~R., {Izzard}, R.~G., {Mohamed}, S.~S., \& {de Mink},
  S.~E. 2013, \aap, 552, A26

\bibitem[{{Abate} {et~al.}(2015{\natexlab{a}}){Abate}, {Pols}, {Karakas}, \&
  {Izzard}}]{Abate2015-2}
{Abate}, C., {Pols}, O.~R., {Karakas}, A.~I., \& {Izzard}, R.~G.
  2015{\natexlab{a}}, \aap, 576, A118

\bibitem[{{Abate} {et~al.}(2015{\natexlab{b}}){Abate}, {Pols}, Stancliffe,
  {Izzard}, {Karakas}, Beers, \& Lee}]{Abate2015-3}
{Abate}, C., {Pols}, O.~R., Stancliffe, R.~J., {et~al.} 2015{\natexlab{b}},
  \aap, {\it in press}

\bibitem[{{Andrievsky} {et~al.}(2009){Andrievsky}, {Spite}, {Korotin}, {Spite},
  {Fran{\c c}ois}, {Bonifacio}, {Cayrel}, \& {Hill}}]{Andrievsky2009}
{Andrievsky}, S.~M., {Spite}, M., {Korotin}, S.~A., {et~al.} 2009, \aap, 494,
  1083

\bibitem[{{Aoki} {et~al.}(2007){Aoki}, {Beers}, {Christlieb}, {Norris}, {Ryan},
  \& {Tsangarides}}]{Aoki2007}
{Aoki}, W., {Beers}, T.~C., {Christlieb}, N., {et~al.} 2007, \apj, 655, 492

\bibitem[{{Arlandini} {et~al.}(1999){Arlandini}, {K{\"a}ppeler}, {Wisshak},
  {Gallino}, {Lugaro}, {Busso}, \& {Straniero}}]{Arlandini1999}
{Arlandini}, C., {K{\"a}ppeler}, F., {Wisshak}, K., {et~al.} 1999, \apj, 525,
  886

\bibitem[{{Asplund} {et~al.}(2009){Asplund}, {Grevesse}, {Sauval}, \&
  {Scott}}]{Asplund2009}
{Asplund}, M., {Grevesse}, N., {Sauval}, A.~J., \& {Scott}, P. 2009, \araa, 47,
  481

\bibitem[{{Bao} {et~al.}(2000){Bao}, {Beer}, {K{\"a}ppeler}, {Voss}, {Wisshak},
  \& {Rauscher}}]{Bao2000}
{Bao}, Z.~Y., {Beer}, H., {K{\"a}ppeler}, F., {et~al.} 2000, Atomic Data and
  Nuclear Data Tables, 76, 70

\bibitem[{{Beers} \& {Christlieb}(2005)}]{BeersChristlieb2005}
{Beers}, T.~C. \& {Christlieb}, N. 2005, \araa, 43, 531

\bibitem[{{Beers} {et~al.}(1992){Beers}, {Preston}, \&
  {Shectman}}]{BeersAJ1992}
{Beers}, T.~C., {Preston}, G.~W., \& {Shectman}, S.~A. 1992, \aj, 103, 1987

\bibitem[{{Bisterzo} {et~al.}(2011){Bisterzo}, {Gallino}, {Straniero},
  {Cristallo}, \& {K{\"a}ppeler}}]{Bisterzo2011}
{Bisterzo}, S., {Gallino}, R., {Straniero}, O., {Cristallo}, S., \&
  {K{\"a}ppeler}, F. 2011, \mnras, 418, 284

\bibitem[{{Bisterzo} {et~al.}(2012){Bisterzo}, {Gallino}, {Straniero},
  {Cristallo}, \& {K{\"a}ppeler}}]{Bisterzo2012}
{Bisterzo}, S., {Gallino}, R., {Straniero}, O., {Cristallo}, S., \&
  {K{\"a}ppeler}, F. 2012, \mnras, 422, 849

\bibitem[{{Boffin} \& {Jorissen}(1988)}]{BoffinJorissen1988}
{Boffin}, H.~M.~J. \& {Jorissen}, A. 1988, \aap, 205, 155

\bibitem[{{Boothroyd} \& {Sackmann}(1999)}]{Boothroyd1999}
{Boothroyd}, A.~I. \& {Sackmann}, I.-J. 1999, \apj, 510, 232

\bibitem[{{Boyer} {et~al.}(2015){Boyer}, {McQuinn}, {Barmby}, {Bonanos},
  {Gehrz}, {Gordon}, {Groenewegen}, {Lagadec}, {Lennon}, {Marengo}, {McDonald},
  {Meixner}, {Skillman}, {Sloan}, {Sonneborn}, {van Loon}, \&
  {Zijlstra}}]{Boyer2015II}
{Boyer}, M.~L., {McQuinn}, K.~B.~W., {Barmby}, P., {et~al.} 2015, \apj, 800, 51

\bibitem[{{Busso} {et~al.}(1999){Busso}, {Gallino}, \&
  {Wasserburg}}]{Busso1999}
{Busso}, M., {Gallino}, R., \& {Wasserburg}, G.~J. 1999, \araa, 37, 239

\bibitem[{{Busso} {et~al.}(1995){Busso}, {Lambert}, {Beglio}, {Gallino},
  {Raiteri}, \& {Smith}}]{Busso1995}
{Busso}, M., {Lambert}, D.~L., {Beglio}, L., {et~al.} 1995, \apj, 446, 775

\bibitem[{{Campbell}(2007)}]{Campbell2007}
{Campbell}, S.~W. 2007, PhD thesis, Monash University

\bibitem[{{Campbell} \& {Lattanzio}(2008)}]{Campbell2008}
{Campbell}, S.~W. \& {Lattanzio}, J.~C. 2008, \aap, 490, 769

\bibitem[{{Charbonnel} {et~al.}(1998){Charbonnel}, {Brown}, \&
  {Wallerstein}}]{Charbonnel1998}
{Charbonnel}, C., {Brown}, J.~A., \& {Wallerstein}, G. 1998, \aap, 332, 204

\bibitem[{{Christlieb} {et~al.}(2001){Christlieb}, {Green}, {Wisotzki}, \&
  {Reimers}}]{Christlieb2001}
{Christlieb}, N., {Green}, P.~J., {Wisotzki}, L., \& {Reimers}, D. 2001, \aap,
  375, 366

\bibitem[{{Cohen} {et~al.}(2003){Cohen}, {Christlieb}, {Qian}, \&
  {Wasserburg}}]{Cohen2003}
{Cohen}, J.~G., {Christlieb}, N., {Qian}, Y.-Z., \& {Wasserburg}, G.~J. 2003,
  \apj, 588, 1082

\bibitem[{{Cohen} {et~al.}(2005){Cohen}, {Shectman}, {Thompson}, {McWilliam},
  {Christlieb}, {Melendez}, {Zickgraf}, {Ram{\'{\i}}rez}, \&
  {Swenson}}]{Cohen2005}
{Cohen}, J.~G., {Shectman}, S., {Thompson}, I., {et~al.} 2005, \apjl, 633, L109

\bibitem[{{Constantino} {et~al.}(2014){Constantino}, {Campbell}, {Gil-Pons}, \&
  {Lattanzio}}]{Constantino2014}
{Constantino}, T., {Campbell}, S., {Gil-Pons}, P., \& {Lattanzio}, J. 2014,
  \apj, 784, 56

\bibitem[{{Fishlock} {et~al.}(2014){Fishlock}, {Karakas}, \&
  {Stancliffe}}]{Fishlock2014}
{Fishlock}, C.~K., {Karakas}, A.~I., \& {Stancliffe}, R.~J. 2014, \mnras, 438,
  1741

\bibitem[{{Frebel} {et~al.}(2006){Frebel}, {Christlieb}, {Norris}, {Beers},
  {Bessell}, {Rhee}, {Fechner}, {Marsteller}, {Rossi}, {Thom}, {Wisotzki}, \&
  {Reimers}}]{Frebel2006}
{Frebel}, A., {Christlieb}, N., {Norris}, J.~E., {et~al.} 2006, \apj, 652, 1585

\bibitem[{{Gallino} {et~al.}(1998){Gallino}, {Arlandini}, {Busso}, {Lugaro},
  {Travaglio}, {Straniero}, {Chieffi}, \& {Limongi}}]{Gallino1998}
{Gallino}, R., {Arlandini}, C., {Busso}, M., {et~al.} 1998, \apj, 497, 388

\bibitem[{{Goriely} \& {Mowlavi}(2000)}]{Goriely2000}
{Goriely}, S. \& {Mowlavi}, N. 2000, \aap, 362, 599

\bibitem[{{Gorlova} {et~al.}(2012){Gorlova}, {Van Winckel}, \&
  {Jorissen}}]{Gorlova2012}
{Gorlova}, N., {Van Winckel}, H., \& {Jorissen}, A. 2012, Baltic Astronomy, 21,
  165

\bibitem[{{Herwig}(2005)}]{Herwig2005}
{Herwig}, F. 2005, \araa, 43, 435

\bibitem[{{Herwig} {et~al.}(2011){Herwig}, {Pignatari}, {Woodward}, {Porter},
  {Rockefeller}, {Fryer}, {Bennett}, \& {Hirschi}}]{Herwig2011}
{Herwig}, F., {Pignatari}, M., {Woodward}, P.~R., {et~al.} 2011, \apj, 727, 89

\bibitem[{{Hollowell} \& {Iben}(1988)}]{Hollowell1988}
{Hollowell}, D. \& {Iben}, Jr., I. 1988, \apjl, 333, L25

\bibitem[{{Iliadis} {et~al.}(2010){Iliadis}, {Longland}, {Champagne}, \&
  {Coc}}]{Iliadis2010}
{Iliadis}, C., {Longland}, R., {Champagne}, A.~E., \& {Coc}, A. 2010, Nuclear
  Physics A, 841, 323

\bibitem[{{Izzard} {et~al.}(2010){Izzard}, {Dermine}, \& {Church}}]{Izzard2010}
{Izzard}, R.~G., {Dermine}, T., \& {Church}, R.~P. 2010, \aap, 523, A10+

\bibitem[{{Izzard} {et~al.}(2006){Izzard}, {Dray}, {Karakas}, {Lugaro}, \&
  {Tout}}]{Izzard2006}
{Izzard}, R.~G., {Dray}, L.~M., {Karakas}, A.~I., {Lugaro}, M., \& {Tout},
  C.~A. 2006, \aap, 460, 565

\bibitem[{{Izzard} {et~al.}(2009){Izzard}, {Glebbeek}, {Stancliffe}, \&
  {Pols}}]{Izzard2009}
{Izzard}, R.~G., {Glebbeek}, E., {Stancliffe}, R.~J., \& {Pols}, O.~R. 2009,
  \aap, 508, 1359

\bibitem[{{Izzard} {et~al.}(2004){Izzard}, {Tout}, {Karakas}, \&
  {Pols}}]{Izzard2004}
{Izzard}, R.~G., {Tout}, C.~A., {Karakas}, A.~I., \& {Pols}, O.~R. 2004,
  \mnras, 350, 407

\bibitem[{{Jonsell} {et~al.}(2006){Jonsell}, {Barklem}, {Gustafsson},
  {Christlieb}, {Hill}, {Beers}, \& {Holmberg}}]{Jonsell2006}
{Jonsell}, K., {Barklem}, P.~S., {Gustafsson}, B., {et~al.} 2006, \aap, 451,
  651

\bibitem[{{Jorissen} {et~al.}(1998){Jorissen}, {Van Eck}, {Mayor}, \&
  {Udry}}]{Jorissen1998}
{Jorissen}, A., {Van Eck}, S., {Mayor}, M., \& {Udry}, S. 1998, \aap, 332, 877

\bibitem[{{Karakas} \& {Lattanzio}(2007)}]{Karakas2007}
{Karakas}, A. \& {Lattanzio}, J.~C. 2007, \pasa, 24, 103

\bibitem[{{Karakas}(2010)}]{Karakas2010}
{Karakas}, A.~I. 2010, \mnras, 403, 1413

\bibitem[{{Karakas} {et~al.}(2010){Karakas}, {Campbell}, \&
  {Stancliffe}}]{Karakas2010-1}
{Karakas}, A.~I., {Campbell}, S.~W., \& {Stancliffe}, R.~J. 2010, \apj, 713,
  374

\bibitem[{{Kashiv} {et~al.}(2010){Kashiv}, {Davis}, {Gallino}, {Cai}, {Lai},
  {Sutton}, \& {Clayton}}]{Kashiv2010}
{Kashiv}, Y., {Davis}, A.~M., {Gallino}, R., {et~al.} 2010, \apj, 713, 212

\bibitem[{{Kipper} \& {Jorgensen}(1994)}]{Kipper1994}
{Kipper}, T. \& {Jorgensen}, U.~G. 1994, \aap, 290, 148

\bibitem[{{Kobayashi} {et~al.}(2011){Kobayashi}, {Karakas}, \&
  {Umeda}}]{Kobayashi2011}
{Kobayashi}, C., {Karakas}, A.~I., \& {Umeda}, H. 2011, \mnras, 414, 3231

\bibitem[{{Lattanzio}(1991)}]{Lattanzio1991}
{Lattanzio}, J.~C. 1991, Meteoritics and Planetary Science, 47, 1998

\bibitem[{{Lee} {et~al.}(2013){Lee}, {Beers}, {Masseron}, {Plez}, {Rockosi},
  {Sobeck}, {Yanny}, {Lucatello}, {Sivarani}, {Placco}, \& {Carollo}}]{Lee2013}
{Lee}, Y.~S., {Beers}, T.~C., {Masseron}, T., {et~al.} 2013, \aj, 146, 132

\bibitem[{{Lucatello} {et~al.}(2006){Lucatello}, {Beers}, {Christlieb},
  {Barklem}, {Rossi}, {Marsteller}, {Sivarani}, \& {Lee}}]{Lucatello2006}
{Lucatello}, S., {Beers}, T.~C., {Christlieb}, N., {et~al.} 2006, \apjl, 652,
  L37

\bibitem[{{Lugaro} {et~al.}(2009){Lugaro}, {Campbell}, \& {de
  Mink}}]{Lugaro2009}
{Lugaro}, M., {Campbell}, S.~W., \& {de Mink}, S.~E. 2009, \pasa, 26, 322

\bibitem[{{Lugaro} {et~al.}(2012){Lugaro}, {Karakas}, {Stancliffe}, \&
  {Rijs}}]{Lugaro2012}
{Lugaro}, M., {Karakas}, A.~I., {Stancliffe}, R.~J., \& {Rijs}, C. 2012, \apj,
  47, 1998

\bibitem[{{Lugaro} {et~al.}(2014){Lugaro}, {Tagliente}, {Karakas}, {Milazzo},
  {K{\"a}ppeler}, {Davis}, \& {Savina}}]{Lugaro2014}
{Lugaro}, M., {Tagliente}, G., {Karakas}, A.~I., {et~al.} 2014, \apj, 780, 95

\bibitem[{{Marsteller} {et~al.}(2005){Marsteller}, {Beers}, {Rossi},
  {Christlieb}, {Bessell}, \& {Rhee}}]{Marsteller2005}
{Marsteller}, B., {Beers}, T.~C., {Rossi}, S., {et~al.} 2005, Nuclear Physics
  A, 758, 312

\bibitem[{{Masseron} {et~al.}(2010){Masseron}, {Johnson}, {Plez}, {van Eck},
  {Primas}, {Goriely}, \& {Jorissen}}]{Masseron2010}
{Masseron}, T., {Johnson}, J.~A., {Plez}, B., {et~al.} 2010, \aap, 509, A93

\bibitem[{{Milam} {et~al.}(2009){Milam}, {Woolf}, \& {Ziurys}}]{Milam2009}
{Milam}, S.~N., {Woolf}, N.~J., \& {Ziurys}, L.~M. 2009, \apj, 690, 837

\bibitem[{{Placco} {et~al.}(2013){Placco}, {Frebel}, {Beers}, {Karakas},
  {Kennedy}, {Rossi}, {Christlieb}, \& {Stancliffe}}]{Placco2013}
{Placco}, V.~M., {Frebel}, A., {Beers}, T.~C., {et~al.} 2013, \apj, 770, 104

\bibitem[{{Pols} {et~al.}(2012){Pols}, {Izzard}, {Stancliffe}, \&
  {Glebbeek}}]{Pols2012}
{Pols}, O.~R., {Izzard}, R.~G., {Stancliffe}, R.~J., \& {Glebbeek}, E. 2012,
  \aap, 547, A76

\bibitem[{{Preston} \& {Sneden}(2001)}]{Preston2001}
{Preston}, G.~W. \& {Sneden}, C. 2001, \aj, 122, 1545

\bibitem[{{Reifarth} {et~al.}(2014){Reifarth}, {Lederer}, \&
  {K{\"a}ppeler}}]{Reifarth2014}
{Reifarth}, R., {Lederer}, C., \& {K{\"a}ppeler}, F. 2014, Journal of Physics G
  Nuclear Physics, 41, 053101

\bibitem[{{Ryan} {et~al.}(2005){Ryan}, {Aoki}, {Norris}, \& {Beers}}]{Ryan2005}
{Ryan}, S.~G., {Aoki}, W., {Norris}, J.~E., \& {Beers}, T.~C. 2005, \apj, 635,
  349

\bibitem[{{Sneden} {et~al.}(2008){Sneden}, {Cowan}, \& {Gallino}}]{Sneden2008}
{Sneden}, C., {Cowan}, J.~J., \& {Gallino}, R. 2008, \araa, 46, 241

\bibitem[{{Stancliffe}(2010)}]{Stancliffe2010}
{Stancliffe}, R.~J. 2010, \mnras, 403, 505

\bibitem[{{Stancliffe} {et~al.}(2007){Stancliffe}, {Glebbeek}, {Izzard}, \&
  {Pols}}]{Stancliffe2007}
{Stancliffe}, R.~J., {Glebbeek}, E., {Izzard}, R.~G., \& {Pols}, O.~R. 2007,
  \aap, 464, L57

\bibitem[{{Starkenburg} {et~al.}(2014){Starkenburg}, {Shetrone}, {McConnachie},
  \& {Venn}}]{Starkenburg2014}
{Starkenburg}, E., {Shetrone}, M.~D., {McConnachie}, A.~W., \& {Venn}, K.~A.
  2014, \mnras, 441, 1217

\bibitem[{{Suda} {et~al.}(2008){Suda}, {Katsuta}, {Yamada}, {Suwa}, {Ishizuka},
  {Komiya}, {Sorai}, {Aikawa}, \& {Fujimoto}}]{Suda2008}
{Suda}, T., {Katsuta}, Y., {Yamada}, S., {et~al.} 2008, \pasj, 60, 1159

\bibitem[{{Suda} {et~al.}(2011){Suda}, {Yamada}, {Katsuta}, {Komiya},
  {Ishizuka}, {Aoki}, \& {Fujimoto}}]{Suda2011}
{Suda}, T., {Yamada}, S., {Katsuta}, Y., {et~al.} 2011, \mnras, 412, 843

\bibitem[{{Thompson} {et~al.}(2008){Thompson}, {Ivans}, {Bisterzo}, {Sneden},
  {Gallino}, {Vauclair}, {Burley}, {Shectman}, \& {Preston}}]{Thompson2008}
{Thompson}, I.~B., {Ivans}, I.~I., {Bisterzo}, S., {et~al.} 2008, \apj, 677,
  556

\bibitem[{{Travaglio} {et~al.}(2001){Travaglio}, {Gallino}, {Busso}, \&
  {Gratton}}]{Travaglio2001}
{Travaglio}, C., {Gallino}, R., {Busso}, M., \& {Gratton}, R. 2001, \apj, 549,
  346

\bibitem[{{Wallerstein} \& {Knapp}(1998)}]{Wallerstein1998}
{Wallerstein}, G. \& {Knapp}, G.~R. 1998, \araa, 36, 369

\bibitem[{{Wanajo} {et~al.}(2006){Wanajo}, {Nomoto}, {Iwamoto}, {Ishimaru}, \&
  {Beers}}]{Wanajo2006}
{Wanajo}, S., {Nomoto}, K., {Iwamoto}, N., {Ishimaru}, Y., \& {Beers}, T.~C.
  2006, \apj, 636, 842

\bibitem[{{Yanny} {et~al.}(2009){Yanny}, {Rockosi}, {Newberg}, {Knapp},
  {Adelman-McCarthy}, {Alcorn}, {Allam}, {Allende Prieto}, {An}, {Anderson},
  {Anderson}, {Bailer-Jones}, {Bastian}, {Beers}, {Bell}, {Belokurov},
  {Bizyaev}, {Blythe}, {Bochanski}, {Boroski}, {Brinchmann}, {Brinkmann},
  {Brewington}, {Carey}, {Cudworth}, {Evans}, {Evans}, {Gates}, {G{\"a}nsicke},
  {Gillespie}, {Gilmore}, {Nebot Gomez-Moran}, {Grebel}, {Greenwell}, {Gunn},
  {Jordan}, {Jordan}, {Harding}, {Harris}, {Hendry}, {Holder}, {Ivans},
  {Ivezi{\v c}}, {Jester}, {Johnson}, {Kent}, {Kleinman}, {Kniazev},
  {Krzesinski}, {Kron}, {Kuropatkin}, {Lebedeva}, {Lee}, {French Leger},
  {L{\'e}pine}, {Levine}, {Lin}, {Long}, {Loomis}, {Lupton}, {Malanushenko},
  {Malanushenko}, {Margon}, {Martinez-Delgado}, {McGehee}, {Monet}, {Morrison},
  {Munn}, {Neilsen}, {Nitta}, {Norris}, {Oravetz}, {Owen}, {Padmanabhan},
  {Pan}, {Peterson}, {Pier}, {Platson}, {Re Fiorentin}, {Richards}, {Rix},
  {Schlegel}, {Schneider}, {Schreiber}, {Schwope}, {Sibley}, {Simmons},
  {Snedden}, {Allyn Smith}, {Stark}, {Stauffer}, {Steinmetz}, {Stoughton},
  {SubbaRao}, {Szalay}, {Szkody}, {Thakar}, {Sivarani}, {Tucker}, {Uomoto},
  {Vanden Berk}, {Vidrih}, {Wadadekar}, {Watters}, {Wilhelm}, {Wyse}, {Yarger},
  \& {Zucker}}]{Yanny2009}
{Yanny}, B., {Rockosi}, C., {Newberg}, H.~J., {et~al.} 2009, \aj, 137, 4377

\bibitem[{{Yong} {et~al.}(2013){Yong}, {Norris}, {Bessell}, {Christlieb},
  {Asplund}, {Beers}, {Barklem}, {Frebel}, \& {Ryan}}]{Yong2013II}
{Yong}, D., {Norris}, J.~E., {Bessell}, M.~S., {et~al.} 2013, \apj, 762, 26

\bibitem[{{Zijlstra}(2004)}]{Zijlstra2004}
{Zijlstra}, A.~A. 2004, \mnras, 348, L23

\end{thebibliography}
\end{document}